\documentclass[letterpaper, 12pt]{article}

\usepackage{amsmath,amssymb,mathrsfs}
\usepackage{graphicx}
\usepackage{color}
\usepackage{ushort}
\usepackage{url}
\usepackage{lscape}
\usepackage{subcaption}
\usepackage{pdflscape}
\usepackage{bbm}
\usepackage{epstopdf}
\usepackage{amsthm}
\usepackage[onehalfspacing]{setspace}

\usepackage[natbibapa]{apacite}

\usepackage{enumerate,enumitem}
\usepackage{libertine}
\usepackage[T2A, T1]{fontenc}
\usepackage[utf8]{inputenc}
\usepackage[english]{babel}
\usepackage{soul}
\usepackage{tcolorbox}
\usepackage{xcolor,colortbl}
\usepackage{threeparttable,multirow}
\usepackage{xfrac}
\usepackage{threeparttable}
\usepackage{rotating}
\usepackage{tabularx,ragged2e,booktabs}
\usepackage{subfiles}
\usepackage{amsmath}
\usepackage{amssymb}
\usepackage[margin=1in]{geometry}
\usepackage{bm}
\usepackage[title]{appendix}
\usepackage{setspace}

\definecolor{red}{rgb}{0.9,0.2,0.6}

\DeclareMathOperator{\E}{\mathbb{E}}
\DeclareMathOperator{\argmax}{argmax}

\DeclareMathOperator{\conv}{conv}
\DeclareMathOperator{\Prob}{Pr}

\newtheorem{proposition}{Proposition}
\newtheorem{corollary}{Corollary}
\newtheorem{lemma}{Lemma}
\newtheorem{theorem}{Theorem}

\newtheorem{definition}{Definition}
\newtheorem{algorithm}{Algorithm}
\newtheorem*{theorem**}{Proposition \theoremnum}
\newenvironment{theorem*}[1][]{%
  \edef\theoremnum{\if\relax\detokenize{#1}\relax\else~#1\fi}% Store theorem number
  \begin{theorem**}
}{%
  \end{theorem**}
}

\definecolor{Blue}{RGB}{0,32,216}
\usepackage[colorlinks=true,allcolors=Blue]{hyperref}
\usepackage{xcolor}
\usepackage{rotating}
\title{Robust Pricing for Quality Disclosure\thanks{An earlier version of the paper circulated under the title ``Robust Advertisement Pricing.'' The authors are grateful to S. Nageeb Ali, Dirk Bergemann, Tilman B\"orgers, Roberto Corrao, Yingni Guo, Nima Haghpanah, Marina Halac, Deniz Kattwinkel, Anton Kolotilin,  Barton Lipman, Elliot Lipnowski, Xiao Lin, Qingmin Liu, Ryota Iijima, Ferdinand Pieroth, Daniel Rappoport, Doron Ravid,  Ludvig Sinander for their helpful discussions and comments. We also thank the participants of SEA 2023, SITE 2024, ACM EC 2024, the 35th Stony Brook conference, GAMES 2024, and seminars at Yale, UCL, UBC, Warwick, Helsinki, and Bonn.}}
\author{Tan Gan\thanks{Tan Gan: Department of Management, London School of Economics, London, WC2A 2AE, UK (email: \href{mailto:t.gan2@lse.ac.uk}{t.gan2@lse.ac.uk}).} \and Hongcheng Li\thanks{
        Hongcheng Li: Department of Economics, Yale University, New Haven, 06511, USA (email: \href{mailto:hongcheng.li@yale.edu}{hongcheng.li@yale.edu}).}}
\date{Jun 15, 2025}

\begin{document}
\onehalfspacing
\maketitle

\begin{abstract}
  \noindent A platform charges a producer for disclosing quality evidence to consumers before trade. It aims to maximize its revenue guarantee across potentially multiple equilibria which arise from the interdependence of producer purchase decisions and consumer beliefs. The platform's optimal pricing strategy entrenches itself as a market gatekeeper: it induces a unique equilibrium in which non-disclosed products’ perceived values are lower than the production cost. To achieve this goal, this pricing strategy iteratively destabilizes under-disclosure equilibria by luring producers to disclose slightly more. Higher-quality producers receive higher rents as their disclosure is prioritized. Despite losing rents, the platform optimally induces socially efficient information transmission for any given evidence structure, and it never benefits from garbling evidence. Compared to the non-robust benchmark, our framework generates more intuitive comparative statics: the platform’s ability to extract surplus increases with its value as an information intermediary.

  %A platform charges a producer for disclosing quality evidence to consumers before trade. Because producer purchase decisions and consumer beliefs are interdependent, platform pricing can yield multiple equilibria. To maximize its revenue guarantee, the platform’s optimal pricing entrenches itself as a market gatekeeper: it induces a unique equilibrium in which non-disclosed products’ perceived values are lower than the production cost. To achieve so, this pricing strategy iteratively destabilizes under-disclosure equilibria by luring producers to disclose slightly more. Moreover, such destabilization prioritizes guaranteeing the disclosure of higher-quality producers and gives them greater rents. Despite losing rents, the platform optimally induces socially efficient information transmission for any given evidence structure, and it never gains from garbling the evidence. Compared to the non-robust benchmark, our framework generates more intuitive comparative statics: the platform’s ability to extract surplus increases with its value as an information intermediary.

\end{abstract}



\newpage

\section{Introduction}

Disclosure services are an important form of information intermediation. Platforms such as search engines, e-commerce sites, and online freelance marketplaces disclose information about the quality of products and services by releasing consumer reviews, presenting historical data, and offering algorithmic recommendations. While improving match efficiency and facilitating trade, these disclosure services also give platforms the possibility to become market gatekeepers. In particular, the market may believe that only products disclosed on the platform are high quality, so any off-platform product is deemed low quality. This can allow the platform to extract much of the trade surplus.

%However, \emph{strategic uncertainty} poses a significant challenge to the platform's surplus extraction practices. This problem arises because the sender’s (i.e., producer's or employee's) choice of disclosure plan is discretionary and unobservable to receivers (i.e., consumers or employers), leading to interdependence between sender actions and receiver beliefs. Specifically, when disclosure does not occur, the receiver’s skepticism belief\footnote{We follow the terminology of \cite{milg1981} who refers to the situation without disclosure as skepticism.} about product value or employee ability depends on his expectations of the sender’s disclosure choices contingent on her quality. This skepticism, in turn, influences the sender’s payoff if she opts not to disclose, affecting her willingness to invest in disclosure contingent on her quality. The resulting interdependence often leads to multiple equilibria in the interaction between the sender and the receiver, making it nontrivial for the platform to achieve its desired outcome.

However, a platform's ability to extract surplus depends on market expectations. Consider a producer who may disclose her product’s quality on the platform, and a consumer who may buy this product. When disclosure does not occur, the consumer’s skepticism belief about the producer’s quality depends on his expectations of the producer’s (quality-contingent) disclosure choices. This skepticism, in turn, influences the producer’s payoff if she opts not to disclose on the platform, affecting her willingness to pay for disclosure given her quality. The resulting interdependence between producer actions and consumer beliefs can lead to multiple equilibria and pose a challenge to the platform.

%When does a platform pay additional rents to senders in order to address strategic uncertainty? What is the platform's robustly optimal pricing strategy? Will the platform optimally distort market efficiency by impeding information transmission? How does strategic uncertainty influence the division of surplus? And how does surplus division vary with market conditions? To address these questions, this paper studies optimal disclosure pricing for a platform that seeks to maximize its \emph{revenue guarantee} across all equilibria.

This paper studies optimal disclosure pricing for a platform that seeks to maximize its revenue guarantee across all equilibria. We solve for the platform’s robustly optimal pricing strategy that addresses strategic uncertainty and examine its implications. Does the platform distort market efficiency by impeding information transmission? How does strategic uncertainty influence the division of surplus? And how does the surplus division vary with market conditions?

In our model, a producer can purchase disclosure services from a platform before deciding whether to produce her product at a cost and sell it to a consumer. A signal about the product's quality (i.e., the product's quality type) is observed by both the platform and the producer, while the consumer holds a prior belief. The signal is contractible hard evidence that may come in the form of awards, test certificates, product reviews/ratings, etc. Hence, the platform can price disclosure based on both this hard evidence and the volume of consumers that the producer wishes to inform. In particular, the platform specifies for each possible signal/type realization a menu of disclosure plans, with each disclosure plan containing a price and a disclosure probability.

Given such a profile of type-dependent menus, a game between the producer and the consumer unfolds as follows. After the producer privately observes her quality type, she can either leave the platform (and thus no disclosure will happen) or pick one plan from the menu exclusive to her type. Crucially, the producer's choice is not observed by the consumer. If a plan is chosen, the platform charges the specified price and discloses the hard evidence with the specified probability. Upon disclosure, the consumer observes the signal. If no disclosure occurs, the consumer will update his belief based on his conjecture of the producer's contingent plan choices. Finally, the producer determines whether to produce the product at a type-independent cost and sell it to the consumer at a take-it-or-leave-it price equal to the consumer's posterior mean. 
%We deliberately consider a stylized disclosure pricing model that is equivalent to other models that match different applications, on which we will elaborate in Section \ref{applications}.
In Section \ref{applications}, we elaborate on how this stylized disclosure pricing model can be matched to various applications.

To delineate how the robustness concern drives platform behavior, we consider a benchmark where the platform selects its preferred equilibrium, as is standard in mechanism design. We show that such a platform can extract full surplus by inducing socially efficient production and trade, while leaving each producer type the surplus of zero. In the best equilibrium, the platform serves as a market gatekeeper: each efficient type (with product value higher than production cost) is disclosed, produced, and sold, whereas any non-disclosed type is inefficient and not worth producing.
This full-extraction pricing strategy is robust to strategic uncertainty \emph{if and only if} the consumer's prior valuation is lower than the production cost, as then trade cannot happen without the platform providing disclosure services. Moreover, Proposition \ref{proposition-benchmark} shows that if the producer and the consumer can trade and generate positive surplus by themselves (i.e., the prior mean is higher than the production cost), then any pricing policy of the platform that achieves close-to-full surplus extraction in the preferred equilibrium also admits a bad equilibrium where the platform’s revenue is close to zero. Such stark revenue disparity highlights the platform’s exposure to strategic uncertainty.

With these motivations, we solve for the robustly optimal menu profile that maximizes the platform's revenue in the worst-case equilibrium of the induced producer-consumer game. 
Our main result, Theorem \ref{main result}, characterizes how the platform achieves robust gatekeeping between producers and consumers. The platform adopts a sequential self-conquering strategy that iteratively guarantees disclosure probability from high producer types to low ones
%the disclosure probability of higher producer types
, which gradually brings down the producer's outside-option profit to zero. To provide intuition, we next describe the structure of optimal pricing.

To start with, the platform induces a \emph{self-conquering} process for the highest-type producer to lure her into full disclosure step by step. Specifically, the platform offers the highest type a plan specifying a positive disclosure probability $\epsilon$ at a sufficiently low marginal price, which destabilizes the undesirable equilibrium with zero disclosure. Building on this, the platform further introduces a second option with probability $2\varepsilon$ and a sufficiently low price premium over the first plan, which tempts the highest type to disclose more. Iteratively in this way, the platform increases the highest-type's purchase in small steps, continuously insulating against outcomes where she does not disclose with probability one.
%outcomes where her product is not disclosed a with positive probability
Crucially, this process creates negative externalities on the producer's outside-option: when the consumer observes no disclosure, yet knowing a high type is likely to disclose, he must attribute this observation to lower types concealing their information, and thus have lower willingness to pay for the product. As the highest type discloses more and more during the above process, her outside-option value
%outside option
keeps declining, progressively enhancing her willingness to pay for marginal disclosure. Consequently, the platform is able to charge a higher price premium at each step of the process.

Given that the highest type fully discloses, the platform then moves on to the second-highest type and triggers a similar self-conquering process to ensure her full disclosure. In fact, the platform finds it optimal to proceed like this in a \emph{sequential} manner down the type hierarchy, for all efficient types (with value higher than production cost). To break under-disclosure in each step, the optimal pricing leaves the producer with a rent equal to exactly her outside-option profit. Moreover, this procedure does not end until the producer's outside option becomes worthless, that is, when the consumer's valuation of a non-disclosed product is even lower than the production cost. At this tipping point, the producer's willingness to pay for disclosure no longer interacts with consumer beliefs, so strategic uncertainty ceases to be a problem. As a result, from then on, the platform can fully extract the rest of surplus with a single full-disclosure option, as in the benchmark where it ignores the robustness concern.

Several interesting implications arise from our result. To begin with, our model generates intuitive patterns regarding how the platform's market power varies with market conditions. From two perspectives, we demonstrate that the platform's surplus-extraction power changes monotonically with its value as an information intermediary. First, recall that if the production cost is higher than the prior mean of the product's value (so trade cannot happen absent the platform), the platform extracts full surplus. On the other hand, when the production cost is so low that the platform becomes dispensable in terms of surplus generation, the platform must leave positive rents to the producer. Proposition \ref{proposition-market power} generalizes this observation by showing that the platform's revenue as a share of total surplus increases with its marginal contribution to social efficiency. In addition, Proposition \ref{proposition-converge0} unveils a similar point from a distinct angle. It shows that as information asymmetry vanishes (the prior belief becomes concentrated on a single type), the platform's revenue converges to zero, though the total surplus remains positive. Intuitively, the platform's social value decreases as information becomes more symmetric, and so should its revenue. Both results here are in contrast with the benchmark where the platform can choose its most preferred equilibrium. In that case, present any degree of positive uncertainty, the platform always extracts full surplus regardless of market parameters. Moreover, the benchmark revenue moves discontinuously when the uncertainty becomes precisely zero: the platform now cannot gain any surplus.
%In that case, the platform always extracts full surplus regardless of market parameters. Moreover, the benchmark revenue is discontinuous: the platform gains no surplus without uncertainty, but can extract all surplus in the presence of any degree of uncertainty.

The next set of implications concerns the platform's informational role. We demonstrate that, along two distinct margins, the platform’s optimal approach to strategic uncertainty does not distort information transmission—even though it allows producers to retain some rents. First, for any given signal structure, the platform’s robustly optimal pricing does not generate distortions. This is because, in the unique induced equilibrium, the platform acts as a gatekeeper who discloses all producer types whose posterior value exceeds the production cost. All types not disclosed are inefficient and do not produce in equilibrium.\footnote{This contrasts with settings involving other economic frictions, such as adverse selection, which typically lead to inefficient communication.} Furthermore, Proposition \ref{proposition-comparative} shows that when the signal structure (that generates the producer's type) becomes more Blackwell-informative, robustly optimal pricing yields a higher revenue guarantee for the platform. This result implies that if the platform could design the signal structure of the hard evidence used in its disclosure service, it would have no incentive to distort the signal’s quality in pursuit of higher revenues. This insight is relevant in applications where the platform controls or generates hard evidence of product value—for example, Amazon’s extraction of keywords from product reviews or YouTube’s distribution of creator awards.

Lastly, we investigate the platform's allocative role. We show that, despite the absence of distortion in information transmission, the platform optimally gives higher-type producers higher rents. Monotonic rents arise because the sequential self-conquering process prioritizes inducing higher types' disclosure by compensating them for giving up the outside option, whose value keeps decreasing during the process.
%due to worsened consumer no-disclosure belief.
%This is a special feature of the optimal pricing design, because many suboptimal menu profiles lead to non-monotonic rents. This differs from a typical adverse-selection setting in which monotonicity always occurs due to incentive compatibility.
We also compare our solution with two other benchmarks where the platform is absent while the producer either controls or does not control disclosure. We show that the presence of the platform makes every producer type worse off than in both benchmarks.

We conclude with three extensions of our model. First, in the main model, the menus are deterministic. Here, we relax this assumption and consider random menus (in the flavor of \cite{halr2021}). We show that it is without loss to consider deterministic menus alone since every random menu is outcome-equivalent to a deterministic one. Nevertheless, random contracts prove to have practical value, as they provide an alternative implementation of our solution: instead of a deterministic menu with continuous options, the platform can offer only the full-disclosure option at a random price. Such implementation extends the interpretation of our model to applications such as certification agencies who can only control certification but not disclosure, so partial disclosure is not common. These agencies can carry out such random pricing by, for instance, distributing dispersed rate offers or discounts via private emails.
%\footnote{Type-dependent discounts have been widely practiced by online certification providers. To give a few examples, \href{https://www.comptia.org/blog/voucher-discount}{CompTIA} offers discounts to various types of customers, \href{https://www.pmi.org/membership}{PMI} distributes discount codes to members, etc.}
Second, we show that under the optimal menu profile, the unique equilibrium outcome is also rationalizable. Our solution thus remains robustly optimal even if the platform worries that market participants may fail to play Bayes Nash. Moreover, the last extension demonstrates that the main features of our characterization are preserved even if disclosure incurs costs for the platform. As a result, the producer faces strictly convex price functions even when the platform bears potentially non-convex disclosure costs.

%\paragraph{Copy Material} List prices together with targeted discounts that vary across buyers are common in applications. Even if this personalization is partly based on customer data, the allocation of discounts is often arbitrary to a significant extent. For instance, in multiplayer gacha games, the use of loot boxes and the so-called “pity system” essentially yield a maximum list price together with random discounts that accrue to only some of the players (see, e.g., Gan, 2023). As another example, the dating app Tinder was found to charge users different prices with much of the differences being unaccounted for by clear explanatory variables (see, e.g., European Commission, 2024). More broadly, numerous merchants use Groupon to provide discounts and build their customer base; selective discounts help them attract consumers via both word-of-mouth marketing and network externalities (see https://www.groupon.com/merchant/working-with-groupon/ merchant-success-stories).

\emph{Outline.} Section \ref{model} sets up the model. Section \ref{preliminary analysis} studies the benchmark in which the platform fails to account for strategic uncertainty. Section \ref{binary type example} illustrates with a binary-type example the optimality of using rich menus to induce self-conquering. Section \ref{optimal menu profile} presents our main result and discusses the implications and proof. Section \ref{extensions} analyzes the extensions. All proofs can be found in the appendix.

\paragraph{Literature Review}

This paper contributes to the growing literature on contracting with externalities that focuses on worst-case selection (or unique implementation).
Following the seminal works of 
\cite{sega2003} and \cite{wint2004}, respectively, one strand of this literature examines settings in which agents’ actions are
bilaterally contractible, while another strand studies moral hazard problems with unobservable actions\footnote{See, for examples, \cite{wint2004}, \cite{bewi2012}, \cite{hakw2020,halr2021,hakw2024}, \cite{camboni2022monitoring}, and \cite{CGP24}.}. Our setting is closer to the former one because our prices can depend on disclosure probabilities. For example, in \cite{sega2003}, the principal uses deterministic contracts to contract with multiple agents who have multiple yet finite available actions. He shows that every menu profile can be mapped to an alternating divide-and-conquer strategy, where agents delete low actions in alternation in the process of iterated deletion of dominated strategies. \cite{halr2025} consider a monopolist who prices goods with network externalities so that buyers' purchasing decisions are complementary. They show that an optimal policy offers personalized discounts to successively insulate against low-demand equilibria and posts a high price to extract revenue from the induced high demand.

We depart from this literature in key ways. In our setting, externalities among agents (types) are generated through the endogenous change of market beliefs, which in turn gives rise to two novel features: indirect and indefinite externalities.
Unlike in previous studies' models where agents' actions directly influence each other's payoffs, in our model, producer types only generate indirect externalities among each other by altering consumer beliefs across different equilibria. This distinct feature of externalities enables our platform to induce self-conquering, which never appears in previous research. For instance, in \cite{sega2003}'s action deletion process, consecutive deletions must involve different agents, rendering the problem trivial for a single-agent scenario. 
In contrast, our self-conquering process successively eliminates a continuum of actions for a single producer type. Thus, even a binary-type example (see Section \ref{binary type example}), essentially a single-agent problem, becomes nontrivial.
The second departure we make from the literature, which predominantly examines supermodular environments, is that our producer types face indefinite externalities: disclosure by higher types exacerbates consumer skepticism, promoting further disclosure, whereas disclosure by lower types impedes additional disclosure.
As a result, the dynamics of equilibria elimination differ from those of traditional divide-and-conquer. Notably, the menu offered to each type does not make full disclosure a dominant strategy: a type may prefer less disclosure if some lower types disclose too aggressively.

%Moreover, Lemma 3 of \cite{sega2003} is related to our Lemma \ref{lemma-path} in showing that every menu profile that robustly induces a certain disclosure outcome induces the outcome as if it adopts an alternating divide-and-conquer strategy. However, due to the presence of indirect externalities, his Round-Robin optimization argument for showing his Lemma 3 is not applicable in our case. After presenting Lemma \ref{lemma-path}, we will provide a more detailed comparison of the two papers in Section \ref{pointwise bounded paths}.

%Another branch of this literature led by \cite{wint2004} also seeks to weaken contractibility by limiting the principal's ability to monitor and incentivize the agent's actions. Many, such as \cite{wint2004}, \cite{bewi2012}, and \cite{hakw2020}, consider binary actions and simple monitoring systems, which results in the principal offering a single plan to each agent. \cite{halr2021}, \cite{halr2022}, and \cite{hakw2023} allow the principal to jointly design communication or monitoring schemes, which enriches the solution structure. In particular, \cite{halr2021} study a principal who can offer random private contracts. Their optimal contracts also alternatingly divide and conquer the probabilities of agents taking actions by designing endogenous ranking types.

Our paper also relates to the literature on verifiable disclosure, starting with \cite{gros1981} and \cite{milg1981}. The key concept incorporated by many models is skepticism, namely, the receiver's belief when no disclosure happens. Skepticism both depends on and shifts the sender's contingent choices of disclosure, which is the source of strategic uncertainty in our model. A classic work on the role of disclosure intermediaries that both design information and price disclosure is \cite{lizz1999}. Unlike us, \cite{lizz1999} considers the mandatory disclosure of signal realizations, and he allows the uninformed party to perfectly observe the informed party's action. Moreover, he does not consider adversarial equilibrium selection. Consequently, his intermediary manages to extract full surplus while inducing no disclosure.
\cite{Rappoport25} studies when a change in the receiver’s prior belief about
the sender’s evidence induces more skepticism, and he fully characterizes receiver optimal equilibrium outcomes in general verifiable disclosure games. \cite{dye1985}, \cite{bedl2018}, \cite{mise2022}, and \cite{whzh2022} consider a receiver who cannot tell whether a sender has acquired evidence. More broadly, \cite{okps1990} and \cite{hakp2014} discuss verifiable information-sharing among multiple players. For other developments in evidence games, see \cite{hakp2017}, \cite{bedl2019}, and \cite{onuchic2023disclosure}.

Finally, our paper is related to \cite{ahls2022}, which explores the robust information design problem faced by certification agencies who flexibly design the signal structure but with a specific pricing rule. In their model, the intermediary (platform) offers only full disclosure at a price uniform across all signal realizations (i.e., producer types), and they show that the robustly optimal signal structure is noisy. 
In contrast, we fix the signal structure and focus on robust pricing design, allowing prices to depend on both the disclosure probability and the signal realizations. Our comparative analysis further shows that the platform's revenue guarantee under optimal pricing increases as the signal structure becomes more precise. This uncovers that when platforms can jointly and flexibly design both the signal structure and pricing strategy, they prefer to influence agent behavior through tailored prices and menu options, rather than by distorting information. In fact, the optimal signal structure is fully revealing.

Another distinctive feature of our model is that the platform plays an active role in welfare creation, as some producer types are inefficient due to production costs. This allows us to meaningfully quantify the quality of the information provided by the platform. Our results contrast with \cite{ahls2022} and indicate that with flexible pricing, the platform does not distort efficient allocation or information transmission for its own benefit.

\section{Model}\label{model}

A risk-neutral producer (she) wants to sell a product to a risk-neutral consumer (he). The product quality, $\omega\geq0$, is drawn from a state space $\Omega\subseteq\mathbb{R}_+$ according to a prior distribution $\nu_0\in\Delta(\Omega)$. We focus on finite-state cases, that is, $\Omega=\{\omega^i\}_{i=1}^I$, where $I=|\Omega|\geq2$ and $\nu_0$ is represented by prior probabilities $\nu_0^1$, $\nu_0^2$, ..., and $\nu_0^I$ that we assume are strictly positive. Reorder the states such that $\omega^1>\omega^2>...>\omega^I$, where $\omega^1$ refers to the highest quality state while $\omega^I$ represents the lowest quality state.
%The producer privately observes a signal $s_P\in S_P$ realized according to an external information structure $\pi_P:\Omega\rightarrow\Delta(S_P)$. Let $\Pi_P$ collect all pairs of signal space and information structure $(S_P,\pi_P)$.
If the product is sold, the producer must pay a production cost $c\geq0$. The producer sells the product with a take-it-or-leave-it price offer. Therefore, absent additional information, the producer charges a price equal to the prior mean $v_0:=\E[\omega]=\sum_{i=1}^I\nu_0^i\omega^i$ if $v_0>c$, and the consumer buys the product; if $v_0\leq c$, no sale happens. Henceforth, $\E$ refers to the expectation under prior, and the expectation under another distribution $\nu$ is denoted $\E[.|\nu]$.

%{\color{red}[Good place to mention social efficiency?]}

\paragraph{Information and disclosure} %Before making the price offer, the producer receives a signal about the quality of her product and can choose a disclosure plan offered by a platform to verifiably disclose the signal to the consumer with a certain probability. Specifically, the signal is generated by a signal structure $(\Theta,\pi)$ where $S$ is a signal space and $\pi:\Omega\rightarrow\Delta(S)$ is a full-support signal-generating process. We focus on finite signal spaces. In addition, each disclosure plan $r=(q,p)$ consists of a disclosure probability $q\in[0,1]$ and a price $p\geq0$. Given a signal structure $(\Theta,\pi)$, the platform designs a profile of signal-dependent menus of disclosure plans: for each signal realization $s\in S$, the platform specifies a menu
%\begin{equation}\label{menu profile}
%    M^s=\{r_j^s\}_j=\{(q_j^s,p_j^s)\}_j.
%\end{equation}
%Since the platform must obtain the signal to perform disclosure, the prices depend on $s$. Namely, each menu $M^s$ is exclusive to the producer who receives the signal $s$, and she cannot choose a plan from other menus. Each menu $M^s$ can be arbitrarily large. However, to ensure the producer's optimal choice is always well-defined, we impose a technical assumption that each menu must induce a compact graph of probabilities and prices. Moreover, since the producer can always ignore the platform, we require that each menu contain $(0,0)$. We let $\Pi$ collect all pairs of (finite) signal space and signal-generating process $(\Theta,\pi)$. Given a signal structure $(\Theta,\pi)$, let $M^\Theta=(M^s)_{s\in S}$ denote a \emph{menu profile} and $\mathcal{M}(\Theta,\pi)$ the space of all such menu profiles. The platform therefore chooses $M^\Theta\in\mathcal{M}(\Theta,\pi)$.

Before the interaction between the producer and the consumer,
the platform can offer a quality disclosure service to the producer. Specifically, the platform obtains a signal $\theta$ of the product quality $\omega$ and offers a menu of signal-dependent disclosure plans that credibly disclose the signal to the consumer. The signal is generated by a signal structure $(\Theta,\pi)$ where $\Theta$ is a signal space and $\pi:\Omega\rightarrow\Delta(\Theta)$ is a full-support signal-generating process.
We focus on finite signal spaces. In addition, each disclosure plan $r=(q,p)$ consists of a disclosure probability $q\in[0,1]$ and a price $p\geq0$. Given a signal structure $(\Theta,\pi)$, the platform designs a profile of signal-dependent menus of disclosure plans: for each signal realization $\theta \in \Theta$, the platform specifies a menu
\begin{equation}\label{menu profile}
    M^\theta=\{r_j^\theta\}_j=\{(q_j^\theta,p_j^\theta)\}_j.
\end{equation}
The producer also observes the signal, and each menu $M^\theta$ is exclusive to the producer who receives $\theta$, so she cannot select a plan from other menus. Each menu $M^\theta$ can be arbitrarily large. However, to ensure the producer's optimal choice is always well-defined, we impose a technical assumption that each menu must induce a compact graph of probabilities and prices. Moreover, we require each menu contain $(0,0)$ to capture the interim individual rationality. %We use $\Pi$ to denote all pairs of (finite) signal space and signal-generating process $(\Theta,\pi)$. 
Given a signal structure $(\Theta,\pi)$, let $M^\Theta=(M^\theta)_{\theta \in \Theta}$ denote a \emph{menu profile}, and let $\mathcal{M}(\Theta,\pi)$ be the space of all such menu profiles. %The platform therefore chooses $M^\Theta\in\mathcal{M}(\Theta,\pi)$.

After observing signal $\theta$, the posterior belief, denoted as $\nu_{\theta}$, is pinned down by the signal structure via the Bayes' rule as follows
\begin{equation}\label{signal posterior}
    \nu_{\theta}(\omega^i)=\frac{\nu_0^i\pi(\theta|\omega^i)}{\sum_{j=1}^I\nu_0^j\pi(\theta|\omega^j)}\text{, for all $i=1,2,...,I$.}
\end{equation}
To keep notations compact, we rename the signal realizations as the associated posterior means, namely $\theta=\E[\omega|\nu_{\theta}]$. Furthermore, let $\Theta=\{\theta^k\}_{k=1}^K$ with $K\geq2$ and reorder the types such that $\theta^1>\theta^2>...>\theta^K\geq0$. Note that if $\theta^k$ is disclosed, consumer's willingness-to-pay will update to exactly $\theta^k$. Therefore, from now on, we also refer to $\Theta$ as the \emph{type space} and each $\theta^k$ as the type $k$ producer. We will say a type $\theta^k$ is \emph{efficient} if $\theta^k>c$; and \emph{ineffcient}, otherwise.

We ask the production cost to satisfy $c\in[\theta^K,\theta^1)$, which allows it to play a nontrvial role. Moreover, choose $\underline{k}$ such that $\theta^{\underline{k}}$ is the lowest type larger than $c$, so $\theta^k\geq\theta^{\underline{k}}$ if and only if $\theta^k$ is efficient.

Moreover, for conciseness, we always use $k$ in place of $\theta^k$. That is, the menu exclusive to each type $\theta^k$ is simply $M^k$, so a menu profile is $M^{\Theta}=(M^k)_{k=1}^K$; \(r^{\Theta} = (r^k)_{k=1}^K\) represents a profile of producer's contingent plan choices, and \(q^{\Theta} = (q^k)_{k=1}^K\) and \(p^{\Theta} = (p^k)_{k=1}^K\) denote her contingent choices of probabilities and prices, respectively. Likewise, we use \(\theta^{-k}\) to refer to all types except \(\theta^k\), and \(r^{-k}\), \(q^{-k}\), and \(p^{-k}\) thus represent profiles excluding the \(k\)-th component. Expressions like \((\widetilde{r}^k, r^{-k})\) denote a full profile where the \(k\)-th component is replaced by \(\widetilde{r}^k\). Additionally, we introduce an incomplete order for comparing probability profiles: we say \(q_1^{\Theta} \leq q_2^{\Theta}\) if \(q_1^k \leq q_2^k\) for all \(k\in\Theta\).

%require that the evidence $s$ is independent of the producer's external signal $s_P$ conditional on the state.

%\footnote{We use superscript $k$ to denote objects that are specific to type $\theta^k$ and superscript $\Theta$ to denote a vector of such type-dependent objects. To address the issue of equilibrium existence, we make a technical assumption that each menu must induce a compact graph in the joint space of probabilities and prices.}

\paragraph{Timing and payoffs} The signal structure $(\Theta,\pi)$ is exogenously given. The platform first commits to a menu profile $M^{\Theta}$ ex ante. A game played by the producer and the consumer then unfolds as follows. To begin with, the state $\omega$ and the signal $\theta^k$ are realized. Nobody observes the state $w$ but the producer privately observes the signal $\theta^k$. Given this information, the producer privately selects a plan $r^k=(q^k, p^k)$ from the menu $M^k$ which is exclusive to her, and pays $p^k$ to the platform. The platform then publicly discloses the signal to the consumer with probability $q^k$. Crucially, the consumer knows $(\Theta,\pi)$ and $M^\Theta$ but not the producer's plan choice $r^k$. He observes the signal $\theta^k$ if and only if disclosure happens. Based on his observation, the consumer forms his posterior belief about the state, $\nu$. Finally, if $\E[\omega|\nu]< c$, the game ends with no trading happens; otherwise, the producer sells the product to the consumer at price $\E[\omega|\nu]$. The platform earns the selected price \(p^k\). The producer's earnings equal $\max\{\E[\omega|\nu]-c,0\}$ minus the price \(p^k\). Meanwhile, the consumer always gets 0 surplus.

%Fixing a signal space $S$, we use the following notational conventions. Let \(r^{S} = (r^s)_{s\in S}\) represent a profile of contingent plan choices and \(q^{S} = (q^s)_{s\in S}\) and \(p^{S} = (p^s)_{s\in S}\) denote contingent choices of probabilities and prices, respectively. We use \(\theta^{-k}\) to refer to all types except \(\theta^k\), with \(r^{-k}\), \(q^{-k}\), and \(p^{-k}\) representing profiles excluding the \(k\)-th component. Expressions like \((\widetilde{r}^k, r^{-k})\) denote a full profile where the \(k\)-th component of $r$ is replaced by \(\widetilde{r}^k\). Additionally, we introduce an incomplete order for comparing probability profiles: \(q_1^{\Theta} \leq q_2^{\Theta}\) if \(q_1^k \leq q_2^k\) for all \(k\).

\paragraph{Disclosure equilibria} 
Given a signal structure $(\Theta,\pi)$ and a menu profile $M^\Theta$, the interaction between the producer and the consumer in the disclosure game is described by the producer's contingent plan choices, $(r^k)_{k=1}^K$, and the consumer's posterior beliefs $(\nu_{\theta^k,e})_{k,e}$. Here, $\nu_{\theta^k,e}\in\Delta(\Omega)$ refers to the consumer's posterior belief when the signal realization is $\theta^k$, and the platform's disclosure action is $e$ where $e=D$ if disclosure happens and $e=N$ if not. This notation simply reflects the fact that the consumer can only observe disclosure of type or nondisclosure, but can never observe the true state or the producer's choice.

We consider pure-strategy weak perfect Bayesian equilibria with a weak refinement, which requires the following: (i) Given consumer beliefs $(\nu_{\theta^k,e})_{k,e}$, after receiving signal $\theta^k$, the producer's choice, $r^k=(q^k,p^k)\in M^k$, maximizes her interim expected payoff
\begin{equation}\label{producer optimality}
    q^k\max\{\E[\omega|\nu_{\theta^k,D}]-c,0\}+(1-q^k)\max\{\E[\omega|\nu_{\theta^k,N}]-c,0\}-p^k;
\end{equation}
(ii) for every $\theta^k$, the disclosure of the signal transmits a hard message so that $\nu_{\theta^k,D}=\nu_{\theta^k}$, pinned down by (\ref{signal posterior});\footnote{This is the refinement assumption, which is satisfied in any sequential equilibrium of finite games.} (iii) given producer strategy profile $(r^k=(q^k,p^k))_{k=1}^K$, for every $\omega^{\ell}$ and $\theta^k$, the belief at nondisclosure, also termed as the \emph{skepticism belief} (following \cite{milg1981}), is given by the Bayes' rule
\begin{equation}\label{bayes-state}
    \nu_{\theta^k,N}(\omega^{\ell})=
    \begin{cases} 
     \frac{\nu^{\ell}_0\sum_{k=1}^K\pi(\theta^k|\omega^{\ell})(1-q^k)}{\sum_{j=1}^{I}\nu^j_0\sum_{k=1}^K\pi(\theta^k|\omega^j)(1-q^k)} & \text{if } \sum_{j=1}^{K}\nu^j_0\sum_{k=1}^K\pi(\theta^k|\omega^j)(1-q^k)>0; \\
    \text{arbitrary} & \text{if }q^j=1\text{, for all }\omega^j\in\Omega.
\end{cases}
\end{equation}

Let $\mathcal{E}(\Theta,\pi,M^{\Theta})$ be the set of all such \emph{disclosure equilibria} induced by a signal-menu-profile structure $(\Theta,\pi,M^{\Theta})$. In addition, let $g=((q^k,p^k)_{k=1}^K,(\nu_{\theta^k,e})_{k,e})\in\mathcal{E}(\Theta,\pi,M^{\Theta})$ denote one generic disclosure equilibrium. We restrict $\mathcal{M}(\Theta,\pi)$ to the subset of menu profiles that induce nonempty equilibrium sets. In general, there is an equilibrium existence issue because we focus on pure strategies, whereas we will argue below that this is without loss for achieving platform optimum (defined immediately).

\paragraph{Robust objectives} This paper investigates the robustly optimal menu profile that maximizes the platform's expected profit guaranteed across all disclosure equilibria.

\begin{definition}\label{MRG-def}
    \emph{The platform's} maximal revenue guarantee \emph{given signal structure $(\Theta,\pi)$ is}
    \begin{equation}\label{MRG}
        R^*(\Theta,\pi):=\sup_{M^{\Theta}\in\mathcal{M}(\Theta,\pi)}\inf_{g\in\mathcal{E}(\Theta,\pi,M^{\Theta})}\sum_{k=1}^Kp^k\sum_{i=1}^I\nu_0^i\pi(\theta^k|\omega^i).
    \end{equation}
\end{definition}

In effect, the infimum above can be replaced with minimum since the equilibrium set $\mathcal{E}(\Theta,\pi,M^{\Theta})$ is compact\footnote{This results from the following facts: (i) we have finitely many menus, each inducing a compact graph; (ii) the equilibrium conditions for disclosure equilibrium are equalities and weak inequalities that involve continuous functions.}. Hence, given each menu profile $M^{\Theta}$, there exists a worst-case disclosure equilibrium whose producer strategy is denoted as $\underline{r}^{\Theta}=(\underline{q}^{\Theta},\underline{p}^{\Theta})$. We thus say $M^{\Theta}$ robustly induces the probability profile $\underline{q}^{\Theta}$ and guarantees revenue $R(\Theta,\pi,M^{\Theta}):=\sum_{k=1}^K\underline{p}^k\sum_{i=1}^I\nu_0^i\pi(\theta^k|\omega^i)$. Sometimes, if the signal structure is obvious from the contexts, we simply write $R(M^\Theta)$ as the revenue above guaranteed by $M^{\Theta}$, and simply $R^*$ as the maximal revenue guarantee.

The supremum in (\ref{MRG}), however, may not be attainable, and we seek to characterize a robustly optimal menu profile. This solution concept is defined as follows
\begin{definition}\label{ROMP-def}
    \emph{A menu profile $M^{*\Theta}$ is }robustly optimal\emph{ given signal structure $(\Theta,\pi)$ if there is a sequence of menu profiles $(M^\Theta_n)_{n=1}^{\infty}$ such that
    \begin{enumerate}[nolistsep]
        \item for all $\theta^k \in \Theta$, the sequence $(M_n^k)_{n=1}^{\infty}$ converges to $M^{*k}$ with respect to the Hausdorff metric;
        \item the sequence $\left(R(M_n^{\Theta})\right)_{n=1}^{\infty}$ converges to $R^*$.
    \end{enumerate}}
\end{definition}

To simplify the terminology we use for stating our results, we say that $M^{*\Theta}$ \emph{robustly induces} a certain probability profile $\underline{q}^{\Theta}$ if there is a sequence $(M^{\Theta}_n)_{n=1}^{\infty}$ that approximates $M^{*\Theta}$ in the sense above and some $n'$ such that for all $n\geq n'$, $M_n^{\Theta}$ robustly induces $\underline{q}^{\Theta}$.

Lastly, we explain why it suffices to consider pure-strategy equilibria. Let the value (\ref{MRG}) be called the pure guarantee, and the counterpart defined when considering mixed-strategy equilibria be the mixed guarantee. On the one hand, every mixed-strategy equilibrium induced by some menu profile, say $M^\Theta$, gives the platform the same expected profit as in a pure-strategy equilibrium induced by the menu profile $\widetilde{M}^\Theta$ with each $\widetilde{M}^k$ being the convex hull of $M^k$.\footnote{This is because (i) given consumer beliefs, the producer's objective (\ref{producer optimality}) is linear in her plan choice; (ii) in mixed-strategy equilibria, the Bayes' rule (\ref{bayes}) is only related to the expected disclosure probability of each type.} Therefore, had $\mathcal{M}(\Theta,\pi)$ been restricted to the subset of menu profiles that induce convex compact graphs, the mixed guarantee would be equal to (\ref{MRG}). As a result, the mixed guarantee is lower than the pure guarantee. On the other hand, our main result will show that the pure guarantee is obtained by a menu profile that induces a unique mixed-strategy equilibrium. This shows the converse that the mixed guarantee is no less than the pure guarantee.

\subsection{Applications}\label{applications}

Our model is built in a stylized manner to highlight the novel trade-off that worst-equilibrium selection introduces to the pricing problem of disclosure service, and to have a general model that connects to multiple applications. In this subsection, we provide a few more concrete examples and discuss how our model is equivalent to other variations.

\paragraph{Example 1} The platform is a contextual advertising platform, such as TV networks including broadcast, cable, and streaming networks. A product has a vertical quality $\omega$ and verifiable hard evidence $\theta$ about the quality, such as awards or third-party certifications. The platform can verify the evidence and disclose the information to users; otherwise, users encounter the product in retail store shelves with no information. By controlling the total time that the product's ads are streamed or directly monitoring ad views, the platform can fine-tune the user volume exposed to the information.

\paragraph{Example 2} The platform is an online merchant platform such as Amazon. There is a unit mass of consumers. Each consumer has a match-specific value $\omega$ with respect to the product with prior distribution $\nu_0$. Using the consumer data it possesses, the platform can divide the market into different segments, represented by consumers' average match value $\theta$ within each segment. An option it offers to the producer is represented by $((q_i)_i,p)$, in which $q_i$ is the probability that the match value $\theta_i$ will be disclosed to the consumers in each segment $i$. The platform hence commits to a single menu of such options. In this specification, the producer, when choosing one option, determines the exposure level $q_i$ of her product in every segment $i$ and pays a total price $p$.
%the producer makes purchase decision before observing the signal, but the option specifies the disclosure on all contingencies.
As discussed in Section \ref{iterated deletion of dominated strategies}, this setting is equivalent to our model after the Harsanyi transformation.

\paragraph{Example 3} The platform is an online freelance marketplace such as Upwork. Each freelancer has an ability state $\omega$. Based on historical performance data (e.g., ratings, completion rates), the platform can generate a prediction of the freelancer's ability $\theta$ that it can disclose to clients (potential employers). The promotion fees the freelancers pay to feature themselves in top search results can depend on their past performances and the amount of views they hope to achieve. In this example, one might think freelancers have some private signal $s$ about their ability. However, our result is robust to such hidden information as long as evidence $\theta$ is sufficiently informative such that $s$ is conditionally independent of the state $\omega$ given $\theta$.\footnote{For example, when the platform's signal is full-revealing. See Section \ref{iterated deletion of dominated strategies} for more discussions.}

\paragraph{Example 4} The platform is a certification body such as BREEAM that evaluates the environmental performance of buildings. A building has an underlying environmental quality $\omega$. The certifier conducts an audit to assess verifiable hard evidence such as architectural designs, energy models, and construction data, which generates a rating $\theta$. The total payment for higher-rated buildings can be higher for certification and for continued label/licensing use, because achieving higher ratings requires more comprehensive assessments and extra services. See Section \ref{random menus} to understand how our model can be applied to such certification agencies who are not even able to control disclosure probability.

\section{Preliminary Analysis}\label{preliminary analysis}

\subsection{Simplifying Disclosure Equilibria}\label{simplifying disclosure equilibria}

In this step, we simplify the notations used to represent each disclosure equilibrium. To start with, notice that both the platform's design and the producer's strategy work with the type space $\Theta$ instead of the states, so we translate the equilibrium concept in terms of the types. For every $\theta^k\in\Theta$, we have that $\theta^k$ equals the posterior mean it induces as a signal realization, and we define
\begin{equation}\label{type prior}
    \mu_0^k=\sum_{i=1}^I\nu_0^i\pi(\theta^k|\omega^i).
\end{equation}
In fact, we show at the beginning of \ref{proof-main result} that the original model is equivalent to the same model but with the following changes: (i) the state space is $\Theta$; (ii) the prior is $(\mu_0^k)_{k=1}^K$; (iii) the signal structure fully reveals the state/type. In particular, let $\mu\in\Delta(\Theta)$ be a generic ``belief'' on the type space. Fixing a menu profile, then the consumer's posterior belief (about types) following a path on which $\theta^k$ is realized and the disclosure action is $e$ is $\mu_{\theta^k,e}$. The skepticism belief (\ref{bayes-state}) is now
\begin{equation}\label{bayes}
    \mu_{\theta^k,N}(\theta^{\ell})=
    \begin{cases} 
     \frac{\mu^{\ell}_0(1-q^{\ell})}{\sum_{j=1}^{K}\mu^j_0(1-q^j)} & \text{if } \sum_{j=1}^{K}\mu^j_0(1-q^j)>0; \\
    \text{arbitrary} & \text{if }q^j=1\text{, for all }\theta^k\in\Theta.
\end{cases}
\end{equation}
Moreover, once disclosure occurs, the consumer posterior $\mu_{\theta^k,D}$ must assign probability one to $\theta^k$. Likewise, we still use $\E$ as the expectation under prior $\mu_0$, and $\E[.|\mu]$ under another belief $\mu$. Finally, note that the revenue term in (\ref{MRG}) now simply becomes $\sum_{k=1}^K\mu_0^kp^k$. Henceforth, we work with this new language in terms of types in place of the old language that works with the state space.

In the following, we argue that it suffices to represent the belief system in each disclosure equilibrium with a single skepticism belief over the type space, denoted $\mu^N$. According to (\ref{bayes}), the only case in which the Bayes' rule cannot be applied is when all types fully disclose. This outcome can only be sustained in equilibrium if the consumer assigns probability 1 to the lowest type when he fails to see any disclosure; otherwise, the lowest type would benefit from skepticism and thus avoids disclosing. Thus, the full-disclosure profile pins down a single skepticism belief. On the other hand, whenever the Bayes' rule can be applied, the skepticism belief \(\mu_{\theta^k, N}\) does not depend on \(\theta^k\). Furthermore, the belief upon observation, \(\mu_{\theta^k, D}\), assigns probability one to \(\theta^k\). Therefore, it suffices to denote an equilibrium as \((r^{\Theta}=(q^{\Theta},p^{\Theta}), \mu^N)\) where \(\mu^N \in \Delta(\Theta)\) is the single skepticism belief determined by the probability profile \(q^{\Theta}\), as given by (\ref{bayes}). Sometimes, we use $\mu^N(q^{\Theta})$ to refer to this dependence.

\paragraph{Skepticism value} Next, we define the \emph{skepticism value} associated with each profile of disclosure probability \(q^{\Theta}\), which equals the producer's gain from transaction when disclosure does not happen, while the consumer anticipates the producer's contingent disclosure probabilities to be $q^{\Theta}$. This concept reflects the outside value of not disclosing in an equilibrium with contingent disclosure \(q^{\Theta}\). Applying (\ref{bayes}), the skepticism value for any $q^{\Theta} \neq 1^{\Theta}$ is given by:

\begin{equation}\label{skepticism value}
    w^N(q^{\Theta}):= \max\{\E[\theta|\mu^N(q^{\Theta})]-c,0\} = \max\left\{\frac{\sum_{k=1}^K \mu_0^k(1 - q^k)\theta^k}{\sum_{k=1}^K \mu_0^k(1 - q^k)}-c,0\right\}.
\end{equation}

We complete the definition above with \(w^N(1^{\Theta}) = \max\{\theta^K-c,0\}=0\). Notice that skepticism value can be zero because the consumer may be so pessimistic about the product's quality that the producer cannot profit from costly production. A key property of this function is that for all disclosure outcome \(q^{\Theta}\), if \(\theta^k > [=;<]\E[\theta|\mu^N(q^{\Theta})]\), then increasing the $k$th type's disclosure probability \(q^k\) will strictly increase [not change; strictly decrease] the skepticism posterior mean $\E[\theta|\mu^N(q^{\Theta})]$, and thus weakly increase [not change; weakly decrease] $w^N(q^{\Theta})$. In other words, more disclosure probability from each type always shifts skepticism away from that type.

Using this function, we can rewrite (\ref{producer optimality}), the optimal choice of each producer type $\theta^k$ as: 
\[(q^k,p^k)\in\argmax_{(q,p)\in M^k}q\max\{\theta^k-c,0\}+(1-q)w^N(q^{\Theta})-p.\]
One can see a type with $\theta^k\leq c$ will not pay for disclosure. Moreover, if we consider a type $\theta^k>c$ and fix the producer's skepticism value $w^N$, this type's indifference curve becomes a straight line. For some $(q',p')$, she is indifferent between this plan and all plans $(q,p)$ that satisfy:
\begin{equation}\label{indifference curve}
    p-p'=(\theta^k-c-w^N)(q-q').
\end{equation}
The slope $\theta^k-c-w^N$ represents the marginal benefit of disclosure with a higher probability. All plans that lie beneath this line become profitable deviations for the producer $\theta^k$ (given the fixed $w^N$).

\subsection{Benchmark Without Strategic Uncertainty}\label{benchmark without strategic uncertainty}

Before considering the robust design, we examine what would happen if the platform does not account for strategic uncertainty. In this problem, the platform selects a menu profile to maximize the expected revenue in the best-case equilibrium, solving  $\max_{M^{\Theta}\in\mathcal{M}(\Theta),g\in\mathcal{E}(M^{\Theta})}\sum_{k=1}^K\mu_0^kp^k$. This problem is solved by the following bang-bang menu profile $\overline{M}^{\Theta}$: for all $\theta^k\in\Theta$,
\[\overline{M}^k=\begin{cases}
    \{(0,0),(1,\theta^k-c)\} &\text{ if }\theta^k>c; \\
    \{(0,0)\} &\text{ if }\theta^k\leq c.
\end{cases}\]
In the best-case equilibrium induced by $\overline{M}^{\Theta}$, all efficient producers ($\theta > c$) choose full disclosure while the inefficient types ($\theta\leq c$) choose zero disclosure, and the platform extracts full surplus
\[\overline{R}(\Theta,\pi):=\E[\max\{\theta-c,0\}].\]
We sometimes use $\overline{R}$ instead of $\overline{R}(\Theta,\pi)$ if the signal structure is obvious from the contexts.

In this scenario of partial implementation, the platform gains massive bargaining power by forming the consumers' beliefs so that all efficient producers would use its disclosure service, and nondisclosure only comes from the inefficient ones. Hence, the consumers' posterior mean of the quality when seeing no disclosure is lower than $c$. Given this punishing belief, consumers are not willing to pay higher than $c$ for a non-disclosed product. As a result, failing to disclose will give producers the outside value of $0$, and thus full disclosure is self-fulfilling in an equilibrium.

This result illustrates that with the ability to select its preferred equilibrium, the platform can achieve social efficiency and extract full surplus: the producers produce if and only if their expected quality $\theta$ is above the production cost; moreover, the platform entrench its market role as a gatekeeper costlessly who punishes any non-disclosed product with the worst outside value.
This simple feature characterizes our model as a frictionless benchmark once strategic uncertainty is ruled out, allowing us to isolate other driving forces beyond strategic uncertainty.

One concerning feature of the best-case design is that the revenue is not continuous with respect to the prior distribution when the prior mean satisfies $v_0:=\E[\theta]>c$. In particular, as long as the prior is not degenerate, the platform's revenue converges to $v_0-c$ when the uncertainty facing the consumer vanishes---that is, the prior assigns most probability to a single type $\theta=v_0$. In sharp contrast, if the consumer knows precisely the quality $\theta=v_0$, the disclosure service has no value to consumers and the platform's revenue will be 0.

Moreover, the platform's best-case design may be highly fragile to strategic uncertainty. In fact, the following proposition demonstrates that strategic uncertainty arises if and only if the producer and the consumer can trade in the absence of the platform, that is $v_0>c$. %Recall that $v_0=\E[\theta]$ is the prior mean, so the trade happens without a platform if and only if $v_0>c$.
Formally, we have

\begin{proposition}\label{proposition-benchmark}
    If $v_0\leq c$, $\overline{M}^{\Theta}$ is also robustly optimal. In contrast, if $v_0>c$, there is $\delta>0$ such that for any small $\varepsilon\geq0$, any menu profile that induces an equilibrium with revenue no less than $\overline{R}-\varepsilon$ also has an equilibrium with revenue no greater than $\delta\varepsilon$
\end{proposition}

Proposition \ref{proposition-benchmark} highlights the case of $v_0>c$, in which the platform faces large degree of strategic uncertainty. Such extreme strategic uncertainty is not just the feature of one particular optimal mechanism, as it cannot be eliminated by using approximately optimal mechanisms.

To understand why zero disclosure can form an equilibrium when $v_0>c$, note that if all types decide to not use the platform's service, the consumer's skepticism belief (belief at nondisclosure) must equal the prior. This suggests the producer can now get a positive surplus of $v_0-c$ by rejecting the platform and selling the non-disclosed product to consumers. Hence, the full-disclosure prices appear too high now, and the bad equilibrium of zero disclosure is also self-fulfilling.

When $v_0\leq c$, however, the no-disclosure equilibrium can be killed by reducing the prices in $\overline{M}^{\Theta}$ by an arbitrarily small amount. Facing the outside value 0, every efficient producer now strictly prefers disclosure. To understand this dichotomy, note that $v_0\leq c$ forbids the producer and the agent to trade absent the platform. Hence, the platform here naturally serves as the market's gatekeeper because any producer type refusing to use its service cannot profit from transaction, \emph{regardless} of consumer skepticism. Thus, there is no ``room'' for the producer's outside value to be coordinated against the platform's objective. In contrast, if we have $v_0>c$ so that the producer and the consumer can trade by themselves, the platform will face a trade-off between surplus extraction and strategic uncertainty. For the rest of the paper, we mainly discuss the case where $v_0>c$.

\section{A Binary-Type Example}\label{binary type example}

As a first step in our analysis of the robustly optimal design, we consider a simple case with binary types $\Theta=\{\underline{\theta},\overline{\theta}\}$, zero production cost $c=0$, and a fully revealing signal structure. This example illustrates the first main feature of the optimal menu profile: 
the revelation principle fails and menus for high types can contain a continuum of plans that together induce a \emph{self-conquering} process to guarantee disclosure.  
%(ii) A continuous menu represented by a tractable price function that is strictly convex in disclosure probability emerges as the optimal menu. 

Slightly different from the standard notation, we let the two types be high type $\overline{\theta}$ and low type $\underline{\theta}$. For this example alone, we use a single scalar $\mu\in[0,1]$ to represent consumer belief, which refers to the probability of the high type. The prior is also re-denoted as $\overline{\mu}$. We further set $\overline{\theta}=1$ and $\underline{\theta}=0$, which brings the convenience that consumer belief coincides with his value expectation. Moreover, since the consumer will not perceive the producer more negatively than the low type, the low-type producer has no incentive to pay for disclosure, so it is without loss to always offer the low type nothing more than the outside option. Thus, the essential choice is a single menu exclusive to the high type. Also, a probability profile $q^{\Theta}$ now degenerates to the high type's probability. Hence, for this example alone, we rewrite the skepticism value (\ref{skepticism value}) as a function of $q\in[0,1]$:
\begin{equation}
    w^N(q)=\frac{\overline{\mu}(1-q)\overline{\theta}+(1-\overline{\mu}
)\underline{\theta}}{\overline{\mu}(1-q)+(1-\overline{\mu}
)}=\frac{\overline{\mu}(1-q)}{\overline{\mu}(1-q)+(1-\overline{\mu}
)}.
\end{equation}

\paragraph{Main insight} The fact that our platform can design a large menu that offers various levels of disclosure indicates a stronger pricing power than, for example, a bang-bang non-discriminative disclosure option. To see why this additional market power can starkly alter platform behavior, first consider the following bang-bang menu offered to the high type (with small $\varepsilon>0$):
\begin{equation}\label{binary optimal}
    M_2=\{(0,0),(1,1-\overline{\mu}-\varepsilon)\}.
\end{equation}
This menu robustly induces full disclosure out of the high type because the price is low enough to \emph{break the candidate equilibrium} where both types choose $(0,0)$. That is, suppose both types choose $(0,0)$ in an equilibrium. Then, the Bayes' rule sets the consumer's skepticism belief equal to the prior $\overline{\mu}$. However, given this prior skepticism, the high type wants to deviate to $(1,1-\overline{\mu}-\varepsilon)$ and earn $\varepsilon>0$. Therefore, this undesirable equilibrium is broken and one can verify that the high type fully discloses in the unique equilibrium. In fact, if the platform can offer only full-disclosure plans, $M_2$ delivers the highest revenue guarantee\footnote{This can be seen immediately once we introduce Lemma \ref{lemma-simplify} in Section \ref{truncated convexifications}, which implies that with only bang-bang plans, it suffices to consider the menus in the form of $M=\{(0,0),(1,p)\}$ with some price $p\geq0$.} as $\varepsilon$ goes to 0.

However, this binary menu can be improved by a three-plan menu given in the following:
\begin{equation}\label{3-plan improvement}
    \begin{aligned}
        M_3&=\{(0,0),(\frac{1}{2},\frac{1-\overline{\mu}}{2}-\frac{\varepsilon}{2}),(1,\frac{1-\overline{\mu}}{2}+\frac{1-w^N(\frac{1}{2})}{2}-\varepsilon)\}.
    \end{aligned}
\end{equation}
Recall that $w^N(\frac{1}{2})$ is the skepticism belief when the high type discloses with probability $\frac{1}{2}$ while the low type has no disclosure. $M_3$ looks similar to $M_2$ in that its second plan is ``half'' of the full disclosure plan in $M_2$. The previous argument again shows that this plan breaks the candidate equilibrium where the high type chooses $(0,0)$. The change is that $M_3$ includes a third plan constructed to break the candidate equilibrium where the high type chooses the second plan of half disclosure. To see this, notice that if, in an equilibrium induced by $M_3$, the high type chooses the half-disclosure plan (and the low type can only choose $(0,0)$), the consumer's skepticism belief is exactly $w^N(\frac{1}{2})$. Now, the high type would rather deviate to the third plan of full disclosure to earn $\varepsilon-\frac{\varepsilon}{2}>0$.

As a result, $M_3$ robustly induces full disclosure from the high type while approximating an expected profit, $\overline{\mu}\left(1-\frac{\overline{\mu}}{2}-\frac{w^N(\frac{1}{2})}{2}\right)$, higher than that of $M_2$, $\overline{\mu}(1-\overline{\mu})$, because $\overline{\mu}>\frac{\overline{\mu}}{2-\overline{\mu}}=w^N(\frac{1}{2})$.

The insight behind such improvement is that a higher disclosure probability out of a high type worsens consumer skepticism, which increases the producer's opportunity cost of choosing low-probability disclosure, making further disclosure even easier to induce. Specifically, in $M_3$, the platform takes advantage of the fact that a candidate equilibrium with more high-type disclosure requires less reduction of price to break. Therefore, the platform benefits from deliberately creating price dispersion to shift the producer's outside value $w^N(q)$ downward, more severely punishing the producer for not disclosing.

%This example highlights that the platform benefits from the two design elements: a large menu and partial disclosure.

\begin{figure}[h]
    \centering
    \includegraphics[width=0.8\linewidth]{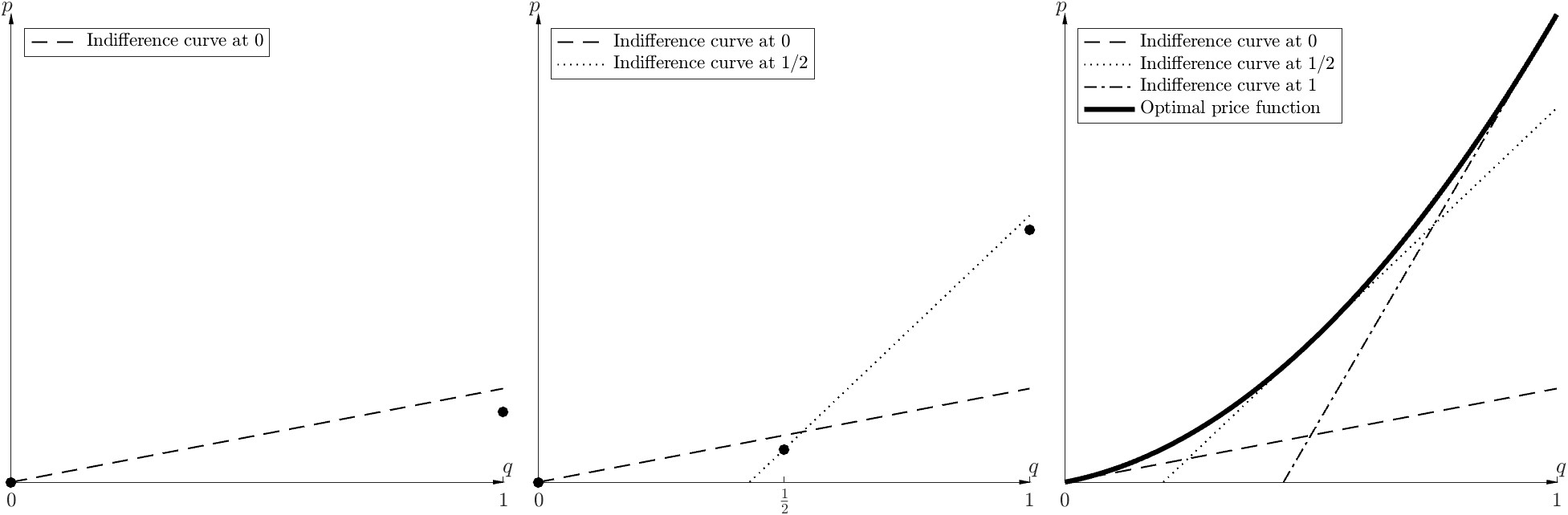}
    \captionsetup{justification=centering}
    \caption{Improvement from $M_2$ (left) to $M_3$ (middle) and $M^*_{\text{binary}}$ (right).}
    \label{fig-improvement}
\end{figure}

\paragraph{Optimal menu.} Next, we demonstrate why the robustly optimal menu is continuous and represented by a strictly convex price function. The idea is to iterate the improvement process we described above by keeping adding plans with different disclosure probabilities until all probabilities fill the entire $[0,1]$ interval. Figure \ref{fig-improvement} depicts two improvement steps toward the eventual optimal menu $M^*_{\text{binary}}$. Each subfigure describes a menu and each black dot represents a plan. Each dashed line is the high type's indifference curve in a candidate equilibrium. In such an equilibrium, the high type wants to deviate to any dot below the curve. For instance, in the middle subfigure of \(M_3\), the “indifference curve at 1/2” refers to the high type's indifference curve when the consumer expects half disclosure from the high type and no disclosure from the low type. In this scenario, the full-disclosure plan becomes a profitable deviation because it lies beneath the indifference curve.

Three features are worth mentioning: (i) the process of equilibrium elimination is driven by a single type, as the low type always chooses zero disclosure, so our problem stays nontrivial even with a single agent; we thus call the process \emph{self-conquering};
%In contrast, in classic frameworks (e.g., \cite{sega2003} and \cite{wint2004}) that focus on multiple agents whose actions have direct externalities on other agents' payoffs, the equilibrium elimination process proceeds among different agents. The optimal contract would be trivial in their settings if there is only one agent.
(ii) the improved menus robustly induce full disclosure by creating \emph{local upward deviations}, meaning that each partial disclosure plan (including \((0,0)\)) is broken by the high type deviating to the next plan with a higher probability; and (iii) the marginal price is \emph{upward-bounded} by the indifference curve of the high type at each plan. Specifically, if \((q_1, p_1)\) and \((q_2, p_2)\) are adjacent plans with \(q_2 > q_1\), then the marginal price \(\frac{p_2 - p_1}{q_2 - q_1}\) must be bounded by the indifference curve slope \(\overline{\theta} - w^N(q_1)\), where \(w^N(q_1)\) is the skepticism belief when the high type selects \((q_1, p_1)\).
%These indifference curve slopes represent the high type’s marginal benefit of disclosure in different candidate equilibria, as disclosing marginally allows the high type to earn her type \(\overline{\theta}\) while forgoing the outside value \(w^N(q_1)\).

As one can imagine, if we continue the improvement process, a continuous menu will emerge. The menu will be equivalent to a price function of disclosure probability, which we denote as $p_{\text{binary}}^*(.)$, such that ${p_{\text{binary}}^*}'(q)\leq\overline{\theta}-w^N(q)$ for all $q\in[0,1]$. By setting the inqualities to equalities, ${p_{\text{binary}}^*}(.)$ becomes the optimal menu and the marginal prices are given by the indifference curve slopes. 
%The key to the construction is to equate the marginal prices with their upper bounds, namely the slopes of indifference curves.
In Figure $\ref{fig-improvement}$, the right-hand-side subfigure depicts three indifference curves that are tangent to the optimal price function. Formally, the optimal price function is given by
\begin{equation}\label{binary type}
    \begin{aligned}
        %M^*_{\text{binary}}&=\{(q,p_{\text{binary}}^*(q)-\varepsilon q):q\in[0,1]\}\text{, where} \\
        p_{\text{binary}}^*(q)&=\int_0^q\left[\overline{\theta}-w^N(\widetilde{q})\right]d\widetilde{q}=\int_0^q\left[\overline{\theta}-\frac{(1-\overline{\mu})\underline{\theta}+\overline{\mu}(1-\widetilde{q})\overline{\theta}}{(1-\overline{\mu})+\overline{\mu}(1-\widetilde{q})}\right]d\widetilde{q}=\int_0^q\frac{\overline{\theta}-\E[\theta]}{1-\overline{\mu}\widetilde{q}}d\widetilde{q} \\
        &=-\frac{\overline{\theta}-\E[\theta]}{\overline{\mu}}\ln(1-\overline{\mu}q),\text{ for all }q\in[0,1].
    \end{aligned}
\end{equation}
%Notice that in (\ref{binary type}), the solution is written without replacing $\overline{\theta}$ or $\underline{\theta}$ with their values. 
%and in interpreting our main result later in Section \ref{main result section}, we will invoke the form $p_{\text{binary}}^*(q)=-\frac{\overline{\theta}-\E[\theta]}{\overline{\mu}}\ln(1-\overline{\mu}q)$.

The price $p_{\text{binary}}^*(.)$ is strictly convex and induces full disclosure in the unique equilibrium. Plugging in $\overline{\theta}=1$ and $\underline{\theta}=0$, the platform's maximal revenue guarantee is thus $\overline{\mu}p^*_{\text{binary}}(1)=-(1-\overline{\mu})\ln(1-\overline{\mu})$, which is strictly lower than full surplus $\overline{\mu}$, and approximates 0 when $\overline{\mu}$ approaches 0 or 1.
%Notably, the maximal revenue guarantee drops to zero when the consumer faces small uncertainty, namely, when $\overline{\mu}$ approximates 0 and 1. This contrasts with the behavior of the best-case revenue that we investigated in Section \ref{benchmark without strategic uncertainty}.

%\emph{Discussion on bargaining power.} The classical approach of assuming the principal can pick the favorite equilibrium is often justified by technique reasons, and in many classical settings, it is innocuous.\footnote{For example, in monopoly pricing and Bayesian persuasion.} However, when this assumption leads to significant differences, it becomes a questionable modeling choice, especially since it further strengthens the principal's bargaining power in a context where the platform already has considerable bargaining leverage due to its ability to commit to a take-it-or-leave-it contract. To the extent that the bargaining position between the platform and the producer is perhaps not totally one-sided in reality and the classical principal-agent approach is sometimes a compromise to the lack of an ideal model of bargaining, we think that the worst-case equilibrium selection is an interesting reduce-form approach to give back some bargaining power to the producer.

\section{Robustly Optimal Menu Profile}\label{optimal menu profile}

We now move to the more general case with multiple types and a nontrivial production cost, and establish our main result that provides the explicit forms of a robustly optimal menu profile and the maximal revenue guarantee. Section \ref{main result section} states the main result, Theorem \ref{main result}, followed by a discussion of its interpretation, implications, and proof sketch. Section \ref{truncated convexifications}-\ref{upper bound for maximal revenue guarantee} elaborates on each proof step.

\subsection{Main Result}\label{main result section}

%Our main result finds that, to maximize its revenue guarantee, the platform chooses to robustly induce full disclosure by (i) offering each type a continuum of plans that together induce a self-conquering process for each producer type; and (ii) prioritizing guaranteeing disclosure probabilities from high types over low types by sequentially inducing the self-conquering process for each type in decreasing order. Specifically, the following menu profile robustly obtains the maximal revenue guarantee:

Our main result demonstrates that, to maximize revenue guarantee, the platform robustly induces full disclosure of the efficient types by prioritizing guaranteeing disclosure probability of higher types over lower types. To do this, the platform sequentially induces a self-conquering process for each type in the descending order. In particular, the platform offers a continuum of plans that first ensure full disclosure of the highest type, and then that of the second highest type, and so on for lower types. This procedure progressively reduces the producer's outside value of not disclosing. Once a tipping point is reached---where the outside value decreases to 0---the platform switches to using bang-bang options to extract full surplus from each remaining low type.

To pin down the tipping point, we introduce a shorthand notation: for all $\theta^k\in\Theta$ and $q\in[0,1]$,
\begin{equation}\label{w^N_k(q)}
    w^N_k(q)=w^N(\underbrace{1,...,1}_{k-1\text{ ones}},q,0,...,0),
\end{equation}
Where skepticism value $w^N(.)$ is defined by (\ref{skepticism value}) as the producer's outside value when disclosure does not occur. In other words, $w_k^N(q)$ is the producer's gain from transaction absent disclosure if the consumer believes that all types higher than $\theta^k$ fully disclose, all types lower than $\theta^k$ choose to not disclose, while $\theta^k$ herself discloses with probability $q$. Thus, there exists a unique\footnote{Uniqueness results from two facts. First, we have the starting point of the sequential self-conquering process $w^N_1(0)=v_0>c$, the ending point $w^N_K(1)=w^N(1^{\Theta})=0<c$, and that the process changes $w^N$ continuously. Second, $w^N_k(q)$ is nonincreasing in $k$ and $q$ because (i) $w_{k}^N(1)=w_{k+1}^N(0)$; and (ii) $\theta^k\geq\E[\theta|\theta\leq\theta^k]$, so raising $q$ will lower $w^N$.} \emph{tipping point} $(k^*,q^*)$ such that
\begin{equation*}
    \begin{aligned}
        \theta^{k^*}&=\max\{\theta^k\in\Theta:\exists q\in[0,1]\text{ s.t. }w^N_k(q)=0\}; \\
        q^*&=\min\{q\in[0,1]:w^N_{k^*}(q)=0\}.
    \end{aligned}
\end{equation*}
In other words, $(k^*,q^*)$ is found via the following process: the platform starts at $0^{\Theta}$ and it first increases the highest type's disclosure probability until $q^1=1$; it next turns to increasing $q^2$ until it depletes the second highest type's probability; the platform then continues for lower types like this until it reaches the first type $k^*$ and this type's first probability $q^*$ such that the skepticism value decreases to 0. To use compact language, we say a pair $(k,q)$ is \emph{before} [\emph{after}] the tipping point if either $k<k^*$, or $k=k^*$ and $q<q^*$ [either $k>k^*$, or $k=k^*$ and $q>q^*$].

In analogy to the process above, the platform optimally proceeds in the same manner to iteratively drive the producer's outside value down to 0. We dub this process \emph{sequential self-conquering}.

Moreover, we need a few other notations for stating the main result. When $\theta\leq\theta^k$ enters expectation, $\E[.|\theta\leq\theta^k]$, and probability, $\Pr(.|\theta\leq\theta^k)$, it refers to the prior distribution $\mu_0$ truncated above at $\theta^k$. We then define the following functions for each $(k,q)$ before the tipping point is reached (so $w^N_k(q)>0$)
\begin{equation}\label{multiple type}
\begin{aligned}
    u_k^*(q)&=\int_0^qw^N_k(s)ds=q(\theta^k-c)-p_k^*(q)\text{, where} \\
    p^*_k(q)&=\frac{\theta^k-\E[\theta|\theta\leq\theta^k]}{\Pr(\theta^k|\theta\leq\theta^k)}\ln\frac{1}{1-\Pr(\theta^k|\theta\leq\theta^k)q}.
\end{aligned}
\end{equation}
Notice that $q(\theta^k-c)=p_k^*(q)+u_k^*(q)$ is the total social gain from disclosing signal $\theta^k$ with probability $q$. Hence, our main result states that the following menu profile robustly obtains the maximal revenue guarantee, in which the platform offers a price function $p_k^*$ to each high type with $\theta^k\geq \theta^{k^*}$, whereas $u_k^*$ represents the rents the platform leaves to each type $k$ producer

\begin{theorem}\label{main result}
    A robustly optimal menu profile $M^{*\Theta}$ is given by $M^{*k}=\{(0,0)\}$ for all $\theta^k\leq c$; and
    
    (i) if $\theta^k>\theta^{k^*}$, $M^{*k}=\{(q,p^*_k(q)):q\in[0,1]\}$;

    (ii) if $\theta^k=\theta^{k^*}$, $M^{*k^*}=\{(q,p_{k^*}^*(q)):q\in[0,q^*]\}\cup\{(1,p_{k^*}^*(q^*)+(1-q^*)(\theta^{k^*}-c))\}$;

    (iii) if $\theta^{k^*}>\theta^k>c$, $M^{*k}=\{(0,0),(1,\theta^k-c)\}$.
    
    Moreover, this profile induces a unique equilibrium where all types with $\theta^k>c$ choose to fully disclose while those with $\theta^k\leq c$ choose zero disclosure. The maximal revenue guarantee is given by
    \begin{equation}\label{MRG*}
        R^*(\Theta,\pi)=\overline{R}(\Theta,\pi)-\sum_{k:\theta^k>\theta^{k^*}}\mu_0^ku_k^*(1)-\mu_0^{k^*}u_{k^*}^*(q^*).
    \end{equation}
\end{theorem}

\begin{figure}[h]
    \centering
    \includegraphics[width=0.8\linewidth]{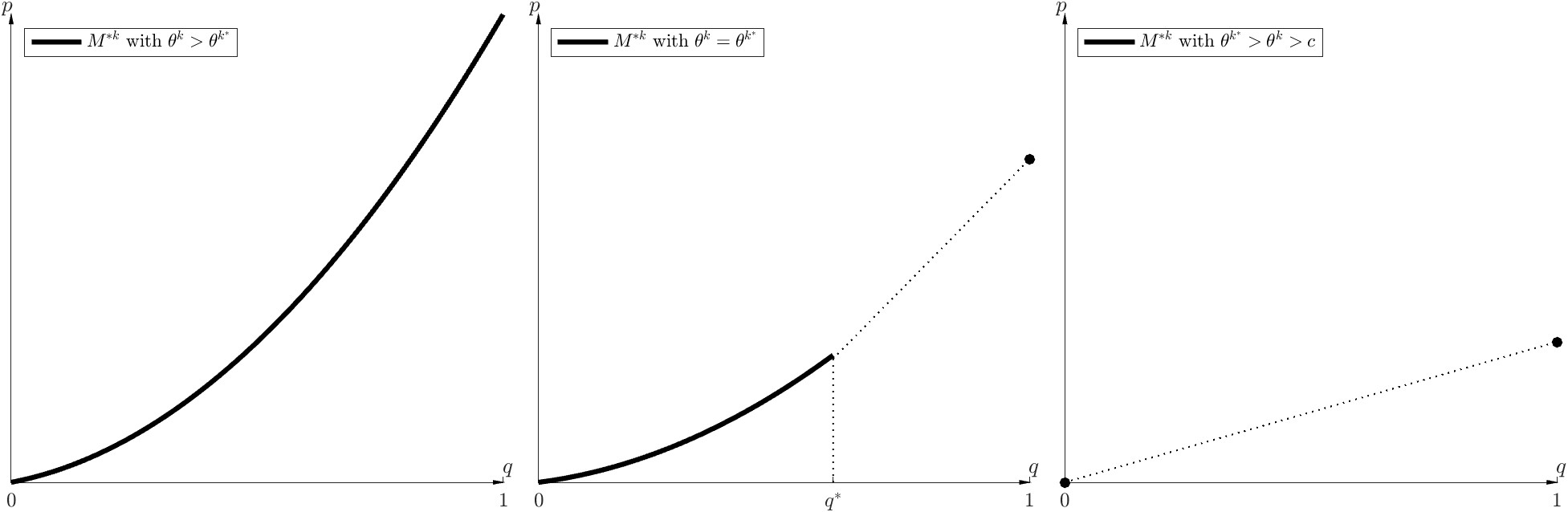}
    \captionsetup{justification=centering}
    \caption{Robustly Optimal Menus for $\theta^k>\theta^{k^*}$, $\theta^k=\theta^{k^*}$, and $\theta^{k^*}>\theta^k>c$.}
    \label{fig-main result}
\end{figure}

One can view the sequential self-conquering process optimally induced by the platform as follows. Assuming the least desirable case of no disclosure occurs, the consumer bears prior skepticism, which pins down the producer's outside value of not disclosing $w_1^N(0)=\max\{\E[\theta]-c,0\}$. To begin with, the platform lures the highest type $\theta^1$ to disclose with a small probability $\delta>0$ by posting a marginal price $\theta^1-c-w_1^N(0)$ equal to her marginal benefit of disclosure, conditional on no other type would disclose. Given this small disclosure of a high type, consumer skepticism deteriorates, and thus the producer's outside value now decreases to $w_1^N(\delta)=\max\{\E[\theta|\text{$\theta^1$ discloses with some probability}]-c,0\}$. Hence, to further induce disclosure of $\theta^1$, the marginal price that can be charged rises to $\theta^1-c-w_1^N(\delta)$. In this way, the platform keeps guaranteeing small probabilities after small probabilities until $\theta^1$ fully discloses. After this, the process moves on to induce full disclosure of the second highest type, while taking the highest type's full disclosure as granted. More generally, for each efficient type $\theta^k>c$, assuming that all types higher than $\theta^k$ fully disclose, the type $\theta^k$ herself discloses with probability $q$, and all types lower than $\theta^k$ choose zero disclosure, the platform tempts $\theta^k$ to disclose marginally by charging the marginal price $\theta^k-c-w_k^N(q)$. Here, $\theta^k-c$ is the social gain from production and transaction, so $w_k^N(q)$ refers to the rent the platform compensates the producer with for giving up the outside value in this candidate equilibrium.

Note that the marginal rent $w_k^N(q)$ drops to 0 after the process reaches the tipping point, which marks the platform has firmly established its market role as the gatekeeper who punishes any non-disclosing producer with the worst outside value. Thereafter, the platform fully extracts the rest of the surplus.

When the sequential self-conquering process reaches some $(k,q)$ before the tipping point, $[0,w_k^N(q)]$ measures the ``room'' for the outside value/consumer beliefs to be coordinated against the platform. The process ensures that, at this point, consumer skepticism cannot be so optimistic that the producer would gain anything higher than $w_k^N(q)$ without disclosure. If the consumer is extremely pessimistic, so the producer faces outside value 0, the platform can fully take away the rest of the surplus. However, if the outside value is anywhere higher, such full extraction will end up with under-disclosure and a low revenue. As a result, although the platform keeps strengthening its surplus-extracting ability as $[0,w_k^N(q)]$ is shrinking, it still suffers from strategic uncertainty; after $[0,w_k^N(q)]$ dwindles to $\{0\}$, the platform finally gains full coordination power and behaves as if it can select equilibrium. 
%(benchmark in Section \ref{benchmark without strategic uncertainty}).

One remarkable feature of the sequential self-conquering process is that a higher type's disclosure probability is always guaranteed before that of a lower type. This may not be anticipated a priori, as \cite{sega2003} shows that for a principal who can offer agent-specific menus to coordinate actions with increasing externalities, she might induce actions alternatingly across agents, instead of luring one agent to play his maximal action before moving on to other agents. In contrast, our platform (principal) prefers the non-alternating manner because the disclosure of producer types (agents) manifests ``increasing'' increasing externalities. That is, to strike down consumer skepticism, thereby charging higher marginal prices later on, it is dominantly more effective to do so by inducing disclosure from a higher type than from a lower type. This structure is dictated simply by the Bayes' rule.

Figure \ref{fig-main result} depicts the robustly optimal menus for the three sorts of efficient types.  As one can observe, before the sequential self-conquering process reaches the tipping point, the platform uses a continuum of plans to continually bring down consumer's skepticism. Since the platform is able to charge a higher marginal price with a lower outside value of the producer, the price functions are strictly convex. After the tipping point, the platform finds it sufficient to offer a bang-bang option to each type.

Our model thus predicts that, for instance, an advertising platform who accounts for strategic risks tends to design sophisticated contracts for high-quality/institutional advertisers, whereas it uses simple linear pricing on low-quality/individual advertisers.

Two extreme cases are worth noticing. If $c\geq v_0$, the tipping point is simply $(1,0)$, so no sophisticated pricing will be employed and the robustly optimal menu profile degenerates to the best-case solution $\overline{M}^{\Theta}$, studied in the benchmark in Section \ref{benchmark without strategic uncertainty}. On the other hand, if the cost is low $c=\theta^K$, the tipping point is just $(K,1)$. In this case, the platform offers a strictly convex price function to every type (except $\theta^K$). More generally, the tipping point under a higher production cost is before that under a lower one.

Finally, we help the reader understand (\ref{multiple type}), the form of the price function $p_k^*$. We build a connection between this multiple-type solution and the binary-type solution (\ref{binary type})
\[p_{\text{binary}}^*(q)=\frac{\overline{\theta}-\E[\theta]}{\overline{\mu}}\ln\frac{1}{1-\overline{\mu}q}.\]
One can see that $p_k^*=p^*_{\text{binary}}$ if we replace above $\overline{\theta}$ with $\theta^k$, $\E[\theta]$ with $\E[\theta|\theta\leq\theta^k]$, and $\overline{\mu}$ with $\Pr(\theta^k|\theta\leq\theta^k)$. In other words, the price function for each type $\theta^k$ can be viewed as the robustly optimal menu for the high type in a binary-type problem, that is constructed by ignoring all types higher than $\theta^k$, seeing $\theta^k$ as the high type, and combining all types lower than $\theta^k$ as the low type.

\subsection{Implications and Comparative Statics}\label{implications and volume-based pricing}

This section highlights several implications of our main result that focus on the platform's informational and allocative role. With the comparative statics results, we demonstrate how these roles of the platform vary with market condition, signal structure, and information asymmetry.

\paragraph{How Market Power Changes with Market Conditions} 
We have seen that the platform can sometimes gain full market power: if $v_0\leq c$, the platform can become the gatekeeper costlessly and take all social surplus in the unique equilibrium. In this case, without the presence of the platform's disclosure service, the producer cannot sell the product to earn a positive profit, and the market does not generate any surplus. However, in the opposite scenario with $v_0>c$, the platform must give the producer positive rents. We are thus interested in how the platform's market power varies with market conditions. To study this, we define the platform's surplus share PSS, the producer's markup MK, and the platform's relative marginal contribution to social efficiency CTE
\[\text{PSS}:=\frac{R^*}{\overline{R}};\qquad\text{MK}:=\max\{v_0-c,0\}; \qquad\text{CTE}:=\frac{\overline{R}-\text{MK}}{\overline{R}}.\]
The markup MK measures the producer's market power absent the platform, and the relative marginal contribution CTE captures how important the platform is for creating additional social efficiency. When $v_0\leq c$, CTE=1, MK=0, and PSS=1, while when $v_0 > c$,  CTE<1, MK>0, and PSS<1. Given this, conjecturing from a bargaining point of view, one may expect that one side of the market will obtain a greater share of surplus (higher PPS) when the party is more valuable in the relationship (higher CTE) and when the other side's outside option becomes worse (lower MK). In fact, we show this is true in general
\begin{proposition}\label{proposition-market power}
    If $v_0>c$, as $c$ increases, PSS and CTE strictly increase, and MK strictly decreases. 
\end{proposition}
By shifting the production cost $c$, Proposition \ref{proposition-market power} identifies a positive correlation between CTE and PSS, and a negative correlation between MK and PSS. Since both total surplus and producer rents decrease in the cost $c$, a priori it is not clear how PSS will change. Thus, perhaps surprisingly, Proposition \ref{proposition-market power} implies that the rent reduction always dominates the decrease in total surplus. This result is in sharp contrast to the best-case benchmark, where the platform always enjoys full bargaining power under all parameters. This exercise highlights how the platform's robustness motives can generate tension among market participants' bargaining power that nontrivially varies with market conditions.

\paragraph{How Revenue Changes as Information Asymmetry Vanishes} In Section \ref{benchmark without strategic uncertainty}, we pointed out that the best-case revenue manifests abnormal behavior as it is discontinuous when the information asymmetry between producer and consumer vanishes. This is abnormal as one might expect the platform to lose its market power as its service has diminishing value on consumers' choices.
We now check whether the worst-case revenue suffers from the same issue or not. When the prior of type approximates a degenerate distribution, uncertainty vanishes, so one would anticipate the disclosure service to have no value. We show this is the case with our robust objective as the platform's maximal revenue guarantee also vanishes
\begin{proposition}
\label{proposition-converge0}
    For any sequence of priors $(\mu_n)_{n=1}^{\infty}$ over a fixed type space $\Theta$, such that $\mu_n^{k_0}$ converges to 1 for some type $\theta^{k_0}\in\Theta$, the associated maximal revenue guarantee $R^*_n$ converges to 0.
\end{proposition}

Proposition \ref{proposition-converge0} hence implies that our robust focus is less sensitive to the change in the informational environment than the benchmark setup.

\paragraph{The Platform's Informational Role} In the presence of a nontrivial production cost $c>\theta^K$, information transmission is no longer neutral (in contrast to, e.g., \cite{ahls2022} who rule out inefficient types). The socially efficient outcome requires a product be produced and consumed if and only if it is efficient. Perhaps surprisingly, despite our platform anticipating the worst-case equilibrium, it optimally induces full disclosure of efficient types, so social efficiency is obtained in the unique equilibrium. The result entails the following observation that a platform able to price-discriminate and commit to partial disclosure will optimally internalize the social gain from information transmission.

To push this observation further, we investigate how the platform's revenue guarantee changes with the signal structure. We say that one signal structure $(\Theta,\pi)$ \emph{Blackwell dominates} another signal structure $(\Theta',\pi')$ if there is $h:\Theta\rightarrow\Delta(\Theta')$ such that $\pi'(\theta'|\omega)=\sum_{\theta\in\Theta}h(\theta'|\theta)\pi(\theta|\omega)$.

\begin{proposition}
\label{proposition-comparative}
    Consider two signal structures $(\Theta,\pi)$ and $(\Theta',\pi')$. If $(\Theta,\pi)$ Blackwell dominates $(\Theta',\pi')$, $R^*(\Theta,\pi)\geq R^*(\Theta',\pi')$. Moreover, the full-revealing signal structure maximizes $R^*(\Theta,\pi)$.
\end{proposition}

Proposition \ref{proposition-comparative} asserts that the platform prefers more informative signal structures. Furthermore, there are applications where the platform also controls the information available to the producer for disclosure. For instance, Amazon extracts keywords from product reviews, and YouTube gives out creator awards; information design and pricing are also widely considered if we interpret our platform as a certification seller. Across these examples, our result predicts that the platform will find it sufficient to design price while causing no harm to information transmission, by choosing perfect information and inducing full disclosure. To compare with a nice benchmark, \cite{ahls2022} consider the same friction while restricting the platform's pricing power. As a result, their platform's optimal behavior typically induces partial disclosure and imperfect signal structure.

To see the insight, we rewrite the revenue guarantee (\ref{MRG*}) in a more generic form
\[\overline{R}(\Theta,\pi)-\text{``summation'' of consumer skepticism/producer outside value ``in a certain order''}.\]
Here, $\overline{R}$ is the total surplus, also the frictionless revenue absent strategic uncertainty (see Section \ref{benchmark without strategic uncertainty}). The latter part corresponds to the total compensation to the producer for giving up outside values in bad equilibria, with the ``certain order'' refering to the order in which bad equilibria are eliminated.

Since we know the full-revealing signal structure maximizes total surplus, the remaining goal is to see why the socially efficient communication also minimizes consumer skepticism. As explained previously, the ``increasing'' increasing externalities across different types' disclosure actions dictate that the platform wants to first induce disclosure from the highest possible type, which is equal to the highest state $\theta^1=\omega^1$. Also, the disclosure of this type should always be prioritized until its probability is depleted. Given the full disclosure of $\theta^1=\omega^1$, the platform aims to find the next highest possible type, which now equals the second highest state $\theta^2=\omega^2$. Likewise, full disclosure is guaranteed to minimize skepticism. Following similar logic, the platform eventually chooses full-revealing $\Theta=\Omega$ and attains full disclosure as the robustly optimal joint design.

%{\color{red}We highlight several implications of Theorem \ref{main result} regarding disclosure pricing. The implications cover the platform's informational and allocative roles. We also discuss the comparative statics that shift the information structure that generates the hard evidence.

%\paragraph{The Platform's Informational Role} Despite the platform anticipating the worst-case equilibrium, it still finds it optimal to induce full disclosure. This feature stems from the platform's ability to offer type-dependent prices. A platform without such ability (e.g., \cite{ahls2022}) typically fails to induce full disclosure. Full disclosure does occur in the classic disclosure models studied by \cite{gros1981} and \cite{milg1981}, where the producer can self-advertise; however, their settings differ significantly from ours in terms of surplus allocation.

\paragraph{The Platform's Allocative Role}
Above, we have studied how surplus changes across markets. Here, we illustrate how surplus is divided within the same market. Denote each producer type $\theta^k$'s surplus in the induced unique equilibrium as \(W(\theta^k) := \int_0^1w_k^N(q)dq\), where the marginal rent $w_k^N(q)$ is defined by (\ref{w^N_k(q)}). Note that according to Theorem \ref{main result}, $W(\theta^k)>0$ if and only if $\theta^k>\theta^{k^*}$, or $k=k^*$ and $q^*>0$. We thus characterize the surplus distribution across producer types

\begin{proposition}\label{proposition-producer surplus}
    For all $\theta^k>\theta^{k^*}$, the producer's surplus is strictly bounded by $\E[\theta|\theta\geq\theta^k]>W(\theta^k)+c>\E[\theta|\theta\geq\theta^{k+1}]$. Hence, $W(\theta^{k_1})\geq W(\theta^{k_2})$ whenever $\theta^{k_1}>\theta^{k_2}$, and the inequality is strict if $\theta^{k_1}>\theta^{k^*}$.
\end{proposition}

Proposition \ref{proposition-producer surplus} highlights that the robustly optimal pricing scheme prioritizes attracting higher types into business by offering them higher rents. One novel feature of our solution is that the platform allocates more rents to the top types while inducing socially efficient action for every type. This is uncommon in models where a principal faces other frictions (for example, one with screening motives).

Under sequential self-conquering, before the tipping point is reached, the skepticism value is increasing in type and decreasing in disclosure probability:
\begin{equation*}
    \begin{aligned}
        &w_k^N(q)>w_{k'}^N(q'),\text{ for all }\theta^k>\theta^{k'}\text{ and } q,q'; \\
        &w_k^N(q)<w_k^N(q'),\text{ for all }q>q'\text{ and }k.
    \end{aligned}
\end{equation*}
This monotonic rent structure is not a result of, for example, incentive compatibility, as in an environment with adverse selection; instead, it stems from the platform's optimal order of "conquering." Namely, to induce disclosure from the highest type first, the platform compensates her for relinquishing significant outside values; nonetheless, once the highest type's disclosure is guaranteed, the outside values diminish, so the compensation required for lower types is reduced.

However, being offered higher rents does not necessarily mean that the higher types favor the platform's presence. Consider $\theta^k>\theta^{k^*}$. Then, Proposition \ref{proposition-producer surplus} says that each producer type is strictly worse off than she is without the platform. To see this, note that the upper bound of \(W(\theta^k)\), \(\mathbb{E}[\theta|\theta \leq \theta^k]-c\), is strictly lower than both \(\theta^k-c\) and the prior skepticism value \(v_0-c\). These three values correspond to the surplus levels in three benchmarks: (i) if the platform was restricted to offering only binary disclosure options, \(W(\theta^k)\) would be \(\mathbb{E}[\theta | \theta \leq \theta^k]-c\); (ii) if the platform was absent and the producer can self-disclose, as in \cite{gros1981} and \cite{milg1981}, \(W(\theta^k)\) would be \(\theta^k-c\); (iii) if the platform was absent and no disclosure can happen, \(W(\theta^k)\) would be \(v_0-c\). Thus, Proposition \ref{proposition-producer surplus} shows that strictly higher type-wise surplus is extracted by our platform than in alternative setups.

%One might expect that the price charged to higher types is also higher. However, this is not necessarily the case. For example, consider a scenario with three types in which the two higher types have similar values but the prior assigns little probability to the highest type and almost all to the middle type. In this case, the middle type pays a higher price than the highest type does. This occurs because the high type's full disclosure has little impact on consumer skepticism, while the middle type's disclosure reduces consumer skepticism drastically, resulting in much lower rent and thus a higher price (recall their values are almost the same).

\subsection{Sketch of Proof}

It is challenging to prove Theorem \ref{main result}: to robustly induce a probability profile and maximize revenue, the platform aims to "optimally" break every under-disclosure equilibrium by creating a profitable deviation for some type; the platform faces two questions: (i) Which type is incentivized to deviate? and (ii) To which plan does the type deviate? For binary types, the answer to (i) is straightforward: the high type, since offering more than the outside option to the low type is unprofitable. With additional arguments, (ii) is answered by local upward deviations, showing the optimality of (\ref{binary type}). However, in the presence of multiple types, the answers to both (i) and (ii) are a priori ambiguous.

%As we previously discussed,

As we will show later in this section, the answer turns out to be that some equilibria are broken by the local upward deviation of the highest type who has not yet fully disclosed in each of these equilibria, while other equilibria are broken by the global downward deviation of some low type who is disclosing too much. Nonetheless, proving this claim remains a challenge.
The entire proof consists of four steps, and the first three steps are supported by three corresponding lemmas, as discussed in Section \ref{truncated convexifications}-\ref{upper bound for maximal revenue guarantee}. The four steps are as follows:

\emph{Step 1} - We show that if a menu profile robustly induces a certain probability profile, it does so even if we convexify each menu and delete all the plans that offer probabilities strictly higher than the robustly induced probability. We hence focus on such upper-truncated convexifications.

\emph{Step 2} - We develop an algorithm that explicitly finds, for every menu profile, an alternating conquering path that creates pointwise upper bounds on the derivatives of each price function. 
To show the existence of such a path, we utilize the fact that every under-disclosing strategy profile cannot form an equilibrium. An integration that sums up the upper bounds on price derivatives %(\ref{pointwise upper-boundedness}) 
yields an upper bound for the revenue guaranteed by the menu profile for which we have found the path.

\emph{Step 3} - We domonstrate that the sequential self-conquering path provides a revenue upper bound that is no less than any alternating conquering path can give. To show this, we construct an improvement for every alternating conquering path by exchanging the order of ``conquering'' to let a high type's disclosure probability be induced before that of a low type. By continuing such improvement, we eventually obtain the sequential self-conquering path. Therefore, the sequential self-conquering path determines an upper bound for the maximal revenue guarantee $R^*$.

\emph{Step 4} - We take a sequence of menu profiles $(M_n^{\Theta})_{n=1}^{\infty}$, where for all $n$ and $k$, we reduce the price of each plan $(q,p)\in M^{*k}$ to $p-\frac{\varepsilon}{n}q$ (with a small $\varepsilon>0$). This sequence converges to $M^{*\Theta}$. We verify that every $M_n^{\Theta}$ induces full disclosure in the unique equilibrium and yields a revenue that approximates the upper bound for $R^*$ found in Step 3. So, $R^*$ equals this upper bound and $M^{*\Theta}$ is robustly optimal.

\subsubsection{Truncated Convexifications}\label{truncated convexifications}

The first step shows that it is without loss to focus on a subset of menu profiles that (i) consist of lower-convexified menus and (ii) robustly induce a plan with maximal probabilities.

To state the result, we define the \emph{price envelope} of a menu $M$ that is denoted by $\check{p}_M:[0,q_M]\rightarrow\mathbb{R}_+$ , where $q_M=\max\{q:(q,p)\in M\}$ is the maximal probability offered in $M$. The price envelope is defined as the highest convex function lying below the menu, given by $\check{p}_M(q)=\min\{p:(q,p)\in co(M)\}$, where $co(M)$ is the convex hull of $M$. We further call the graph of the price envelope the \emph{(lower-)convexification} of $M$, denoted as $\conv M$. Due to convexity, $\check{p}_M(.)$ has nondecreasing left and right derivatives at all interior points, denoted by $\check{p}_M(.-0)$ and $\check{p}_M(.+0)$, respectively. For simplicity, we sometimes denote by $p_k(.)$ the price envelope of menu $M^k$ when a menu profile $M^{\Theta}$ is specified in the contexts.

%Lemma \ref{lemma-simplify} says that if a menu profile $M^{\Theta}$ robustly induces some choice profile, it remains a worst-case equilibrium play even if, for every menu, we delete (i) all plans that lie above the convexification, and (ii) all plans that have higher probabilities than the robustly induced one. We call the resulting menu profile $\widehat{M}^{\Theta}$ defined in Lemma \ref{lemma-simplify} the \emph{truncated convexification} of $M^{\Theta}$, or sometimes \emph{convexified menus} in short.

\begin{lemma}\label{lemma-simplify}
    If a menu profile $M^{\Theta}$ robustly induces some contingent choices $\underline{r}^{\Theta}$, the menu profile $\widehat{M}^{\Theta}$ given by, for all $k$, $\widehat{M}^k=\{(q,p)\in M^k\cap\conv M^k:q\leq\underline{q}^k\}$ also robustly induces $\underline{r}^{\Theta}$.
\end{lemma}

Lemma \ref{lemma-simplify} says that if a menu profile $M^{\Theta}$ robustly induces some choice profile, it remains a worst-case equilibrium play even if, for every menu, we delete (i) all plans that lie above the convexification and (ii) all plans that have higher probabilities than the robustly induced one. We call the resulting menu profile $\widehat{M}^{\Theta}$ defined in Lemma \ref{lemma-simplify} the \emph{truncated convexification} of $M^{\Theta}$, or  \emph{convexified menus} for short.

\subsubsection{Pointwise Bounded Paths}\label{pointwise bounded paths}

%In fact, an important lemma discussed later, Lemma \ref{lemma-path}, implies that every menu profile corresponds to such a partial and alternating path such that the points on the path provide upper bounds for all marginal prices induced by the menus. Loosely speaking, for all $k$ and $\widetilde{q}\in[0,1]$, there is a probability profile $q^{\Theta}$ on the path such that its $k$th component $q^k=\widetilde{q}$ and:
%\begin{equation}\label{pointwise upper-boundedness}p_k'(\widetilde{q})\leq\theta^k-w^N(q^{\Theta}),
%\end{equation}
%Notice that (\ref{local upper bound}) satisfies the complete divide-and-conquer version of (\ref{pointwise upper-boundedness}). We say that a path that satisfies (\ref{pointwise upper-boundedness}) has the property of \emph{pointwise upper-boundedness}.

The second step in proving Theorem \ref{main result} is to show our central lemma that maps every menu profile to an alternating conquering path that offers an upper bound for the revenue guaranteed by the menu profile. We define (and rename) an alternating conquering path as follows:

\begin{definition}\label{nondecreasing alternating path}
    \emph{A }nondecreasing alternating path\emph{ from probability profile $\widetilde{q}_S^{\Theta}$ to profile $\widetilde{q}_E^{\Theta}$ with $\widetilde{q}_S^{\Theta}\leq\widetilde{q}_E^{\Theta}$ is a sequence of tuples $(q_t^{\Theta},k_t)_{t=0}^T$, where $k_t\in\Theta$ and $T$ can be infinite. The sequence satisfies
    \begin{enumerate}[nolistsep]
        \item $q_0^{\Theta}=\widetilde{q}_S^{\Theta}$ and $\lim_{t\rightarrow T}q_t^{\Theta}=\widetilde{q}_E^{\Theta}$;
        \item for all $t<T$, $q_t^{k_t}<q_{t+1}^{k_t}$ and $q_t^{-k_t}=q_{t+1}^{-k_t}$.
    \end{enumerate}}
\end{definition}

In other words, a nondecreasing alternating path specifies the step number $T$ and the moving direction $k_t$ for each step $t<T$. At every step $t$, the probability profile must increase its $k_t$th component and leave the other components unchanged. We also call $k_t$ the \emph{direction type} and the interval $[q_t^{k_t},q_{t+1}^{k_t}]$ the \emph{path stride} of step $t$. 

\begin{figure}[h]
    \centering
    \includegraphics[width=0.6\linewidth]{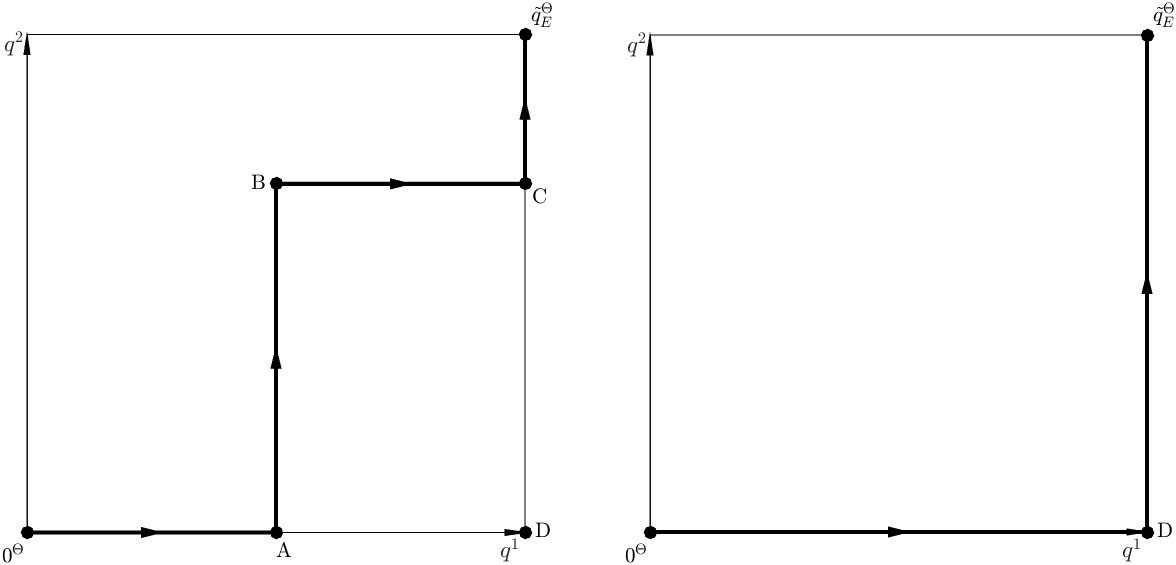}
    \captionsetup{justification=centering}
    \caption{A Typical Nondecreasing Alternating Path (left) and\\the Sequential Self-Conquering Path (right).}
    \label{fig-nondecreasing alternating path}
\end{figure}

To illustrate, we consider a three-type example. We start with zero disclosure, $\widetilde{q}_S^{\Theta}=0^{\Theta}$. Let $\widetilde{q}_E^K=0$ as the lowest type will not disclose, so it suffices to consider moving in two directions. The left subfigure of Figure \ref{fig-nondecreasing alternating path} depicts a possible nondecreasing alternating path from $0^{\Theta}$ to $\widetilde{q}_E^{\Theta}$. In this case, the step number is $T=4$, and the direction types are $k_0=k_2=1$ and $k_1=k_3=2$. The path is always nondecreasing as it moves only by increasing either $q^1$ or $q^2$ while holding the other directions unchanged.

We now state the following important lemma:

\begin{lemma}\label{lemma-path}
    If a menu profile $M^{\Theta}$ robustly induces some probability profile $\underline{q}^{\Theta}$, there exists a nondecreasing alternating path $(q^{\Theta}_t,k_t)_{t=0}^T$ from $0^{\Theta}$ to some $\widehat{q}^{\Theta}\leq\underline{q}^{\Theta}$ such that $\theta^{k_t}\geq c$ for all $t<T$, and
    \begin{equation}\label{upper bound}
        R(M^{\Theta})\leq\sum_{t=0}^{T-1}\mu_0^{k_t}\int_{q^{k_t}_t}^{q^{k_t}_{t+1}}\left[\theta^{k_t}-c-w^N(\widetilde{q}^{k_t},q^{-k_t}_t)\right]d\widetilde{q}^{k_t}.
    \end{equation}
\end{lemma}

In the proof of Lemma \ref{lemma-path}, we in fact show a stronger version of the result which says that there is a pointwise upper bound for every marginal price. In particular, each step $t$ has
\begin{equation}\label{path local upper bound}
 p_{k_t}'(\widetilde{q})\leq\theta^{k_t}-c-w^N(\widetilde{q},q_t^{-k_t})\text{, for all }\widetilde{q}\in[q_t^{k_t},q_{t+1}^{k_t}].
\end{equation}
At each step, the path raises the disclosure probability of the direction type that wants to deviate locally upward at each probability profile in the path stride; the other types' probabilities remain unchanged. Put differently, the left-hand side and the right-hand side of (\ref{path local upper bound}) represent the marginal cost and benefit of disclosure, respectively, for type $\theta^k$ in the candidate equilibrium where the probability profile is $(\widetilde{q},q_t^{-k_t})$. By integrating the left-hand side of (\ref{path local upper bound}) along the path weighted by prior probability $\mu_0^k$, we can obtain the revenue in the worst-case equilibrium $R(M^{\Theta})$, so the integral with respect to the right-hand side of (\ref{path local upper bound}) produces an upper bound for $R(M^{\Theta})$.

The proof of Lemma \ref{lemma-path} relies on an algorithm that explicitly finds a nondecreasing alternating path recursively. Initially, the path starts at $q_0^{\Theta}=0^{\Theta}$. At every step $t$, as long as the path's current position has not reached the worst-case equilibrium outcome, $\underline{q}^{\Theta}$, we find a type $\theta^k$ that wants to deviate locally upward, which is shown to be always possible given our knowledge that any disclosure outcomes other than the worst case, $\underline{q}^{\Theta}$, cannot form an equilibrium. We then let $k$ be this new direction type and push the path toward this direction until it meets the first point where this type no longer wants to deviate locally upward. After this, the next step begins. In this way, we guarantee (\ref{path local upper bound}) holds along the whole path. Finally, we show the algorithm must end within countable steps, exploiting the fact that the path cannot converge to a point that produces a revenue lower than $R(M^{\Theta})$ since, otherwise, this limit point would form a worse equilibrium than the worst-case equilibrium.

\subsubsection{An Upper Bound for Maximal Revenue Guarantee}\label{upper bound for maximal revenue guarantee}

We have shown that the revenue guarantee of every menu profile is controlled by an upper bound determined by a nondecreasing alternating path. Each such path has a fixed starting point $0^{\Theta}$ and a flexible endpoint $\underline{q}^{\Theta}\leq1^{\Theta}$. To further bound the maximal revenue guarantee $R^*(\Theta,\pi)$, we search among all such paths to maximize the path-specific bound (\ref{upper bound}).

The following result states that the sequential self-conquering path generates the maximal bound:

\begin{lemma}\label{lemma-upper bound}
    The bound (\ref{upper bound}) is maximized by the endpoint $\underline{q}^{*\Theta}:=(1,...,1,0,...,0)$ with $\underline{k}$ ones, where $\theta^{\underline{k}}$ is the lowest type that is larger than $c$, and the sequential self-conquering path $(q_t^{\Theta},k_t)_{t=0}^{\underline{k}}$, where for all $t<\underline{k}$, we have $k_t=t+1$ and $q_t^{\Theta}=(1,...,1,0,...,0)$ with $t$ ones. Moreover, the maximal value of (\ref{upper bound}) equals (\ref{MRG*}).
\end{lemma}

In Figure \ref{fig-nondecreasing alternating path}, the sequential self-conquering path is depicted by the subfigure on the right-hand side.

The insight for the optimality of sequential self-conquering is related to the intuition we gave in the binary-type example: in a candidate equilibrium where high types have more disclosure, the consumer becomes more skeptical when seeing no disclosure, which makes all producer types' disclosure easier to induce in ``future'' candidate equilibria because they face worse outside value of not disclosing.

To see this more formally, consider finding the optimal path while fixing the endpoint to be $\underline{q}^{*\Theta}$. The total ex-ante disclosure probability of $\underline{q}^{*\Theta}$ is $Q_M:=\sum_{k=1}^{\underline{k}}\mu_0^k$. Therefore, we use a quantile $Q\in[0,Q_M]$ to parameterize each such path $(q_t^{\Theta},k_t)_{t=1}^T$. That is, for every $Q$, we can find the step $t'$ such that
\begin{equation*}
    \underline{Q}_{t'}:=\sum_{t=0}^{t'-1}\mu^{k_t}_0(q_{t+1}^{k_t}-q_t^{k_t})\leq Q\leq\overline{Q}_{t'}:=\sum_{t=0}^{t'}\mu^{k_t}_0(q_{t+1}^{k_t}-q_t^{k_t}).
\end{equation*}
Let $t'$ vary with $Q$ as above, and re-denote the path by
\begin{equation*}
    q^{\Theta}_Q=(q^{k_{t'}}_{t'}+(Q-\underline{Q}_{t'})/\mu^{k_{t'}}_0,q^{-k_{t'}}_{t'})\text{, for all }Q\in[0,Q_M].
\end{equation*}
In other words, $Q$ denotes the total ex-ante probability of disclosure that has been robustly induced along the path. $t'$ represents the path step in which $Q$ is reached, and $q_Q^{\Theta}$ marks the position by which the path has induced total ex-ante probability $Q$. The bound (\ref{upper bound}) can then be rewritten as
\begin{equation}\label{upper bound-reformulate}
    \begin{aligned}
        \sum_{t=0}^{T-1}\mu_0^{k_t}\int_{q^{k_t}_t}^{q^{k_t}_{t+1}}&\left[\theta^{k_t}-c-w^N(\widetilde{q}^{k_t},q^{-k_t}_t)\right]d\widetilde{q}^{k_t} \\
        &=\sum_{t=0}^{T-1}\mu_0^{k_t}\int_{q^{k_t}_t}^{q^{k_t}_{t+1}}(\theta^{k_t}-c)d\widetilde{q}^{k_t}-\sum_{t=0}^{T-1}\mu_0^{k_t}\int_{q^{k_t}_t}^{q^{k_t}_{t+1}}w^N(\widetilde{q}^{k_t},q^{-k_t}_t)d\widetilde{q}^{k_t} \\
        &=\E[\theta-c|\theta\geq c]-\sum_{t=0}^{T-1}\int_{\underline{Q}_t}^{\overline{Q}_t}w^N(q_Q^{\Theta})dQ \\
        &=\overline{R}-\int_0^{Q_M}w^N(q_Q^{\Theta})dQ.
    \end{aligned}
\end{equation}
Hence, we decompose the revenue bound into two simple parts: the full surplus $\overline{R}$ and $\int_0^{Q_M}w^N(q_Q^{\Theta})dQ$. Only the latter part is path-specific and determined by how consumer skepticism changes along the path. This form expresses the previous insight formally: an optimal path constructs a mapping $q_Q^{\Theta}$ that drives the skepticism value down as ``fast'' as possible (regard $Q$ as ``time'') to minimize the total compensation to the producer $\int_0^{Q_M}w^N(q_Q^{\Theta})dQ$ for giving up outside values in the candidate equilibria along the path.

%The most effective strategy is to prioritize the highest type whose probability has not yet been depleted.

The proof of Lemma \ref{lemma-upper bound} investigates the following \emph{quantile-skepticism value} each path induces
\begin{equation}\label{quantile-skepticism value}
    v^N(Q):=\E[\theta|\mu^N(q_Q^{\Theta})],
\end{equation}
Which equals the consumer's skepticism posterior mean when the path arrives at $q_Q^{\Theta}$. Note that $w^N(q_Q^{\Theta})=\max\{v^N(Q),c\}-c$, so (\ref{upper bound-reformulate}) indicates that the platform's goal is to minimize $\int_0^{Q_M}\max\{v^N(Q),c\}dQ$. To show the sequential self-conquering path achieves the minimum, we take any path that is not sequential self-conquering, which means we can always find two consecutive steps in which the direction type of the first step is lower than that of the second step, meaning the platform ``conquers'' a low type before a high type. We show that by exchanging these two steps so as to ``conquer'' the high type first, the path ends up producing a pointwise weakly lower quantile-skepticism value. In fact, all paths can be improved with a sequence of such exchanges until they become (or approximate) sequential self-conquering. This implies the sequential self-conquering path yields a quantile-skepticism value that is pointwise weakly lower than any other path's quantile-skepticism value.

\section{Extensions}\label{extensions}

We discuss three extensions of our main model. %Section \ref{joint design of evidence structure} analyzes a platform that can jointly design the information structure and the menu profile. We show that the platform benefits from choosing a more informative signal and the optimal choice fully reveals product quality.
%, based on which we compare our model with that of \cite{ahls2022}.
Section \ref{random menus} allows the platform to offer random menus (in the sense of \cite{halr2021}). We show that it suffices to consider deterministic menus for achieving the maximal revenue guarantee, whereas a class of random menus is equivalent to the robustly optimal menu profile and takes a simple form that merely offers full-disclosure plans.
Section \ref{iterated deletion of dominated strategies} demonstrates that the unique equilibrium induced by the robustly optimal menu profile can also be solved from the iterated deletion of strictly dominated strategies. Thus, our solution is robust under weaker assumptions on player behavior. Section \ref{advertising costs} shows that the main features of our solution are preserved even when the platform faces disclosure costs with arbitrary functional form.

\subsection{Random Menus}\label{random menus}

\cite{halr2021} point out that introducing contractual uncertainty may be able to improve the principal's payoff guarantee when contracting in the presence of externalities. This section illustrates that, in our framework, however, making menus random does not improve the platform's revenue guarantee. This contends that our platform wields pricing power that is sufficient for extracting surplus maximally. On the other hand, the format of random menus provides an alternative implementation of the robustly optimal menu profile: the platform offers a type-dependent random price for full disclosure alone. This implementation may be more applicable to settings in which the platform finds it difficult to commit to partial disclosure. For example, certification agencies only decide whether to issue certification for the buyer; whereas, the disclosure decision is within the buyer's discretion. We relegate the formal definitions and analysis to Online Appendix and leave only two informal statements.

\begin{proposition}[Informal]\label{proposition-random menus}
    Every random menu profile is outcome-equivalent to a deterministic one.
\end{proposition}
The key to such equivalence is the fact that when computing the skepticism belief, only each type's expected disclosure probability matters. An immediate implication of Proposition \ref{proposition-random menus} is that it suffices to achieve the maximal revenue guarantee by using deterministic menu profiles, as in the main model.

Nonetheless, allowing for random menus enriches the possible forms of robustly optimal menu profiles. One particularly interesting case is that the platform only provides full-disclosure but with a type-dependent random price. To imagine how this format can correspond to a useful implementation, consider a platform that publicly sets a single price for guaranteeing that the consumer is informed while privately offering dispersed discounts to all producers.

\begin{proposition}[Informal]\label{proposition-bang bang price dispersion}
There is a robustly optimal random menu profile that only sells full disclosure but with a type-dependent random discount. In the unique equilibrium all producers with $\theta \geq c$ purchase full disclosure regardless of the realization of the random discount.
\end{proposition}

\subsection{Rationalizable Outcomes}\label{iterated deletion of dominated strategies}

The term ``strategic uncertainty'' may be used to refer to the presence of multiple equilibria or multiple rationalizable outcomes. This section shows that the choice of definition does not matter: the menu profile in Theorem \ref{main result} remains robustly optimal even if the platform seeks to maximize its revenue guarantee among all rationalizable outcomes.

In Section \ref{simplifying disclosure equilibria}, we showed that a disclosure equilibrium induced by a menu profile simply consists of the producer's contingent choices of disclosure plans $r^{\Theta}$ and the consumer's skepticism belief $\mu^N$. The implicit strategy of the consumer is his decision about, facing the producer's price offer, whether to purchase the product when observing the type $\theta\in\Theta$ or not observing disclosure, which we simply denote as $\theta=N$. Thus, the consumer's formal strategy is a mapping $a: (\Theta\cup \{N\})\times\mathbb{R}\to\{0,1\}$, where $a=1$ means the consumer purchases the product.

With this extended model, we consider the simultaneous move game between the producer and the consumer, where the consumer's strategy is a function $a$ defined above and the producer's strategy $r^{\Theta}$ maps her types to a (potentially mixed) choice of plan. Both parties choose their strategies to maximize their ex-ante expected utility. According to the Harsanyi transformation, a Nash equilibrium of this game is also an equilibrium in our main specification.
To make the main text compact, we relegate the formal definitions and analysis to Appendix \ref{rationalizable outcomes formal} and leave only an informal statement.
\begin{proposition}[Informal]\label{proposition-iterated strict dominance}
The robustly optimal menu profile in Theorem \ref{main result} also induces a unique rationalizable outcome in this producer--consumer simultaneous move game.
\end{proposition}

\paragraph{Private Signals}
As a final remark, the Harsanyi transformation, together with Propositions \ref{proposition-random menus} and \ref{proposition-iterated strict dominance}, also allows us to generalize our result to settings in which the producer observes a private signal $s$ about her quality. In this case, the producer’s disclosure plan $r(\theta, s)$ can depend on both the evidence $\theta$ and the signal $s$. However, if the evidence $\theta$ is sufficiently informative such that $s$ is conditionally independent of the state $\omega$ given $\theta$, then any disclosure strategy $r(\theta, s)$ can be replicated by a random disclosure plan $\tilde{r}(\theta, \epsilon)$ that depends only on $\theta$ and a random variable $\epsilon$ that has the same conditional distribution as $s$ given $\theta$.

\subsection{Disclosure Costs}\label{advertising costs}

In our previous analysis, we assumed that disclosure is costless for the platform. We generalize our result to a setting where the platform incurs a probability-dependent cost $c(q)$ when implementing a plan with disclosure probability $q\in[0,1]$. We do not impose any structure on the costs except that $c(.)$ must be continuous to guarantee solution existence. We leave the detail to Online Appendix.

%Furthermore, in many cases such as $c(1)=\infty$ (e.g., \cite{butt1977}), full disclosure is clearly no longer optimal.

%\section{Conclusion}

\bibliographystyle{apacite}
\bibliography{bib}

\begin{thebibliography}{}

\bibitem [\protect \citeauthoryear {%
Ali%
, Haghpanah%
, Lin%
\BCBL {}\ \BBA {} Siegel%
}{%
Ali%
\ \protect \BOthers {.}}{%
{\protect \APACyear {2022}}%
}]{%
ahls2022}
\APACinsertmetastar {%
ahls2022}%
\begin{APACrefauthors}%
Ali, S\BPBI N.%
, Haghpanah, N.%
, Lin, X.%
\BCBL {}\ \BBA {} Siegel, R.%
\end{APACrefauthors}%
\unskip\
\newblock
\APACrefYearMonthDay{2022}{}{}.
\newblock
{\BBOQ}\APACrefatitle {How to sell hard information} {How to sell hard information}.{\BBCQ}
\newblock
\APACjournalVolNumPages{The Quarterly Journal of Economics}{137}{1}{619--678}.
\PrintBackRefs{\CurrentBib}

\bibitem [\protect \citeauthoryear {%
Ben-Porath%
, Dekel%
\BCBL {}\ \BBA {} Lipman%
}{%
Ben-Porath%
\ \protect \BOthers {.}}{%
{\protect \APACyear {2018}}%
}]{%
bedl2018}
\APACinsertmetastar {%
bedl2018}%
\begin{APACrefauthors}%
Ben-Porath, E.%
, Dekel, E.%
\BCBL {}\ \BBA {} Lipman, B\BPBI L.%
\end{APACrefauthors}%
\unskip\
\newblock
\APACrefYearMonthDay{2018}{}{}.
\newblock
{\BBOQ}\APACrefatitle {Disclosure and choice} {Disclosure and choice}.{\BBCQ}
\newblock
\APACjournalVolNumPages{The Review of Economic Studies}{85}{3}{1471--1501}.
\PrintBackRefs{\CurrentBib}

\bibitem [\protect \citeauthoryear {%
Ben-Porath%
, Dekel%
\BCBL {}\ \BBA {} Lipman%
}{%
Ben-Porath%
\ \protect \BOthers {.}}{%
{\protect \APACyear {2019}}%
}]{%
bedl2019}
\APACinsertmetastar {%
bedl2019}%
\begin{APACrefauthors}%
Ben-Porath, E.%
, Dekel, E.%
\BCBL {}\ \BBA {} Lipman, B\BPBI L.%
\end{APACrefauthors}%
\unskip\
\newblock
\APACrefYearMonthDay{2019}{}{}.
\newblock
{\BBOQ}\APACrefatitle {Mechanisms with evidence: Commitment and robustness} {Mechanisms with evidence: Commitment and robustness}.{\BBCQ}
\newblock
\APACjournalVolNumPages{Econometrica}{87}{2}{529--566}.
\PrintBackRefs{\CurrentBib}

\bibitem [\protect \citeauthoryear {%
Bernstein%
\ \BBA {} Winter%
}{%
Bernstein%
\ \BBA {} Winter%
}{%
{\protect \APACyear {2012}}%
}]{%
bewi2012}
\APACinsertmetastar {%
bewi2012}%
\begin{APACrefauthors}%
Bernstein, S.%
\BCBT {}\ \BBA {} Winter, E.%
\end{APACrefauthors}%
\unskip\
\newblock
\APACrefYearMonthDay{2012}{}{}.
\newblock
{\BBOQ}\APACrefatitle {Contracting with heterogeneous externalities} {Contracting with heterogeneous externalities}.{\BBCQ}
\newblock
\APACjournalVolNumPages{American Economic Journal: Microeconomics}{4}{2}{50--76}.
\PrintBackRefs{\CurrentBib}

\bibitem [\protect \citeauthoryear {%
Camboni%
\ \BBA {} Porcellacchia%
}{%
Camboni%
\ \BBA {} Porcellacchia%
}{%
{\protect \APACyear {2025}}%
}]{%
camboni2022monitoring}
\APACinsertmetastar {%
camboni2022monitoring}%
\begin{APACrefauthors}%
Camboni, M.%
\BCBT {}\ \BBA {} Porcellacchia, M.%
\end{APACrefauthors}%
\unskip\
\newblock
\APACrefYearMonthDay{2025}{}{}.
\newblock
{\BBOQ}\APACrefatitle {Monitoring Team Members: Information Waste and the Transparency Trap} {Monitoring team members: Information waste and the transparency trap}.{\BBCQ}
\newblock
\APACjournalVolNumPages{American Economic Journal: Microeconomics}{}{}{forthcoming}.
\PrintBackRefs{\CurrentBib}

\bibitem [\protect \citeauthoryear {%
Cusumano%
, Gan%
\BCBL {}\ \BBA {} Pieroth%
}{%
Cusumano%
\ \protect \BOthers {.}}{%
{\protect \APACyear {2024}}%
}]{%
CGP24}
\APACinsertmetastar {%
CGP24}%
\begin{APACrefauthors}%
Cusumano, C\BPBI M.%
, Gan, T.%
\BCBL {}\ \BBA {} Pieroth, F.%
\end{APACrefauthors}%
\unskip\
\newblock
\APACrefYearMonthDay{2024}{}{}.
\newblock
{\BBOQ}\APACrefatitle {Misaligning Incentives in Teams} {Misaligning incentives in teams}.{\BBCQ}
\newblock

\PrintBackRefs{\CurrentBib}

\bibitem [\protect \citeauthoryear {%
Dye%
}{%
Dye%
}{%
{\protect \APACyear {1985}}%
}]{%
dye1985}
\APACinsertmetastar {%
dye1985}%
\begin{APACrefauthors}%
Dye, R\BPBI A.%
\end{APACrefauthors}%
\unskip\
\newblock
\APACrefYearMonthDay{1985}{}{}.
\newblock
{\BBOQ}\APACrefatitle {Disclosure of nonproprietary information} {Disclosure of nonproprietary information}.{\BBCQ}
\newblock
\APACjournalVolNumPages{Journal of accounting research}{}{}{123--145}.
\PrintBackRefs{\CurrentBib}

\bibitem [\protect \citeauthoryear {%
Grossman%
}{%
Grossman%
}{%
{\protect \APACyear {1981}}%
}]{%
gros1981}
\APACinsertmetastar {%
gros1981}%
\begin{APACrefauthors}%
Grossman, S\BPBI J.%
\end{APACrefauthors}%
\unskip\
\newblock
\APACrefYearMonthDay{1981}{}{}.
\newblock
{\BBOQ}\APACrefatitle {The informational role of warranties and private disclosure about product quality} {The informational role of warranties and private disclosure about product quality}.{\BBCQ}
\newblock
\APACjournalVolNumPages{The Journal of Law and Economics}{24}{3}{461--483}.
\PrintBackRefs{\CurrentBib}

\bibitem [\protect \citeauthoryear {%
Hagenbach%
, Koessler%
\BCBL {}\ \BBA {} Perez-Richet%
}{%
Hagenbach%
\ \protect \BOthers {.}}{%
{\protect \APACyear {2014}}%
}]{%
hakp2014}
\APACinsertmetastar {%
hakp2014}%
\begin{APACrefauthors}%
Hagenbach, J.%
, Koessler, F.%
\BCBL {}\ \BBA {} Perez-Richet, E.%
\end{APACrefauthors}%
\unskip\
\newblock
\APACrefYearMonthDay{2014}{}{}.
\newblock
{\BBOQ}\APACrefatitle {Certifiable pre-play communication: Full disclosure} {Certifiable pre-play communication: Full disclosure}.{\BBCQ}
\newblock
\APACjournalVolNumPages{Econometrica}{82}{3}{1093--1131}.
\PrintBackRefs{\CurrentBib}

\bibitem [\protect \citeauthoryear {%
Halac%
, Kremer%
\BCBL {}\ \BBA {} Winter%
}{%
Halac%
\ \protect \BOthers {.}}{%
{\protect \APACyear {2020}}%
}]{%
hakw2020}
\APACinsertmetastar {%
hakw2020}%
\begin{APACrefauthors}%
Halac, M.%
, Kremer, I.%
\BCBL {}\ \BBA {} Winter, E.%
\end{APACrefauthors}%
\unskip\
\newblock
\APACrefYearMonthDay{2020}{}{}.
\newblock
{\BBOQ}\APACrefatitle {Raising capital from heterogeneous investors} {Raising capital from heterogeneous investors}.{\BBCQ}
\newblock
\APACjournalVolNumPages{American Economic Review}{110}{3}{889--921}.
\PrintBackRefs{\CurrentBib}

\bibitem [\protect \citeauthoryear {%
Halac%
, Kremer%
\BCBL {}\ \BBA {} Winter%
}{%
Halac%
\ \protect \BOthers {.}}{%
{\protect \APACyear {2024}}%
}]{%
hakw2024}
\APACinsertmetastar {%
hakw2024}%
\begin{APACrefauthors}%
Halac, M.%
, Kremer, I.%
\BCBL {}\ \BBA {} Winter, E.%
\end{APACrefauthors}%
\unskip\
\newblock
\APACrefYearMonthDay{2024}{}{}.
\newblock
{\BBOQ}\APACrefatitle {Monitoring teams} {Monitoring teams}.{\BBCQ}
\newblock
\APACjournalVolNumPages{American Economic Journal: Microeconomics}{16}{3}{134--161}.
\PrintBackRefs{\CurrentBib}

\bibitem [\protect \citeauthoryear {%
Halac%
, Lipnowski%
\BCBL {}\ \BBA {} Rappoport%
}{%
Halac%
\ \protect \BOthers {.}}{%
{\protect \APACyear {2021}}%
}]{%
halr2021}
\APACinsertmetastar {%
halr2021}%
\begin{APACrefauthors}%
Halac, M.%
, Lipnowski, E.%
\BCBL {}\ \BBA {} Rappoport, D.%
\end{APACrefauthors}%
\unskip\
\newblock
\APACrefYearMonthDay{2021}{}{}.
\newblock
{\BBOQ}\APACrefatitle {Rank uncertainty in organizations} {Rank uncertainty in organizations}.{\BBCQ}
\newblock
\APACjournalVolNumPages{American Economic Review}{111}{3}{757--786}.
\PrintBackRefs{\CurrentBib}

\bibitem [\protect \citeauthoryear {%
Halac%
, Lipnowski%
\BCBL {}\ \BBA {} Rappoport%
}{%
Halac%
\ \protect \BOthers {.}}{%
{\protect \APACyear {2025}}%
}]{%
halr2025}
\APACinsertmetastar {%
halr2025}%
\begin{APACrefauthors}%
Halac, M.%
, Lipnowski, E.%
\BCBL {}\ \BBA {} Rappoport, D.%
\end{APACrefauthors}%
\unskip\
\newblock
\APACrefYearMonthDay{2025}{}{}.
\newblock
{\BBOQ}\APACrefatitle {Pricing for Coordination} {Pricing for coordination}.{\BBCQ}
\newblock

\PrintBackRefs{\CurrentBib}

\bibitem [\protect \citeauthoryear {%
Hart%
, Kremer%
\BCBL {}\ \BBA {} Perry%
}{%
Hart%
\ \protect \BOthers {.}}{%
{\protect \APACyear {2017}}%
}]{%
hakp2017}
\APACinsertmetastar {%
hakp2017}%
\begin{APACrefauthors}%
Hart, S.%
, Kremer, I.%
\BCBL {}\ \BBA {} Perry, M.%
\end{APACrefauthors}%
\unskip\
\newblock
\APACrefYearMonthDay{2017}{}{}.
\newblock
{\BBOQ}\APACrefatitle {Evidence games: Truth and commitment} {Evidence games: Truth and commitment}.{\BBCQ}
\newblock
\APACjournalVolNumPages{American Economic Review}{107}{3}{690--713}.
\PrintBackRefs{\CurrentBib}

\bibitem [\protect \citeauthoryear {%
Lizzeri%
}{%
Lizzeri%
}{%
{\protect \APACyear {1999}}%
}]{%
lizz1999}
\APACinsertmetastar {%
lizz1999}%
\begin{APACrefauthors}%
Lizzeri, A.%
\end{APACrefauthors}%
\unskip\
\newblock
\APACrefYearMonthDay{1999}{}{}.
\newblock
{\BBOQ}\APACrefatitle {Information revelation and certification intermediaries} {Information revelation and certification intermediaries}.{\BBCQ}
\newblock
\APACjournalVolNumPages{The RAND Journal of Economics}{}{}{214--231}.
\PrintBackRefs{\CurrentBib}

\bibitem [\protect \citeauthoryear {%
Migrow%
\ \BBA {} Severinov%
}{%
Migrow%
\ \BBA {} Severinov%
}{%
{\protect \APACyear {2022}}%
}]{%
mise2022}
\APACinsertmetastar {%
mise2022}%
\begin{APACrefauthors}%
Migrow, D.%
\BCBT {}\ \BBA {} Severinov, S.%
\end{APACrefauthors}%
\unskip\
\newblock
\APACrefYearMonthDay{2022}{}{}.
\newblock
{\BBOQ}\APACrefatitle {Investment and information acquisition} {Investment and information acquisition}.{\BBCQ}
\newblock
\APACjournalVolNumPages{American Economic Journal: Microeconomics}{14}{3}{480--529}.
\PrintBackRefs{\CurrentBib}

\bibitem [\protect \citeauthoryear {%
Milgrom%
}{%
Milgrom%
}{%
{\protect \APACyear {1981}}%
}]{%
milg1981}
\APACinsertmetastar {%
milg1981}%
\begin{APACrefauthors}%
Milgrom, P\BPBI R.%
\end{APACrefauthors}%
\unskip\
\newblock
\APACrefYearMonthDay{1981}{}{}.
\newblock
{\BBOQ}\APACrefatitle {Good news and bad news: Representation theorems and applications} {Good news and bad news: Representation theorems and applications}.{\BBCQ}
\newblock
\APACjournalVolNumPages{The Bell Journal of Economics}{}{}{380--391}.
\PrintBackRefs{\CurrentBib}

\bibitem [\protect \citeauthoryear {%
Okuno-Fujiwara%
, Postlewaite%
\BCBL {}\ \BBA {} Suzumura%
}{%
Okuno-Fujiwara%
\ \protect \BOthers {.}}{%
{\protect \APACyear {1990}}%
}]{%
okps1990}
\APACinsertmetastar {%
okps1990}%
\begin{APACrefauthors}%
Okuno-Fujiwara, M.%
, Postlewaite, A.%
\BCBL {}\ \BBA {} Suzumura, K.%
\end{APACrefauthors}%
\unskip\
\newblock
\APACrefYearMonthDay{1990}{}{}.
\newblock
{\BBOQ}\APACrefatitle {Strategic information revelation} {Strategic information revelation}.{\BBCQ}
\newblock
\APACjournalVolNumPages{The Review of Economic Studies}{57}{1}{25--47}.
\PrintBackRefs{\CurrentBib}

\bibitem [\protect \citeauthoryear {%
Onuchic%
\ \BBA {} Ramos%
}{%
Onuchic%
\ \BBA {} Ramos%
}{%
{\protect \APACyear {2025}}%
}]{%
onuchic2023disclosure}
\APACinsertmetastar {%
onuchic2023disclosure}%
\begin{APACrefauthors}%
Onuchic, P.%
\BCBT {}\ \BBA {} Ramos, J.%
\end{APACrefauthors}%
\unskip\
\newblock
\APACrefYearMonthDay{2025}{}{}.
\newblock
{\BBOQ}\APACrefatitle {Disclosure by groups} {Disclosure by groups}.{\BBCQ}
\newblock

\PrintBackRefs{\CurrentBib}

\bibitem [\protect \citeauthoryear {%
Rappoport%
}{%
Rappoport%
}{%
{\protect \APACyear {2025}}%
}]{%
Rappoport25}
\APACinsertmetastar {%
Rappoport25}%
\begin{APACrefauthors}%
Rappoport, D.%
\end{APACrefauthors}%
\unskip\
\newblock
\APACrefYearMonthDay{2025}{}{}.
\newblock
{\BBOQ}\APACrefatitle {Evidence and Skepticism in Verifiable Disclosure Games} {Evidence and skepticism in verifiable disclosure games}.{\BBCQ}
\newblock
\APACjournalVolNumPages{Theoretical Economics}{}{}{forthcoming}.
\PrintBackRefs{\CurrentBib}

\bibitem [\protect \citeauthoryear {%
Segal%
}{%
Segal%
}{%
{\protect \APACyear {2003}}%
}]{%
sega2003}
\APACinsertmetastar {%
sega2003}%
\begin{APACrefauthors}%
Segal, I.%
\end{APACrefauthors}%
\unskip\
\newblock
\APACrefYearMonthDay{2003}{}{}.
\newblock
{\BBOQ}\APACrefatitle {Coordination and discrimination in contracting with externalities: Divide and conquer?} {Coordination and discrimination in contracting with externalities: Divide and conquer?}{\BBCQ}
\newblock
\APACjournalVolNumPages{Journal of Economic Theory}{113}{2}{147--181}.
\PrintBackRefs{\CurrentBib}

\bibitem [\protect \citeauthoryear {%
Whitmeyer%
\ \BBA {} Zhang%
}{%
Whitmeyer%
\ \BBA {} Zhang%
}{%
{\protect \APACyear {2022}}%
}]{%
whzh2022}
\APACinsertmetastar {%
whzh2022}%
\begin{APACrefauthors}%
Whitmeyer, M.%
\BCBT {}\ \BBA {} Zhang, K.%
\end{APACrefauthors}%
\unskip\
\newblock
\APACrefYearMonthDay{2022}{}{}.
\newblock
{\BBOQ}\APACrefatitle {Costly Evidence and Discretionary Disclosure} {Costly evidence and discretionary disclosure}.{\BBCQ}
\newblock

\PrintBackRefs{\CurrentBib}

\bibitem [\protect \citeauthoryear {%
Winter%
}{%
Winter%
}{%
{\protect \APACyear {2004}}%
}]{%
wint2004}
\APACinsertmetastar {%
wint2004}%
\begin{APACrefauthors}%
Winter, E.%
\end{APACrefauthors}%
\unskip\
\newblock
\APACrefYearMonthDay{2004}{}{}.
\newblock
{\BBOQ}\APACrefatitle {Incentives and discrimination} {Incentives and discrimination}.{\BBCQ}
\newblock
\APACjournalVolNumPages{American Economic Review}{94}{3}{764--773}.
\PrintBackRefs{\CurrentBib}

\end{thebibliography}

\appendix
\renewcommand{\thesection}{Appendix \Alph{section}}
\renewcommand{\thesubsection}{\Alph{section}.\arabic{subsection}}

\section{}\label{proof-main result}

Before we start proving the main result, we first show the equivalence of our model and the same model but with the following changes: (i) the state space is $\Theta$; (ii) the prior is $(\mu_0^k)_{k=1}^K$; (iii) the signal structure fully reveals the state/type. The equivalence is built in the sense that, under the two languages, every two equivalent menu profiles induce equivalent equilibrium sets. \\

\noindent\emph{Proof.} In the original model, the state space is $\Omega$, the prior is $(\nu^i_0)_{i=1}^N$, and the signal structure is $(\Theta,\pi)$ with each $\theta\in\Theta$ equal to the posterior mean it induces. To show the equivalence claimed above, let $M_0^{\Theta}$ be a menu profile in the old language, and $M^{\Theta}$ be the equivalent menu profile in the new language that satisfies, for all $\theta^k\in\Theta$, $M^k=M^k_0$. Now, we compare the equilibrium conditions given these two menu profiles in the two languages, respectively.

First, if the consumer observes signal $\theta^k$, his posterior mean is $\theta^k$ in both cases. Second, if expecting the same skepticism posterior mean of the consumer $v^N$, the type $\theta^k$ producer faces the same problem
\begin{equation*}
    \begin{aligned}
        \max_{(q,p)\in M^k_0}&q\max\{\theta^k-c,0\}+(1-q)\max\{v^N-c,0\}-p \\
        \Leftrightarrow\max_{(q,p)\in M^k}&q\max\{\theta^k-c,0\}+(1-q)\max\{v^N-c,0\}-p.
    \end{aligned}
\end{equation*}
Lastly, we verify that, given equivalent contingent choices of the producer, the Bayes' rule yields identical skepticism beliefs whenever it can be applied. Given a choice profile $(q^k,p^k)_{k=1}^K$ with $\sum_{k=1}^Kq^k<1$, the old language produces a skepticism belief, given by for all $\omega^i\in\Omega$
\begin{equation}\label{B-old language skepticism}
    \mu^N_0(\omega^i)=\frac{\nu^i_0\sum_{k=1}^K\pi(\theta^k|\omega^i)(1-q^k)}{\sum_{j=1}^{I}\nu^j_0\sum_{k=1}^K\pi(\theta^k|\omega^j)(1-q^k)}=\frac{\nu^k_0\sum_{k=1}^K\pi(\theta^k|\omega^i)(1-q^k)}{\sum_{k=1}^K\mu_0^k(1-q^k)},
\end{equation}
Where the second equality uses (\ref{type prior}), the definition of $\mu_0$. On the other hand, given the same choice profile $(q^k,p^k)$, the new language produces a skepticism belief over types (but not yet states) according to (\ref{bayes}), namely for all $\theta^k\in\Theta$
\begin{equation}
    \mu^N(\theta^k)=\frac{\mu^k_0(1-q^k)}{\sum_{k'=1}^K\mu^{k'}_0(1-q^{k'})}.
\end{equation}
We further translate it as the skepticism belief over states, given by for all $\omega^i\in\Omega$
\begin{equation}\label{B-new language skepticism}
    \begin{aligned}
        \mu^N(\omega^i)&=\sum_{k=1}^K\nu_{\theta^k}(\omega^i)\mu^N(\theta^k)=\sum_{k=1}^K\frac{\nu^i_0\pi(\theta^k|\omega^i)}{\sum_{j=1}^I\nu^j_0\pi(\theta^k|\omega^j)}\frac{\mu^k_0(1-q^k)}{\sum_{k'=1}^K\mu^{k'}_0(1-q^k)} \\
        &=\sum_{k=1}^K\frac{\nu^i_0\pi(\theta^k|\omega^i)}{\mu^k_0}\frac{\mu^k_0(1-q^k)}{\sum_{k'=1}^K\mu^{k'}_0(1-q^k)}=\frac{\nu^k_0\sum_{k=1}^K\pi(\theta^k|\omega^i)(1-q^k)}{\sum_{k=1}^K\mu^k_0(1-q^k)}.
    \end{aligned}
\end{equation}
By comparing (\ref{B-old language skepticism}) and (\ref{B-new language skepticism}), one can see that the skepticism beliefs are equivalent, and they thus produce the same posterior mean. Therefore, one should be able to see that with equivalent menu profiles, every disclosure equilibrium in one language corresponds to one in the other, and vice versa.

\subsection{Proof of Lemma \ref{lemma-simplify}}

\emph{Proof.} Notice that choosing $(0,0)$ is the dominant strategy for a producer type $\theta^k<c$, so any change of his menu will not have any effect. Thus, we establish the lemma for types $\theta^k\geq c$ below.

Let the skepticism value associated with $\underline{r}^{\Theta}=(\underline{q}^{\Theta},\underline{p}^{\Theta})$ be $\underline{w}^N$ (and the belief $\underline{\mu}^N$). The producer's optimal choices give that for all $k$, for all $(q,p)\in M^k$, $\underline{q}^k(\theta^k-c)+(1-\underline{q}^k)\underline{w}^N-\underline{p}^k\geq q(\theta^k-c)+(1-q)\underline{w}^N-p$, which can be rewritten as $(\theta^k-c-\underline{w}^N)(\underline{q}^k-q)\geq\underline{p}^k-p$. This means $\underline{r}^k=(\underline{q}^k,\underline{p}^k)\in\conv M^k$. Since $\underline{q}^k\leq\underline{q}^k$, we also have $\underline{r}^k\in\widehat{M}^k$. Moreover, $\widehat{M}^{\Theta}$ is constructed by deleting points in $M^{\Theta}$, so it is even easier for $\underline{r}^N$ to satisfy the equilibrium conditions. Thus, we have $(\underline{r}^{\Theta},\underline{\mu}^N)\in\mathcal{E}(\widehat{M}^{\Theta})$.

Next, take any $(q^{\Theta},p^{\Theta})$ such that for all $k$, $(q^k,p^k)\in\widehat{M}^k$ and $\sum_{k=1}^K\mu^k_0p^k<\sum_{k=1}^K\mu^k_0\underline{p}^k$. In other words, this $(q^{\Theta},p^{\Theta})$ under-discloses and thus cannot correspond to an equilibrium given $M^{\Theta}$. For all $k$, let $p_k(.)$ be the price envelope of $M^k$. Let the skepticism value associated with $q^{\Theta}$ be $w^N$. Due to the fact that $\underline{r}^{\Theta}$ is robustly induced by $M^{\Theta}$, there must be some $\widetilde{k}$ and deviation $(q,p)\in M^{\widetilde{k}}$ such that $q(\theta^{\widetilde{k}}-c)+(1-q)w^N-p>q^{\widetilde{k}}(\theta^{\widetilde{k}}-c)+(1-q^{\widetilde{k}})w^N-p^{\widetilde{k}}$.

We show by contradiction that $(q^{\Theta},p^{\Theta})$ also cannot correspond to an equilibrium given $\widehat{M}^{\Theta}$. Suppose there exists some skepticism belief $\mu^N$ such that $((q^{\Theta},p^{\Theta}),\mu^N)\in\mathcal{E}(\widehat{M}^{\Theta})$, then repeating previous arguments, we have that for all $k$, $(\theta^k-c-w^N)(q^k-q)\geq p^k-p$ for all $(q,p)\in\widehat{M}^k$. We construct a contradiction in the following three steps. Once this is done, we know $\underline{r}^{\Theta}$ also delivers the worst-case equilibrium given $\widehat{M}^{\Theta}$ and the proof is completed.

First, because of the convexity of $p_k(.)$, the separating hyperplane expression above is equivalent to that: if $q^k=0$, $\theta^k-c-w^N\leq p_k'(0+0)$; if $0<q^k<\underline{q}^k$, $p_k'(q^k-0)\leq\theta^k-c-w^N\leq p_k'(q^k+0)$; if $q^k=\underline{q}^k$, $p_k'(q^k-0)\leq\theta^k-c-w^N$. Note that as long as $q^k<\underline{q}^k$, these conditions also imply the equilibrium conditions given $M^k$, namely $(\theta^k-c-w^N)(q^k-q)\geq p^k-p$ for all $(q,p)\in M^k$. Thus, given $M^{\Theta}$, the equilibrium must be broken by a type $\widetilde{k}$ that satisfies $q^{\widetilde{k}}=\underline{q}^{\widetilde{k}}$. Also, type $\widetilde{k}$'s deviation $(q,p)\in M^{\widetilde{k}}$ must be upward with $q>q^{\widetilde{k}}$ because, otherwise, if $q=q^{\widetilde{k}}$, then $(q^{\widetilde{k}},p^{\widetilde{k}})\in\conv M^{\widetilde{k}}$ implies $p\geq p^{\widetilde{k}}$, which cannot be a profitable deviation; or, if $q<q^{\widetilde{k}}$, then $(q,p)$ being a downward deviation means $(\theta^{\widetilde{k}}-c-w^N)(q^{\widetilde{k}}-q)<p^{\widetilde{k}}-p$ and hence $\frac{p^{\widetilde{k}}-p}{q^{\widetilde{k}}-q}>\theta^{\widetilde{k}}-c-w^N$, while the fact that $(q^{\widetilde{k}},p^{\widetilde{k}})$ lies on a convex function $p_{\widetilde{k}}(.)$ implies $\frac{p^{\widetilde{k}}-p}{q^{\widetilde{k}}-q}\leq p_{\widetilde{k}}'(q^{\widetilde{k}}-0)$, which together contradicts the equilibrium condition $p_{\widetilde{k}}'(q^{\widetilde{k}}-0)\leq\theta^{\widetilde{k}}-c-w^N$. This implies $q^{\widetilde{k}}=\underline{q}^{\widetilde{k}}$ is an interior point in the domain of $p_{\widetilde{k}}(.)$, and the fact that $(\underline{r}^{\Theta},\underline{\mu}^N)\in\mathcal{E}(M^{\Theta})$ hence gives equilibrium condition $p_k'(\underline{q}^{\widetilde{k}}-0)\leq\theta^k-c-\underline{w}^N\leq p_k'(\underline{q}^{\widetilde{k}}+0)$.

Second, we compare $\underline{w}^N$ and $w^N$. To do this, notice that there must be some $k'$ such that $q^{k'}<\underline{q}^{k'}$ since, otherwise, $q^{\Theta}=\underline{q}^{\Theta}$ and thus $p^k=p_k(q^k)=p_k(\underline{q}^k)=\underline{p}^k$ for all $k$, contradicting $\sum_{k=1}^K\mu^k_0p^k<\sum_{k=1}^K\mu^k_0\underline{p}^k$. Then, the equilibrium conditions and the fact that $p_k(.)$ is convex together imply $\theta^{k'}-c-w^N\leq p_k'(q^{k'}+0)\leq p_k'(\underline{q}^{k'}-0)\leq\theta^{k'}-c-\underline{w}^N$, so we have $\underline{w}^N\leq w^N$.

To summarize, we reach a contradiction: (i) $(q,p)$ being an upward deviation for type $\widetilde{k}$ means $\frac{p-p^{\widetilde{k}}}{q-q^{\widetilde{k}}}<\theta^{\widetilde{k}}-c-w^N$; (ii) $(q^{\widetilde{k}},p^{\widetilde{k}})$ lying on a convex function $p_{\widetilde{k}}(.)$ means $\frac{p-p^{\widetilde{k}}}{q-q^{\widetilde{k}}}\geq p_{\widetilde{k}}'(q^{\widetilde{k}}+0)$; (iii) we have shown $\theta^{\widetilde{k}}-c-w^N\leq\theta^{\widetilde{k}}-c-\underline{w}^N\leq p_{\widetilde{k}}'(\underline{q}^{\widetilde{k}}+0)=p_{\widetilde{k}}'(q^{\widetilde{k}}+0)$ (the last equality results from $q^{\widetilde{k}}=\underline{q}^{\widetilde{k}}$). As a result, (i) and (ii) imply $\theta^{\widetilde{k}}-c-w^N>p_{\widetilde{k}}'(q^{\widetilde{k}}+0)$, contradicting (iii). So, there cannot be any $\mu^N$ such that $((q^{\theta},p^{\theta}),\mu^N)\in\mathcal{E}(\widehat{M}^{\Theta})$. Consequently, $\underline{r}^{\Theta}$ also delivers the worst-case equilibrium given $\widehat{M}^{\Theta}$.

\subsection{Proof of Lemma \ref{lemma-path}}

\emph{Proof.} Before laying out our path-finding algorithm, we process some preliminary steps first. Lemma \ref{lemma-simplify} allows us to focus on the truncated convexification of $M^{\Theta}$, denoted by $\widehat{M}^{\Theta}$, and for simplicity, we replace $M^{\Theta}$ with $\widehat{M}^{\Theta}$ and suppress the hat accent. Thus, $M^k\subseteq\conv M^k$ for all $k$. We now fix one such menu profile and the probability profile $\underline{q}^{\Theta}$ it robustly induces. For all $k$, since $(0,0)\in M^k$, the price envelope has $p_k(0)=0$. Also, $p_k(\underline{q}^k)=\underline{p}^k$. We denote the expected revenue at $q^{\Theta}$ by $R(q^{\Theta})=\sum_{k=1}^K\mu^k_0p_k(q^k)$. We define marginal price $p_k'(.)$ by setting $p_k'(q)$ to be the left-derivative $p_k'(q-0)$ at each $q>0$ and $p_k'(0)$ to be the right-derivative $p_k'(0+0)$ at 0. We must have $p_k'(0+0)\geq0$ because $p_k(.)\geq0$, and convexity means $p_k(.)$ and its derivatives are all nondecreasing. We then define the following two sets for all $k$:
\begin{equation}\label{A-AB sets}
    \begin{aligned}
        A_k&=\{q^{\Theta}\leq\underline{q}^{\Theta}:q^k<\underline{q}^k\text{ and }p_k'(q^k+0)<\theta^k-c-w^N(q^{\Theta})\},\text{ and} \\
        B_k&=\{q^{\Theta}\leq\underline{q}^{\Theta}:q^k>0\text{ and }p_k'(q^k-0)>\theta^k-c-w^N(q^{\Theta})\}.
    \end{aligned}
\end{equation}
Notice that $A_k$ collects all the probability profiles where $\theta^k$ wants to deviate upward while $B_k$ is the set for downward deviations. $C_k:=(A_k\cup B_k)^c$ is the set on which $\theta^k$ reaches optimum. Due to convexity of $p_k(.)$, all the deviations and optimum here are in the global sense. One important property of $A_k$ is  ``right-openness'' which says that for all $q^{\Theta}\in A_k$, there exists some small $\varepsilon>0$ such that $(q^k+\varepsilon',q^{-k})\in A_k$ for all $\varepsilon'\in[0,\varepsilon]$. This results from the following facts: (i) the right-derivative of a price envelope is nondecreasing and right-continuous; (ii) $\theta^k-c-w^N(q^{\Theta})$ is continuous in $q^k$; (ii) the right-derivative cannot be defined on the boundary with $q^k=\underline{q}^k$, so $A_k$ contains no such boundary points.

Our proof consists of four consecutive parts.
First, we describe an algorithm for finding the path stated in the lemma. Let $\Theta^A(q^{\Theta})=\{k:q^{\Theta}\in A_k\}$ collect all types that are willing to deviate upward at $q^{\Theta}$. For some $k$ and $q^{\Theta}$ such that $q^{\Theta}\in A_k$, we say that to \emph{increase $q^k$ to its closest boundary of $A_k$} is to increase $q^k$ to some $\overline{q}^{k}\leq\underline{q}^k$ such that (i) $(\widetilde{q}^k,q^{-k})\in A_k$ for all $\widetilde{q}^k\in[q^k,\overline{q}^k)$, and (ii) $(\overline{q}^k,q^{-k})\notin A_k$. In other words, $\overline{q}^k$ is the first point above $q^k$ that falls out of $A_k$. Since $A_k$ is right-open and the upper boundary $\underline{q}^k$ is never contained in $A_k$, such a $\overline{q}^k$ always exists.
The path-finding algorithm is given by:
\begin{algorithm}\label{A-algorithm}
    \emph{Start with $\widetilde{q}^{\Theta}=0^{\Theta}$, path $P=(q^{\Theta}_t,k_t)_{t=0}^T$, and $T=-1$:
    \begin{enumerate}[nolistsep]
        \renewcommand{\labelenumi}{}
        \item \textbf{Step 1}: Set $T=T+1$ and $q^{\Theta}_T=\widetilde{q}^{\Theta}$. If $R(q_T^{\Theta})=R(\underline{q}^{\Theta})$, set $k_T$ to be any $k$ and exit the algorithm. Otherwise, go to Step 2.
        \item \textbf{Step 2}: Randomly pick from $\Theta^A({\widetilde{q}^{\Theta}})$ some $k$ with equal probabilities, and set $k_{T-1}=k$. Increase $\widetilde{q}^k$ to its closest boundary of $A_k$. Go to Step 1.
    \end{enumerate}}
\end{algorithm}

Second, we show that Algorithm \ref{A-algorithm} is well-defined, that is, every step, once reached under any possible circumstance, can be implemented. Step 1 is obviously well-defined. %because $R(.)$ is.

The challenging part is to prove the well-definedness of Step 2 by showing $\Theta^A(\widetilde{q}^{\Theta})\neq\emptyset$. Step 2 is reached only when the previous step is Step 1 and $R(\widetilde{q}^{\Theta})\neq R(\underline{q}^{\Theta})$. Since $p_k(.)$ is nondecreasing, we must have $R(\widetilde{q}^{\Theta})< R(\underline{q}^{\Theta})$, so $\widetilde{q}^{\Theta}$ cannot correspond to an equilibrium because, otherwise, $\underline{q}^{\Theta}$ would not be the worst case.

We start by proving one claim: conditional on Step 2  being reached at some $T$, then for all $t\leq T$, for all $k$ such that $q_t^k>0$, we have $p_k'(q_t^k-0)\leq\theta^k-c-w^N(q_t^{\Theta})$.

We show the claim by induction. First, this must be true for $t=1$ since $k_1\in\Theta^A(0^{\Theta})$ and $q_1^{k_1}$ is found such that every point lower than it belongs to $A_{k_1}$ meaning $p_{k_1}'(q_1^{k_1}-\varepsilon+0)<\theta^{k_1}-c-w^N(q_1^{k_1}-\varepsilon,q_1^{-k_1})$ for small $\varepsilon>0$, whose limit form is $p_{k_1}'(q_1^{k_1}-0)\leq\theta^{k_1}-c-w^N(q_1^{\Theta})$. Next, suppose the claim is true for all $t<t'$. For any $k$ such that $q_{t'}^k>0$, we have two cases to discuss. If $q_{t'-1}^{\Theta}$ changes to $q_{t'}^{\Theta}$ by increasing the $k$th component in Step 2, the same argument as for the $t=1$ case applies. If the increase happens for some other $k'\neq k$, then the induction assumption implies $\theta^{k'}-c-w^N(q_{t'-1}^{\Theta})\geq p_{k'}'(q^{k'}_{t'-1}-0)\geq0$ and $\theta^k-c-w^N(q_{t'-1}^{\Theta})\geq p_k'(q^k_{t'-1}-0)$. The first inequality further implies $\theta^{k'}-c\geq w^N(q_{t'-1}^{\Theta})$. By the definition of skepticism value (\ref{skepticism value}), increasing the disclosure level of such a type can only worsen the consumer's skepticism and thus drive the skepticism value weakly lower, so we have $w^N(q_{t'}^{\Theta})\leq w^N(q_{t'-1}^{\Theta})$. We thus have $\theta^k-c-w^N(q_{t'}^{\Theta})\geq\theta^k-c-w^N(q_{t'-1}^{\Theta})\geq p_k'(q^k_{t'-1}-0)=p_k'(q^k_{t'}-0)$. Hence, the claim is also true for $t=t'$. By induction, the claim is true for all $t$.

Now, suppose $\Theta^A(\widetilde{q}^{\Theta})=\emptyset$ when Step 2 is reached. For each $k$ we have three cases to consider. Case 1: for $k$ such that $\widetilde{q}^k=\underline{q}^k$,  from the claim above, we have $p_k'(\widetilde{q}^k-0)\leq\theta^k-c-w^N(\widetilde{q}^{\Theta})$, which means $\theta^k$ has reached its optimum by choosing plan $(\underline{q}^k,\underline{p}^k)\in M^k$. 

Case 2: for  $k$ such that $\widetilde{q}^k\in(0,\underline{q}^k)$, the claim above again gives $p_k'(\widetilde{q}^k-0)\leq\theta^k-c-w^N(\widetilde{q}^{\Theta})$, so $\widetilde{q}^{\Theta}\notin B_k$. Since $\Theta^A(\widetilde{q}^{\Theta})=\emptyset$, $\widetilde{q}^{\Theta}\in(A_k\cup B_k)^c=C_k$. Hence, as long as $(\widetilde{q}^k,p_k(\widetilde{q}^k))\in M^k$, $\theta^k$ reaches its optimum by choosing this plan.
Now, we show $(\widetilde{q}^k,p_k(\widetilde{q}^k))\in M^k$. Suppose this is not true, then by compactness of $M^k$, there is a small $\varepsilon>0$ such that $p_k$ is differentiable on $[\widetilde{q}^k-\varepsilon,\widetilde{q}^k+\varepsilon]$ and $p_k'(q)$ is constant on this closed interval. Suppose type $k$'s disclosure level reaches $\widetilde{q}^k$ in some previous step $t'$ and back then we have $(\widetilde{q}^k-\varepsilon,q^{-k}_{t'})\in A_k$. Notice that the proof of the claim above also shows that the skepticism value obtained at the end of each step, $w^N(q^{\Theta}_t)$, is nonincreasing in $t$. So, by the definition of $A_k$, if $(\widetilde{q}^k-\varepsilon,q^{-k}_{t'})\in A_k$ in a previous step, we also have $(\widetilde{q}^k-\varepsilon,\widetilde{q}^{-k})\in A_k$ in the current step. Moreover, the assumption that $\Theta^A(\widetilde{q}^{\Theta})=\emptyset$ means $\widetilde{q}^{\Theta}=(\widetilde{q}^k,\widetilde{q}^{-k})\notin A_k$. The fact that $p_k'(q^k+0)$ is locally constant around $\widetilde{q}^k$ implies that these can only happen when $w^N(\widetilde{q}^k-\varepsilon,\widetilde{q}^{-k})<w^N(\widetilde{q}^k,\widetilde{q}^{-k})$. However, this contradicts $\widetilde{q}^k>0$ since our previous claim requires that $\theta^k-c\geq w^N(\widetilde{q}^{\Theta})+p_k'(\widetilde{q}^k-0)\geq w^N(\widetilde{q}^{\Theta})$, so increasing $q^k$ from $\widetilde{q}^k-\varepsilon$ to $\widetilde{q}^k$ cannot drive the skepticism value higher. Thus, $(\widetilde{q}^k,p_k(\widetilde{q}^k))$ must be offered in $M^k$, and $\theta^k$ reaches its optimum by choosing this plan.

Case 3: for all $k$ such that $\widetilde{q}^k=0$, $\Theta^A(\widetilde{q}^{\Theta})=\emptyset$ implies $p_k'(0+0)\geq\theta^k-c-w^N(\widetilde{q}^{\Theta})$, which means $\theta^k$ reaches its optimum by choosing $(0,0)\in M^k$. To summarize all three cases above, we conclude that $\widetilde{q}^{\Theta}$ corresponds to an equilibrium that yields worse revenue than $\underline{q}^{\Theta}$ does, a contradiction. This shows $\Theta^A(\widetilde{q}^{\Theta})\neq\emptyset$ conditional on Step 2 being reached and the algorithm is well-defined.

Third, we show that Algorithm \ref{A-algorithm} indeed generates a desired path within countable steps with probability one (probability arises because in Step 2 we randomly select $k\in\Theta^A(\widehat{q}^{\theta})$. If the algorithm stops within finite steps, the ending step Step 1 guarantees $R(q_T^{\Theta})=R(\underline{q}^{\Theta})$. If the algorithm does not stop within finite steps, then due to the fact that $q_t^{\Theta}$ is nondecreasing in $t$ and bounded, we know a limit exists, denoted $\widehat{q}^{\Theta}:=\lim_{t\rightarrow\infty}q_t^{\Theta}\leq\underline{q}^{\Theta}$. 

We show that $R(\widehat{q}^{\Theta})=R(\underline{q}^{\Theta})$ with probability one. We first prove $\Theta^A(\widehat{q}^{\Theta})=\emptyset$. To see this, suppose there is some $k\in\Theta^A(\widehat{q}^{\Theta})$, then due to three facts that (i) the inequality in the definition of $A_k$ is strict, (ii) $p_k'(.)$ is nondecreasing, and (iii) the skepticism value $w^N(.)$ is continuous, we have $k\in\Theta^A(q^{\Theta})$ for all $q^{\Theta}\leq\widehat{q}^{\Theta}$ close enough to $\widehat{q}^{\Theta}$. Therefore, there is some $\underline{t}$ such that for all $t>\underline{t}$, $k\in\Theta^A(q_t^{\Theta})$. But this is impossible because among the infinite times of reaching Step 2, $k$ is chosen with probability one, and when this is the case in some step $t'>\underline{t}$, $q_{t'}^k$ will be increased to $q_{t'+1}^k$ such that $q_{t'+1}^{\Theta}\notin A_k$, which also means $k\notin\Theta^A(q_{t'+1}^{\Theta})$, a contradiction. Next, given $\Theta^A(\widehat{q}^{\Theta})=\emptyset$, following the arguments in the second part of the proof, we conclude that $\widehat{q}^{\Theta}$ corresponds to an equilibrium. If $R(\widehat{q}^{\Theta})=R(\underline{q}^{\Theta})$, the proof is done. So, suppose $R(\widehat{q}^{\Theta})<R(\underline{q}^{\Theta})$. We then find an equilibrium that offers worse revenue than $\underline{q}^{\Theta}$ does, a contradiction. So, we must have $R(\widehat{q}^{\Theta})=R(\underline{q}^{\Theta})$.

Hence, we have shown that the algorithm does yield a nondecreasing alternating path from $0^{\Theta}$ to some $\widehat{q}^{\Theta}\leq\underline{q}^{\Theta}$ with the revenue guarantee satisfying $R(\widehat{q}^{\Theta})=R(\underline{q}^{\Theta})=R(M^{\Theta})$.

Fourth, we show that (\ref{upper bound}) indeed bounds $R(M^{\Theta})$ from above. Note that Algorithm \ref{A-algorithm} guarantees that every step from $q_t^{\Theta}$ to $q_{t+1}^{\Theta}$ is chosen such that the following local bound exists:
\begin{equation}\label{A-local upper bound}
    p_{k_t}'(\widetilde{q}^{k_t})\leq\theta^{k_t}-c-w^N(\widetilde{q}^{k_t},q_t^{-k_t})\text{, for all $\widetilde{q}^{k_t}\in[q_t^{k_t},q_{t+1}^{k_t}]$}.
\end{equation}
Then, by multiplying both sides by $\mu^{k_t}_0$, integrating both sides along the step, and summing up all such integrals for all steps, we complete the proof by obtaining the bound for the revenue guaranteed by $M^{\Theta}$:
\begin{equation}\label{A-total upper bound}
    \begin{aligned}
        \begin{aligned}
            R(M^{\Theta})&=R(\widehat{q}^{\Theta})=\sum_{k=1}^K\mu^k_0p_k(\widehat{q}^k)=\sum_{t=1}^{T-1}\mu^{k_t}_0\int_{q_t^{k_t}}^{q_{t+1}^{k_t}}p_k'(\widetilde{q}^k)d\widetilde{q}^k \\
            &\leq\sum_{t=1}^{T-1}\mu^{k_t}_0\int_{q_t^{k_t}}^{q_{t+1}^{k_t}}\left[\theta^{k_t}-c-w^N(\widetilde{q}^{k_t},q_t^{-k_t})\right]d\widetilde{q}^k.
        \end{aligned}
    \end{aligned}
\end{equation}

\subsection{Proof of Lemma \ref{lemma-upper bound}}

\emph{Proof.} We prove the lemma in three steps. First, we claim that, fixing any endpoint $\underline{q}^{\Theta}\leq 1^{\Theta}$, the bound (\ref{upper bound}) is maximized by the sequential self-conquering path $(q_t^{\Theta},k_t)_{t=0}^K$ where for all $t<K$, $k_t=t+1$ and $q_t^{\Theta}=(\underline{q}^1,\underline{q}^2,...,\underline{q}^t,0,...,0)$. Take any nondecreasing alternating path $(\widehat{q}_t^{\Theta},\widehat{k}_t)_{t=0}^T$ from $0^{\Theta}$ to $\underline{q}^{\Theta}$. We then show the claim by proving that if this path has $\widehat{k}_t>\widehat{k}_{t+1}$ for some $t$, namely it induces some disclosure probability from a low type right before it induces some disclosure probability from a high type, then it can always be (weakly) improved by exchanging the order. That is, instead of moving $\widehat{q}_t^{\widehat{k}_t}$ to $\widehat{q}_{t+1}^{\widehat{k}_t}$ in step $t+1$ and moving $\widehat{q}_{t+1}^{\widehat{k}_{t+1}}$ to $\widehat{q}_{t+2}^{\widehat{k}_{t+1}}$ in step $t+2$, we consider doing the latter in step $t+1$ and the former in step $t+2$. To simplify notations, let $\widehat{k}_t=j>i=\widehat{k}_{t+1}$, $\underline{q}^j:=\widehat{q}_t^{\widehat{k}_t}$, $\overline{q}^j:=\widehat{q}_{t+1}^{\widehat{k}_t}$, $\underline{q}^i:=\widehat{q}_{t}^{\widehat{k}_{t+1}}$, and $\overline{q}^i:=\widehat{q}_{t+2}^{\widehat{k}_{t+1}}$. Let $\widehat{q}_t^{-\widehat{k}_t,\widehat{k}_{t+1}}$ denote the probability profile $\widehat{q}_t^{\Theta}$ without the $\widehat{k}_t$th and the $\widehat{k}_{t+1}$th components. Notice that the stated exchange will not affect the quantile-skepticism value (defined in (\ref{quantile-skepticism value}) as the skepticism posterior mean at each point on a path) on the part of the path other than the two exchanged steps. Therefore, we rewrite the quantile-skepticism value as a function of only $q^i\in[0,\widehat{q}_{t+2}^{\widehat{k}_{t+1}}-\widehat{q}_{t+1}^{\widehat{k}_{t+1}}]$ and $q^j\in[0,\widehat{q}_{t+1}^{\widehat{k}_t}-\widehat{q}_t^{\widehat{k}_t}]$
\begin{equation}\label{A-skepticism value}
    v^N(q^i,q^j)=\E[\theta|\mu^N(\widehat{q}_{t}^{\widehat{k}_{t+1}}+q^i,\widehat{q}_t^{\widehat{k}_t}+q^j,\widehat{q}_t^{-\widehat{k}_t,\widehat{k}_{t+1}})]=\frac{\E[\theta]-\mu_0^iq^i\theta^i-\mu_0^jq^j\theta^j-\sum_{k\neq i,j}^K\mu^k_0\widehat{q}_t^k\theta^k}{1-\mu_0^iq^i-\mu_0^jq^j-\sum_{k\neq i,j}^K\mu^k_0\widehat{q}_t^k}.
\end{equation}

\begin{comment}
Notice that the stated exchange will not affect the values of integrals in the summation of (\ref{upper bound}) other than the two integrals associated with the exchanged steps $t+1$ and $t+2$. Therefore, the improvement it brings upon the original path $(\widehat{q}_t^{\Theta},\widehat{k}_t)_{t=0}^T$ is given by the following:
\begin{equation}\label{A-imp}
    \begin{aligned}
        &\int_{\underline{q}^i}^{\overline{q}^i}\mu^i_0\left[\theta^i-w^N(\widetilde{q}^i,\underline{q}^j)\right]d\widetilde{q}^i+\int_{\underline{q}^j}^{\overline{q}^j}\mu^j_0\left[\theta^j-w^N(\overline{q}^i,\widetilde{q}^j)\right]d\widetilde{q}^j \\
        &\;\;\;\;\;\;\;\;\;\;-\int_{\underline{q}^j}^{\overline{q}^j}\mu^j_0\left[\theta^j-w^N(\underline{q}^i,\widetilde{q}^j)\right]d\widetilde{q}^j-\int_{\underline{q}^i}^{\overline{q}^i}\mu^i_0\left[\theta^i-w^N(\widetilde{q}^i,\overline{q}^j)\right]d\widetilde{q}^i \\
        =&\int_{\underline{q}^j}^{\overline{q}^j}\mu^j_0w^N(\underline{q}^i,\widetilde{q}^j)d\widetilde{q}^j+\int_{\underline{q}^i}^{\overline{q}^i}\mu^i_0w^N(\widetilde{q}^i,\overline{q}^j)d\widetilde{q}^i-\int_{\underline{q}^i}^{\overline{q}^i}\mu^i_0w^N(\widetilde{q}^i,\underline{q}^j)d\widetilde{q}^i-\int_{\underline{q}^j}^{\overline{q}^j}\mu^j_0w^N(\overline{q}^i,\widetilde{q}^j)d\widetilde{q}^j.
    \end{aligned}
\end{equation}
\end{comment}

To make things more aligned with definition (\ref{quantile-skepticism value}), we further rewrite (\ref{A-skepticism value}) in terms of the total ex-ante disclosure probability. By letting $x:=\mu^i_0q^i$, $y:=\mu^j_0q^j$, $\alpha:=\E[\theta]-\sum_{k=1}^K\mu^k_0\widehat{q}_t^k\theta^k$, $\beta:=1-\sum_{k=1}^K\mu^k_0\widehat{q}_t^k$, $Q^i=\mu^i_0(\overline{q}^i-\underline{q}^i)$, and $Q^j=\mu^j_0(\overline{q}^j-\underline{q}^j)$, and defining $\widehat{v}^N(x,y)=v^N(q^i,q^j)=\frac{\alpha-x\theta^i-y\theta^j}{\beta-x-y}$, we define
% two quantile-skepticism values (\ref{A-skepticism value i-first}) and (\ref{A-skepticism value j-first}):
\begin{gather}
    \widehat{v}_i^N(Q)=
    \begin{cases}
        \widehat{v}^N(Q,0) & \text{if } Q\in[0,Q^i]; \\
        \widehat{v}^N(Q^i,Q-Q^i) & \text{if } Q\in[Q^i,Q^i+Q^j].
    \end{cases} \label{A-skepticism value i-first} \\
    \widehat{v}_j^N(Q)=
    \begin{cases}
        \widehat{v}^N(0,Q) & \text{if } Q\in[0,Q^j]; \\
        \widehat{v}^N(Q-Q^j,Q^j) & \text{if } Q\in[Q^j,Q^i+Q^j].
    \end{cases} \label{A-skepticism value j-first}
\end{gather}
Note that (\ref{A-skepticism value i-first}) and (\ref{A-skepticism value j-first}), respectively, correspond to the quantile-skepticism values (only the part regarding the two steps $t+1$ and $t+2$) of the improved path and the original path. For either path, in total, we have probability $Q^i$ induced from $\theta^i$ and probability $Q^j$ is induced from $\theta^j$. $\widehat{v}^N_i(Q)$ denotes the quantile-skepticism value for the new path where $q^i$ increases before $q^j$ does when the total ex-ante probability is $Q$. In particular, for $Q\in[0,Q^i]$, $\widehat{v}_i^N(Q)$ refers to the skepticism value when $q^i=\underline{q}^i+Q/\mu_0^i$ and $q^{-i}=\widehat{q}_t^{-i}$, and for $Q\in[Q^i,Q^i+Q^j]$, $\widehat{v}^N_i(Q)$ refers to the skepticism value when $q^i=\overline{q}^i$, $q^j=\underline{q}^j+Q/\mu^j_0$, and $q^{-i,j}=\widehat{q}_t^{-i,j}$. Likewise, $\widehat{v}_j^N(.)$ corresponds to the quantile-skepticism value of the original path where $q^j$ moves before $q^i$ does.

\begin{comment}
What we eventually want to do with all the fuss above is to rewrite the benefit of exchanging path order in the following concise manner:
\begin{equation}\label{A-imp cont}
    \begin{aligned}
        \text{RHS of (\ref{A-imp})}&=\int_0^{Q^j}\widehat{w}^N(Q,0)dQ+\int_0^{Q^i}\widehat{w}^N(Q^j,Q)dQ-\int_0^{Q^i}\widehat{w}^N(0,Q)dQ-\int_0^{Q^j}\widehat{w}^N(Q,Q^i)dQ \\
        &=\int_0^{Q^j}\widehat{w}^N_j(Q)dQ+\int_{Q^j}^{Q^i+Q^j}\widehat{w}^N_j(Q)dQ-\int_0^{Q^i}\widehat{w}^N_i(Q)dQ-\int_{Q^i}^{Q^i+Q^j}\widehat{w}^N_i(Q)dQ \\
        &=\int_0^{Q^i+Q^j}\left[\widehat{w}_j^N(Q)-\widehat{w}_i^N(Q)\right]dQ.
    \end{aligned}
\end{equation}
\end{comment}

Our goal is to show $\widehat{v}_j^N(.)\geq\widehat{v}_i^N(.)$ for all $Q\in(0,Q^i+Q^j)$. Applying the argument we developed in Section \ref{upper bound for maximal revenue guarantee}, this implies that exchanging the two steps weakly decreases the total compensation to the producer, $\int_0^{Q_M}w^N(q_Q^{\Theta})dQ$, derived in (\ref{upper bound-reformulate}). One can see the claim above by noticing the following:
\begin{gather}
    (\beta-Q)\widehat{v}_i^N(Q)=
    \begin{cases}
        \alpha-Q\theta^i & \text{if } Q\in[0,Q^i]; \\
        \alpha-Q\theta^j-(\theta^i-\theta^j)Q^i & \text{if } Q\in[Q^i,Q^i+Q^j].
    \end{cases} \label{A-trick i} \\
    (\beta-Q)\widehat{v}_j^N(Q)=
    \begin{cases}
        \alpha-Q\theta^j & \text{if } Q\in[0,Q^j]; \\
        \alpha-Q\theta^i+(\theta^i-\theta^j)Q^j & \text{if } Q\in[Q^j,Q^i+Q^j].
    \end{cases} \label{A-trick j}
\end{gather}
Knowing that $\beta-Q>0$ and $\theta^i>\theta^j$, one should be able to verify the claim above. Hence, we construct an alternative path that does no worse than $(\widehat{q}_t^{\Theta},\widehat{k}_t)_{t=0}^T$.

Any finite-step path can be improved sequentially by exchanging all the ``unreasonable'' orders (with $\widehat{k}_t>\widehat{k}_{t+1}$) to finally become the sequential self-conquering path. For a countable-step path, given every integer $\widetilde{T}>0$, the truncated finite path $(\widehat{q}_t^{\Theta},\widehat{k}_t)_{t=0}^{\widetilde{T}}$ can be similarly improved to the sequential self-conquering path toward the endpoint $\widehat{q}_{\widetilde{T}}^{\Theta}$. Denote by $R(q_{\widehat{T}}^{\Theta})$ the revenue bound (\ref{upper bound}) produced by this very sequential self-conquering path. Because the original path is nondecreasing and bounded, it must converge to some $\widehat{q}_{\infty}^{\Theta}\geq\widehat{q}_{\widehat{T}}^{\Theta}$. Thus, the revenue guaranteed by the path is upward bounded by $R(\widehat{q}_{\widehat{T}}^{\Theta})+\theta^1\sum_{k=1}^K\mu^k_0(\widehat{q}^k_{\infty}-\widehat{q}^k_{\widehat{T}})$. Since every $\widehat{T}$ generates such a bound and $R(.)$ is continuous, this is also a bound when $\widehat{T}\rightarrow\infty$, and the limit bound is $R(\widehat{q}_{\infty}^{\Theta})$, which is produced by the sequential self-conquering path toward the endpoint $\widehat{q}_{\infty}^{\Theta}$. So, the same conclusion holds for any infinite-step path.

Second, we show that given the optimal path being sequential self-conquering, (\ref{upper bound}) is maximized by the full-disclosure endpoint $\underline{q}^{*\Theta}$. Given the optimal path, we rewrite (\ref{upper bound}) for every endpoint $\underline{q}^{\Theta}$
\begin{equation}\label{A-upper bound with endpoint}
    \text{(\ref{upper bound})}=\sum_{k=1}^K\mu^k\int_0^{\underline{q}^k}\left[\theta^k-c-w^N(\underline{q}^1,\underline{q}^2,...,\underline{q}^{k-1},\widetilde{q},0,...,0)\right]d\widetilde{q}
\end{equation}
A nice property of the sequential self-conquering path is that changing the probability of a low type $\underline{q}^k$ will not affect the first $k-1$ integrals in the summation of (\ref{A-upper bound with endpoint}) that are associated with types higher than $\theta^k$. Thus, it is weakly beneficial to set $\underline{q}^k=0$ for all $\theta^k\leq c$ because these low types' associated integrands in (\ref{A-upper bound with endpoint}) are always nonpositive. Next, consider first moving $\underline{q}^1$ upward. This increases the value of the first integral in the summation of (\ref{A-upper bound with endpoint}) since (i) $\theta^1-c\geq w^N(q^{\Theta})$ for all possible $q^{\Theta}$; and (ii) a higher $\underline{q}^1$ expands the integration interval with the integrand being nonnegative. Meanwhile, this change of $\underline{q}^1$ also weakly increases the other integrals. To see this, note that for all $q^{\Theta}\neq 1^{\Theta}$ and $k$:
\begin{equation}\label{A-derivative of w^N}
    \frac{\partial w^N(q^{\Theta})}{\partial q^k}=
    \begin{cases}
        -\frac{\mu^k_0\sum_{j=1}^K\mu^j_0(1-q^j)(\theta^k-\theta^j)}{\left(1-\sum_{j=1}^K\mu^j_0q^j\right)^2} &\text{ if }w^N(q^{\Theta})>0; \\
        0 &\text{ otherwise.}
    \end{cases}
\end{equation}
Hence, with $k=1$, the skepticism value is nonincreasing in $\underline{q}^1$ and so it is (at least weakly) improving to set $\underline{q}^1=1$. Furthermore, once $\underline{q}^1=1$, we can consider next moving $\underline{q}^2$ upward. In fact, suppose for some $k$ with $\theta^k>c$, $\underline{q}^j=1$ for all $j<k$, then moving $\underline{q}^k$ up increases the value of the $k$th integral in (\ref{A-upper bound with endpoint}) by also expanding the integration interval of a nonnegative function, $\theta^k-w^N(1,...,1,\widetilde{q},0,...,0)$ with $k-1$ ones. The derivative (\ref{A-derivative of w^N}) now becomes $-\mu^k_0\sum_{j=k+1}^K\mu^j_0(1-q^j)(\theta^k-\theta^j)/(1-\sum_{j=1}^K\mu^j_0q^j)^2\leq0$, still improving the integrals that come after the $k$th. All these show that the endpoint $\underline{q}^{*\Theta}$ together with the sequential self-conquering path delivers the maximal value of (\ref{upper bound}).

Third, we show the maximal bound obtained under the path and the endpoint chosen above equals exactly (\ref{MRG*}). By comparing (\ref{MRG*}) and (\ref{upper bound-reformulate}), one can see this immediately by recalling that $\int_0^{Q_M}w^N(q_Q^{\Theta})dQ=\sum_{k=1}^{\underline{k}}\int_0^1\mu_0^ku_k^*(q)dq$ and the fact that $u_k^*(q)=0$ once $(k,q)$ has passed the tipping point $(k^*,q^*)$.

We complete the proof by verifying the form of each $p_k^*$ given in (\ref{multiple type}): for all $(k,q)$ with $w^N_k(q)>0$,
\begin{equation}\label{A-explicit form}
    \begin{aligned}
        p_k^*(q)&=\int_0^q[\theta^k-c-w^N_k(q)]d\widetilde{q}=\int_0^q(\theta^k-\E[\theta|\mu^N(\underbrace{1,...,1}_{k-1\text{ ones}},\widetilde{q},0,...,0)])d\widetilde{q} \\
        &=\int_0^q\frac{\sum_{j=k}^K\mu^j_0(\theta^k-\theta^j)}{\sum_{j=k}^K\mu^j_0-\mu^k_0\widetilde{q}}d\widetilde{q}=\int_0^q\frac{\sum_{j=k}^K\mu^j_0(\theta^k-\theta^j)\big/\sum_{j=k}^K\mu^j_0}{1-\mu^k_0\widetilde{q}\big/\sum_{j=k}^K\mu^j_0}d\widetilde{q} \\
        &=\int_0^q\frac{\theta^k-\E[\theta|\theta\leq\theta^k]}{1-\Pr(\theta^k|\theta\leq\theta^k)\widetilde{q}}d\widetilde{q}=\frac{\theta^k-\E[\theta|\theta\leq\theta^k]}{\Pr(\theta^k|\theta\leq\theta^k)}\ln\frac{1}{1-\Pr(\theta^k|\theta\leq\theta^k)q}.
    \end{aligned}
\end{equation}

\subsection{Proof of Theorem \ref{main result}}

\emph{Proof.} Let $\widetilde{R}$ denote the maximal value of the bound (\ref{upper bound}) provided in Lemma \ref{lemma-upper bound}. Since $R^*\leq\widetilde{R}$, it suffices to show that there exists a sequence of menu profiles $(M_n^{\Theta})_{n=1}^{\infty}$ such that (i) it converges to $M^{*\Theta}$ with respect to the Hausdorff metric, and (ii) $\lim_{n\rightarrow\infty}R(M_n^{\Theta})=\widetilde{R}$. With this, we can conclude $R^*=\widetilde{R}$ and $M^{*\Theta}$ is a robustly optimal menu profile. To do this, take a small $\varepsilon>0$ and consider the menu profile sequence given by for all $n$, for all $k$, $M^k_n=\{(q,p-\frac{\varepsilon}{n}q):(q,p)\in M^{*k}\text{ and }p>0\}$. One can easily see that this sequence satisfies (i) and contains menu profiles that are slightly lower than those in $M^{*\Theta}$. To show (ii), it suffices to show that for all $n$, $M_n^{\Theta}$ robustly induces the full-disclosure profile $\underline{q}^{*\Theta}$, at which the induced revenue guarantee will converge to $\widetilde{R}$ as implied by the proof of Lemma \ref{lemma-upper bound}.

Thus, we need to show that every $M_n^{\Theta}$ satisfies the claim above. Notice that every possible candidate equilibrium must have a probability profile $\widehat{q}^{\Theta}\leq\underline{q}^{*\Theta}$. In the first step, we prove that if $\widehat{q}^{\Theta}$ is an equilibrium, $\widehat{q}^1=1$. Suppose $\widehat{q}^1<1$ and $\widehat{q}^{\Theta}$ corresponds to an equilibrium. Denote the supporting types of $\widehat{q}^{\Theta}$ by $\widehat{\Theta}=\{k:\widehat{q}^k>0\}$. 
Let $\widetilde{q}^{\Theta}=(\widehat{q}^1,0,...,0)$.
First, we consider the case with $w^N(\widehat{q}^{\Theta}) \leq w^N(\widetilde{q}^{\Theta})$ and show now $\widehat{q}^{\Theta}$ cannot be an equilibrium. Abusing the notations a little, let $p_k^*$ be the price envelope of $M^{*k}$. By the construction in (\ref{multiple type}) and Theorem \ref{main result}, we have ${p_k^*}'(q)=\theta^k-c-w^N_k(q)$ for all $\theta^k>c$ (this is because after the tipping point is reached, $w_k^N(q)=0$ and the marginal price of $M^{*k}$ now equals $\theta^k-c$).
%These $\widehat{q}^{\Theta}$ are candidate equilibria close to the polyline created by the hybrid divide-and-conquer scheme.
Thus, given $M_n^{\Theta}$, the marginal price type $\theta^1$ faces is $p_1^{*'}(\widehat{q}^1)-\frac{\varepsilon}{n}=\theta^1-c-w^N(\widetilde{q}^{\Theta})-\frac{\varepsilon}{n}<\theta^1-c-w^N(\widehat{q}^{\Theta})$, which means $\theta^1$ wants to deviate upward, a contradiction to $\widehat{q}^{\Theta}$ forming an equilibrium. Second, we consider the case where $w^N(\widehat{q}^{\Theta}) > w^N(\widetilde{q}^{\Theta})$ and show it cannot form an equilibrium. Such $\widehat{q}^{\Theta}$ corresponds to a candidate equilibrium where some lower type discloses too aggressively. Because the highest type discloses with the same probability in both candidate equilibria and $w^N(\widehat{q}^{\Theta}) > w^N(\widetilde{q}^{\Theta})$, $\widehat{q}^{\Theta}$ must include a positive probability of some type $k'$ such that $\theta^{k'}-c<w^N(\widetilde{q}^{\Theta})$, which further implies $\theta^{k'}-c< w^N(\widehat{q}^{\Theta})$. However, because the price at $\widehat{q}^{k'}>0$ is strictly positive, type $k'$ is strictly better off by deviating to $(0,0)$. To conclude the paragraph, if $\widehat{q}^{\Theta}$ is an equilibrium, $\widehat{q}^1=1$.

The next step is to do the same thing as above for $\theta^2$ and sequentially down to $\theta^{\underline{k}}$. We show this by induction. Given $2\leq k\leq\underline{k}$, suppose $\widehat{q}^j=1$ for all $j<k$, $\widehat{q}^k<1$, and $\widehat{q}^{\Theta}$ corresponds to an equilibrium. The argument is almost the same as before but with $\widetilde{q}^{\Theta}=(1,...,1,\widehat{q}^k,0,...,0)$ with $k-1$ ones. In this way, we show that $\underline{q}^{*\Theta}$ is the only candidate profile for equilibrium.

Lastly, we verify that $\underline{q}^{*\Theta}$ corresponds to an equilibrium, and thus is the unique equilibrium. On the one hand, each price envelop $p_k^*$ is convex, so for all $\theta^k>c$, $\underline{q}^k=1$ is globally optimal if and only if ${p_k^*}'(1)-\frac{\varepsilon}{n}=\theta^k-c-w_k^N(1)-\frac{\varepsilon}{n}\leq\theta^k-c-w^N(\underline{q}^{*\Theta})$, which is true because $w^N(\underline{q}^{*\Theta})=\max\{\E[\theta|\theta\leq c]-c,0\}=0\leq w_k^N(1)$. On the other hand, every $\theta^k\leq c$ always reaches optimum as she only has a single choice. Thus, $\underline{q}^{\Theta}$ forms an equilibrium. Back to the logit of first paragraph, we complete the proof.

\section{}

\subsection{Proof of Proposition \ref{proposition-benchmark}}

We first prove the result for the case where $v_0\leq c$. To show robust optimality, our goal is to construct a sequence of menu profiles $(M_n^{\Theta})_{n=1}^{\infty}$ that approximates $\overline{M}^{\Theta}$ in the sense of Definition \ref{ROMP-def}. Take a small $\varepsilon>0$, and for all $n$ and $k$, let $M_n^k=\{(0,0),(1,\theta^k-c-\frac{\varepsilon}{n})\}$ if $\theta^k>c$; and $M_n^k=\{0,0\}$ if $\theta^k\leq c$. The sequence $(M_n^k)_{n=1}^{\infty}$ obviously converges to $\overline{M}^k$ with respect to the Hausdorff metric, and if $M_n^k$ robustly induces all types $\theta^k>c$ to fully disclose, the sequence $(R(M_n^k))_{n=1}^{\infty}$ also converges to $\overline{R}$. Thus, it remains to show each $M_n^k$ fully implements full disclosure (of types $\theta^k>c$). Since the prices are lower than in $\overline{M}^{\Theta}$, following the argument in Section \ref{benchmark without strategic uncertainty}, we can show this full-disclosure equilibrium exists. Then, we suppose there is another equilibrium. Note that the consumer's skepticism posterior mean in this equilibrium must be no greater than $v_0$ since (i) the posterior mean is $v_0$ when no disclosure happens; (ii) any type $\theta^k$ willing to disclose must have $\theta^k\geq c\geq v_0$, so the consumer's skepticism here can only be worse than the prior. Hence, if no disclosure happens, the producer will not produce and earn 0. To differ from the full-disclosure equilibrium, there must be some $\theta^k>c$ who does not fully disclose, which is not possible here because given the outside value 0, she wants to deviate to $(1,\theta^k-c-\frac{\varepsilon}{n})$ to earn $\frac{\varepsilon}{n}$. This part of the proof is therefore done.

Next, we focus on the case with $v_0>c$. A first remark is that for this result to hold, the equilibrium concept needs to be extended to mixed-strategy equilibrium. This is a technical adjustment for guaranteeing equilibrium existence. We didn't bother reporting this subtlety in the main text because our main results also hold when considering the optimal robust design that maximizes revenue guarantee across all mixed equilibria. Another remark is that we only examine the producer types with $\theta^k\geq c$ whenever we take some $k$, since choosing $(0,0)$ is dominant for a type $\theta^k<c$.

Now, we fix some menu profile $M^{\Theta}$, a small $\varepsilon\geq0$, and an induced equilibrium $\overline{g}$ where the revenue is no less than $\overline{R}-\varepsilon$. In $\overline{g}$, denote the skepticism value by $\overline{w}^N$, let $q^k$ and $p^k$ be the expected probability and expected price chosen on path by type $\theta^k$ (we take expectation to deal with mixed strategies), and let the indifference curve of $\theta^k$ be:
\begin{equation}\label{C^k indiff curve}
    \mathcal{C}^k:=\{(q,p):(q-q^k)(\theta^k-c-\overline{w}^N)=p-p^k\}.
\end{equation}
Even if type $\theta^k$ randomizes among plans, all on-path plans must stay on $\mathcal{C}^k$. Since $(0,0)\in M^k$, it cannot lie beneath $\mathcal{C}^k$ because, otherwise, it would be a profitable deviation of $\theta^k$. The chosen probability satisfies $q^k\leq1$. Moreover, the lower bound for revenue $\overline{R}-\varepsilon=\sum_{\theta^k\geq c}\mu_0^k(\theta^k-c)-\varepsilon$ requires the chosen price have $p^k\geq\theta^k-c-\frac{\varepsilon}{\mu^k_0}$ for all $\theta^k\geq c$. This is because, with any single type $k'$ violating this, the revenue is at most $\sum_{\theta^k\geq c}\mu_0^kp^k<\mu_0^{k'}(\theta^k-c)-\varepsilon+\sum_{\theta^k\geq c,k\neq k'}\mu_0^k(\theta^k-c)=\overline{R}-\varepsilon$. The fact that $p^k>0$ implies $q^k>0$, so the slope of $\mathcal{C}^k$, $\theta^k-c-\overline{w}^N$, is strictly greater than 0. 

With all the observations above, we construct $\delta$ and show the existence of an equilibrium where the revenue is no greater than $\delta\varepsilon$. We start by fixing any small $\delta>0$ and then find conditions it needs to satisfy. Let $(q,\delta\varepsilon/K)$ be the intersection of $\mathcal{C}^k$ and the horizontal line $p=\delta\varepsilon/K(<p^k)$. We then claim that $q\leq a^k\varepsilon$ where $a^k:=\frac{\delta/K+1/\mu_0^k}{\theta^k-c}$. To see this, we solve for the value of $q$ according to (\ref{C^k indiff curve}) and show the following
\begin{equation}
\label{akLieAbove}
    q=q^k-\frac{p^k-\delta\varepsilon/K}{\theta^k-c-\overline{w}^N}\leq 1-\frac{p^k-\delta\varepsilon/K}{\theta^k-c-\overline{w}^N}\leq 1-\frac{p^k-\delta\varepsilon/K}{\theta^k-c}\leq 1-\frac{(\theta^k-c-\varepsilon/\mu_0^k)-\delta\varepsilon/K}{\theta^k-c}=a^k\varepsilon.
\end{equation}
Here, the three inequalities apply $q^k\leq1$, $\overline{w}^N\geq0$, and $p^k\geq\theta^k-c-\frac{\varepsilon}{\mu^k_0}$, respectively.

Next, we consider a new disclosure game induced by the following menu profile $M_{\delta}^{\Theta}$ given by, for all $k$, $M_{\delta}^k=\{(q,p)\in M^k:p\leq\delta\varepsilon/K\}$. Since $M_{\delta}^{\Theta}$  has closed graph (because $M^{\Theta}$ has closed graph) and $(0,0)\in M_{\delta}^k$ (so the menus are nonempty), there always exists a (mixed-strategy) equilibrium in this ``small'' game, and we denote it by $g_{\delta}$. Our goal is thus to find $\delta$ such that $g_{\delta}$ is also an equilibrium induced by $M^{\Theta}$, and then by construction, the revenue in $g_{\delta}$ cannot exceed $K\times(\delta\varepsilon/K)=\delta\varepsilon$. If $\delta$ does not depend on $\varepsilon$, the proof is done. 

Given $M^{\Theta}$, $\overline{g}$ being an equilibrium means all plans in $M^k$ lie weakly above $\mathcal{C}^k$. Likewise, we denote by $\mathcal{C}^k_{\delta}$ the indifference curve of $\theta^k$ in $g_{\delta}$, and $M^k_{\delta}$ lies weakly above $\mathcal{C}^k_{\delta}$. Thus, to show $g_{\delta}$ is an equilibrium given $M^{\Theta}$, we must show $M^k$ also lies weakly above $\mathcal{C}_{\delta}^k$. In fact, it suffices to show that every $(q,p)\in M^k$ that lies below $\mathcal{C}^k_{\delta}$ satisfies $p\leq\delta\varepsilon/K$. This is sufficient because, in $g_{\delta}$, given $M^{\Theta}$, whenever type $\theta^k$ wants to deviate to some $(q,p)\in M^k$, then the claim above implies we also have $(q,p)\in M^k_{\delta}$; however, since $g_{\delta}$ is an equilibrium given $M^{\Theta}_{\delta}$, such $(q,p)$ cannot exist, so $g_{\delta}$ is also an equilibrium given $M^{\Theta}$.

To show the claim above, we recall that every plan in $M^k_{\delta}$ (and thus in $g_{\delta}$) cannot have a probability larger than $\delta\varepsilon/K$. Hence, the continuity of the skepticism value (\ref{skepticism value}) implies that the skepticism value in $g_{\delta}$, denoted as $w^N_{\delta}$, must be $\max\{v_0-c+o(\varepsilon),0\}=v_0-c+o(\varepsilon)$ (here is where we use $v_0>c$) where $o(\varepsilon)$ is a term that is of magnitude $\varepsilon$. We draw an auxiliary line, denoted as $\mathcal{C}^k_0$, that crosses $(0,0)$ and has slope $\theta^k-c-w_{\delta}^N$. Notice that $\mathcal{C}^k_0$ is parallel to $\mathcal{C}^k_\delta$ and cannot lie beneath $\mathcal{C}^k_\delta$ since, otherwise, $(0,0)\in \mathcal{C}_0^k$ lies beneath $\mathcal{C}^k_\delta$, a contradiction to $g_{\delta}$ being an equilibrium given $M^{\Theta}_{\delta}$. 

Now, we show the claim above. Take any $(q,p)\in M^k$ that lies beneath $\mathcal{C}^k_{\delta}$, and thus also lie beneath $\mathcal{C}^k_0$. $(q,p)\in M^k$ also means it must lie weakly above $\mathcal{C}^k$. We will analyze the relationship of $\mathcal{C}^k$ and $\mathcal{C}^k_0$ to locate $(q,p)$. On the one hand, $\mathcal{C}^k$ crosses some point $(q',0)$ with $q'\geq0$ because $(0,0)\in M^k$ cannot lie beneath $\mathcal{C}^k$. Also, inequality (\ref{akLieAbove}) shows that $(a^k\varepsilon,\delta\varepsilon/K)$ lies weakly beneath $\mathcal{C}^k$. On the other hand, $\mathcal{C}^k_0$ crosses $(0,0)$. Thus, if point $(a^k\varepsilon,\delta\varepsilon/K)$ lies weakly above $\mathcal{C}^k_0$, the curve $\mathcal{C}^k_0$ will always be strictly beneath the curve $\mathcal{C}^k$ for any $q> a^k \varepsilon$. Consequently, the pair $(q,p)$, which lies weakly above $\mathcal{C}^k$ but lies beneath $\mathcal{C}^k_0$, must satisfy $q\leq a^k \varepsilon$ and $p\leq\delta\varepsilon/K$, implying $(q,p)\in M_{\delta}^k$ as well.

To conclude the previous paragraph, one sufficient condition for $(a^k\varepsilon,\delta\varepsilon/K)$ to lie weakly above $\mathcal{C}^k_0$ is that the intersection of $\mathcal{C}^k_0$ and the vertical line $q=a^k\varepsilon$, denoted as $(a^k\varepsilon,p')$, satisfies $p'\leq\delta\varepsilon/K$. This is equivalent to the following
\begin{equation}
    \begin{aligned}
        p'=a^k\varepsilon&(\theta^k-c-w^N_{\delta})\leq\frac{\delta\varepsilon}{K} \Leftrightarrow \delta\geq Ka^k[\theta^k-c-(v_0-c+o(\varepsilon))]=(\theta^k-v_0)\frac{\delta+K/\mu_0^k}{\theta^k-c}+o(\varepsilon) \\
        \Leftrightarrow \delta&\geq\frac{\theta^k-v_0}{v_0-c}\frac{K}{\mu_0^k}+o(\varepsilon)\Rightarrow\text{it suffices to set }\delta>\frac{K}{v_0-c}\max_k\frac{\theta^k-v_0}{\mu_0^k}.
    \end{aligned}
\end{equation}
The constructed $\delta$ above is independent of $\varepsilon$ and relies on $v_0>c$. Therefore, we complete the proof.

\subsection{Proof of Proposition \ref{proposition-market power}}

%We consider increasing the production cost from $c$ to $c'$, with both $c,c'<v_0$.
%Let the tipping point under $c$ be $(k^*,q^*)$, and that under $c'$ be $({k^*}',{q^*}')$. By how we construct the tipping point, the sequential self-conquering process drives the producer's skepticism value to hit a higher cost before a lower one, so $({k^*}',{q^*}')$ is before $(k^*,q^*)$.
\emph{Proof.} Let $\mu^N_k(q)$ be the consumer's skepticism belief given the producer's profile in (\ref{w^N_k(q)}) that we used to define $w_k^N(q)$. Moreover, for every efficient type $\theta^k\geq \theta^{\underline{k}}>c$ and $q\in[0,1]$, let $l_c(k,q)$ be
\[l_c(k,q):=\frac{\theta^k-c-w^N(q)}{\theta^k-c}=\frac{\theta^k-c-\max\{\E[\theta|\mu^N_k(q)]-c,0\}}{\theta^k-c}.\]
Then, by Theorem \ref{main result}, the maximal revenue guarantee can be written as
\[R^*(c):=\sum_{k=1}^{\underline{k}}\mu_0^k\int_0^1\left[\theta^k-c-\max\{\E[\theta|\mu^N_k(q)]-c,0\}\right]dq=\sum_{k=1}^{\underline{k}}\mu_0^k\int_0^1l_c(k,q)(\theta^k-c)dq.\]
Note that if $\E[\theta|\mu^N_k(q)]>c$ (namely $w_k^N(q)>0$, so $(k,q)$ is before the tipping point), we have $l_c(k,q)=(\theta^k-\E[\theta|\mu^N_k(q)])/(\theta^k-c)$ and its derivative is $\frac{\partial l_c(k,q)}{\partial c}=\frac{l_c(k,q)}{\theta^k-c}>0$; otherwise, $l_c(k,q)=1$, not varying with $c$. Hence, when $c<v_0$, the tipping point $(k^*,q^*)$ cannot be $(1,0)$, so $l_c(k,q)$ increases strictly in $c$ for a nonzero-measure of $(k,q)$. Next, let $\overline{q}(k)=1$ if $\theta^k>\theta^{k^*}$; $\overline{q}(k^*)=q^*$; and $\overline{q}(k)=0$ if $\theta^k<\theta^{k^*}$. In addition, we have $\overline{R}(c)=\sum_{\theta^k>c}\mu^k_0(\theta^k-c)$, so $\overline{R}'(c)=-\Pr(\theta^k>c)$ and $-\frac{\overline{R}(c)}{\overline{R}'(c)}=\E[\theta-c|\theta>c]>\E[\theta-c]=v_0-c$. We then examine the derivative of $R^*$ with respect to $c$, for $c<v_0$
\begin{equation*}
    \begin{aligned}
        {R^*}'(c)&=\sum_{k=1}^{\underline{k}}\mu_0^k\int_0^1\frac{\partial l_c(k,q)}{\partial c}(\theta^k-c)dq-\sum_{k=1}^{\underline{k}}\mu_0^k\int_0^1l_c(k,q)dq \\
        %&=\sum_{k=1}^{k^*}\mu_0^k\int_0^{\overline{q}(k)}\frac{l_c(k,q)}{\theta^k-c}(\theta^k-c)dq-\sum_{k=1}^{\underline{k}}\mu_0^k\int_0^1l_c(k,q)dq \\
        &=\sum_{k=1}^{k^*}\mu_0^k\int_0^{\overline{q}(k)}l_c(k,q)dq-\sum_{k=1}^{\underline{k}}\mu_0^k\int_0^1l_c(k,q)dq=-\sum_{k=k^*}^{\underline{k}}\mu_0^k\int_{\overline{q}(k)}^1l_c(k,q)dq \\
        &=-\sum_{k=k^*}^{\underline{k}}\mu_0^k\int_{\overline{q}(k)}^1dq=-\Pr(\theta^k>c)+\sum_{k=1}^{k^*}\mu_0^k\int_0^{\overline{q}(k)}dq=\overline{R}'(c)+\sum_{k=1}^{k^*}\mu_0^k\int_0^{\overline{q}(k)}\frac{v_0-c}{v_0-c}dq \\
        &>\overline{R}'(c)+\sum_{k=1}^{k^*}\mu_0^k\int_0^{\overline{q}(k)}\frac{v_0-c}{-\overline{R}(c)/\overline{R}'(c)}dq>\overline{R}'(c)+\sum_{k=1}^{k^*}\mu_0^k\int_0^{\overline{q}(k)}\frac{w_k^N(q)}{-\overline{R}(c)/\overline{R}'(c)}dq \\
        &=\frac{\overline{R}(c)-\sum_{k=1}^{k^*}\mu_0^k\int_0^{\overline{q}(k)}w_k^N(q)dq}{\overline{R}(c)/\overline{R}'(c)}=\frac{\overline{R}'(c)R^*(c)}{\overline{R}(c)}.
    \end{aligned}
\end{equation*}
Above, the fifth equality applies $l_c(k,q)=1$ for all $(k,q)$ after $(k^*,q^*)$; the sixth and seventh equalities use $\sum_{k=1}^{\underline{k}}\mu_0^k\int_0^1dq=\Pr(\theta^k>c)=-\overline{R}'(c)$; the first inequality uses $-\overline{R}(c)/\overline{R}'(c)>v_0-c$; the second inequality uses $w_k^N(q)\leq v_0-c$, which we will show in the proof of Proposition \ref{proposition-producer surplus}, and the inequality is strict when $(k,q)$ is after $(1,0)$. As a result, PSS is strictly increasing in $c>v_0$ because
\begin{equation*}
    \begin{aligned}
        \frac{d}{dc}\frac{R^*(c)}{\overline{R}(c)}=\frac{{R^*}'(c)\overline{R}(c)-\overline{R}'(c)R^*(c)}{\left[\overline{R}(c)\right]^2}>\frac{\frac{\overline{R}'(c)R^*(c)}{\overline{R}(c)}\overline{R}(c)-\overline{R}'(c)R^*(c)}{\left[\overline{R}(c)\right]^2}=0.
    \end{aligned}
\end{equation*}
Moreover, the statement for MK is trivial. Whereas, CTE strictly increases in $c>v_0$ because
\begin{equation*}
    \begin{aligned}
        \frac{d}{dc}\left[1-\frac{v_0-c}{\overline{R}(c)}\right]=\frac{\overline{R}(c)+\overline{R}'(c)(v_0-c)}{\left[\overline{R}(c)\right]^2}>\frac{\overline{R}(c)+\overline{R}'(c)\left[-\frac{\overline{R}(c)}{\overline{R}'(c)}\right]}{\left[\overline{R}(c)\right]^2}=0.
    \end{aligned}
\end{equation*}
Here, we utilize $v_0-c<-\frac{\overline{R}(c)}{\overline{R}'(c)}$ and $\overline{R}'(c)<0$. At this point, we complete the proof.

\begin{comment}
\subsection{Proof of Proposition \ref{proposition-comparative}}
\begin{proof}
This proof relies on the proof of Proposition \ref{proposition-joint design} and the language in Section \ref{joint design of evidence structure}.

   In the proof of Proposition \ref{proposition-joint design}, we show that, given a signal $\pi$, the choice of menu profile is equivalent to the same choice in the main model where the types are essentially the posterior means, namely for all $s\in S$. Using this observation we can prove Proposition \ref{proposition-comparative} via Proposition  \ref{proposition-joint design}. 
    
    To do so, notice that $\mu_1$ is a mean-preserving spread of $\mu_2$. Consider a model with prior $\mu_1$, then there exists a signal structure $\pi$ such that the posterior distribution coincides with $\mu_2$. We show in the proof of Proposition \ref{proposition-joint design} that the full-revealing signal makes the platform strictly better off than any noisy signal structure $\pi$. This joint with the previous observation completes the proof.
\end{proof}
\end{comment}

\subsection{Proof of Proposition \ref{proposition-converge0}}

\emph{Proof.} Denote by $\theta^{k_0}$ the type equal to the prior mean $v_0$. If $v_0\leq c$, as the prior converges to a degenerate inefficient type, the total surplus goes to zero, and so will the platform's revenue. Hence, we focus on $v_0>c$ below. Fixing each prior $\mu_n$, we denote the associated expectation and probability with $\E_n$ and $\Pr_n$. Also, let the associated tipping point be $(k^*_n,q^*_n)$. We first show the following claim: for every $\varepsilon>0$, there is $n_{\varepsilon}$ such that for all $n>n_{\varepsilon}$, $(k^*_n,q^*_n)$ is after $(k_0,1-\varepsilon)$. This is because we can always find a prior close enough to the degenerate limit $v_0$ with $\Pr_n(\theta<v_0)<\delta$ and $\Pr_n(\theta=v_0)>1-\delta$, where $\delta>0$ satisfies that when the sequential self-conquering process proceeds to $(k_0,1-\varepsilon)$, the consumer skepticism posterior mean, which is lower-bounded by the following, is still greater than $c$
    \[\frac{\varepsilon(1-\delta)v_0+\delta\theta^K}{\varepsilon(1-\delta)+\delta}>c\text{, namely }\delta<\frac{\varepsilon(v_0-c)}{(c-\theta^K)+\varepsilon(v_0-c)}>0.\]
    This implies that the process cannot stop here, so the tipping point must be after $(k_0,1-\varepsilon)$. In turn, we define $\varepsilon_n:=\inf\{\varepsilon>0:n_{\varepsilon}<n\}$ with $\varepsilon_n\rightarrow0$ as $n\rightarrow\infty$. In this way, we construct an upper bound for the platform's maximal revenue guarantee
\begin{equation*}
    R^*_n\leq\sum_{k=1}^{k_0}\mu_n^k \frac{\theta^k-\E_n[\theta|\theta\leq\theta^k]}{\Pr_n(\theta^k|\theta\leq\theta^k)}\ln\frac{1}{1-\Pr_n(\theta^k|\theta\leq\theta^k)}+\left(\mu_n^{k_0}\varepsilon_n+\Prob_n(\theta<v_0)\right)(v_0-c).
\end{equation*}
The latter part above converges to 0 as $n\rightarrow\infty$ since $\varepsilon_n$ and $\Prob_n(\theta<v_0)\rightarrow0$. In the former summation above, for each $\theta^k\geq\theta^{k_0}$, as $n \to \infty$, we know that $\Pr_n(\theta^k|\theta\leq\theta^k)\to 1$ if $k=k_0$, and 0, otherwise. Therefore, for the prior type $\theta^{k_0}$, we have the following
\begin{align*}
     \mu_n^{k_0} \frac{\theta^{k_0}-\E_n[\theta|\theta\leq\theta^{k_0}]}{\Pr_n(\theta^{k_0}|\theta\leq\theta^{k_0})}&\ln\frac{1}{1-\Pr_n(\theta^{k_0}|\theta\leq\theta^{k_0})}
     \leq\big( \theta^{k_0}-\E_n[\theta|\theta\leq\theta^{k_0}] 
 \big)\ln\frac{1}{1-\Pr_n(\theta^{k_0}|\theta\leq\theta^{k_0})}  \\
 &\leq \theta^{k_0} ( 1-\Prob_n(\theta^{k_0}|\theta\leq\theta^{k_0}) ) \ln\frac{1}{1-\Pr_n(\theta^{k_0}|\theta\leq\theta^{k_0})} 
 \overset{n\to \infty}{\longrightarrow} 0,
\end{align*}
Where the first inequality uses $\mu_n^k=\Pr_n(\theta^k)\leq\Pr_n(\theta^k|\theta\leq\theta^k)$; the second inequality applies that when $n$ is large and thus $\Pr_n(\theta^{k_0}|\theta\leq\theta^{k_0})\approx1$, we have $\E_n[\theta|\theta\leq\theta^{k_0}]\leq\theta^{k_0}\approx\theta^{k_0}\Pr_n(\theta^{k_0}|\theta\leq\theta^{k_0})$; the last step relies on the fact that for $x>0$, $x\ln x$ converges to 0 as $x\rightarrow0$. Next, for each $\theta^k>\theta^{k_0}$
\begin{align*}
     \mu_n^k \frac{\theta^k-\E_n[\theta|\theta\leq\theta^k]}{\Pr_n(\theta^k|\theta\leq\theta^k)}&\ln\frac{1}{1-\Pr_n(\theta^k|\theta\leq\theta^k)}
     \leq\big( \theta^k-\E_n[\theta|\theta\leq\theta^k] 
 \big)\ln\frac{1}{1-\Pr_n(\theta^k|\theta\leq\theta^k)}  \\
 &\approx (\theta^k-\theta^{k_0}) \ln\frac{1}{1-\Pr_n(\theta^k|\theta\leq\theta^k)} 
 \overset{n\to \infty}{\longrightarrow} 0,
\end{align*}
Where the second step utilizes $\E[\theta|\theta\leq\theta^k]\approx\theta^{k_0}$ for a large $n$, and the last step uses $\ln 1=0$. By further noticing $R_n^*\geq0$, at this point, we complete the proof.

\begin{comment}
\subsection{Proof of Corollary \ref{corollary-blackwell dominance MRG}}
\begin{proof}
    In the proof of Proposition \ref{proposition-joint design}, we show that, given a signal $\pi$, the choice of menu profile is equivalent to the same choice in the main model where the types are essentially the posterior means, namely for all $s\in S$. Using this observation we can proof the corollary via Proposition  \ref{proposition-joint design}. 
    
    To do so, one simply regards the posterior distribution induced by $\pi_1$ as the new prior distribution and sees the garbling $\sigma$ as the new signal, so the new posterior distribution coincides with the posterior distribution induced by $\pi_2$. We show in the proof of Proposition \ref{proposition-joint design} that the full-revealing signal makes the platform strictly better off than any other signal not Blackwell equivalent to it does.
\end{proof}
\end{comment}

\subsection{Proof of Proposition \ref{proposition-comparative}}

\emph{Proof.} Crucially, note that the result is equivalent to saying if the model has the state space $\Theta$, then the full-revealing signal structure generates a higher maximal revenue guarantee than the garbling $h$ does.
%, and it is strictly higher if $h$ produces a posterior distribution that is different from the prior on the region $[c,\infty)$.
Hence, it suffices to show that in any model with state space $\Omega$, the signal structure $(\Theta,\pi)$ that maximizes $R^*(\Theta,\pi)$ is the fully revealing one. %(with the strict relation above).

With everything above, we note that Theorem \ref{main result} relates every signal to a sequential self-conquering path that generates a robustly menu profile. Notice that every nondecreasing alternating path induces a quantile-skepticism value (\ref{quantile-skepticism value}), so for every signal, we can pin down such a function $v^N:[0,Q_M]\rightarrow\mathbb{R}$, where $Q_M$ is the total ex-ante disclosure probability induced by this path. For convenience, we extend $v^N$ to $[0,1]$ by continuing the sequential self-conquering process for all inefficient types ($\theta^k\leq c$). Note that the extended part of $v^N$ is no greater than $\E[\theta|\theta\leq c]\leq c$, which will not influence revenue (see below). Therefore, (\ref{upper bound-reformulate}) illustrates that the platform's revenue robustly induced by this path is
\[\overline{R}(\Theta,\pi)-\int_0^1\max\{v^N(Q)-c,0\}dQ.\]
On the one hand, we know the full surplus $\overline{R}(\Theta,\pi)$ is maximized by the fully revealing signal structure because it allows production and sale to happen if and only if product quality is higher than cost. On the other hand, we will show below that the full-revealing signal structure generates the pointwise lowest $v^N$, so the rents given to producer is minimized. These two claims above suffice to deliver the result.

To show the second claim above that full-revealing minimizes $v^N$ pointwise, we fix any $(\Theta,\pi)$ and identify a necessary condition that the induced quantile-skepticism value needs to satisfy, and therefore construct a non-signal-specific pointwise lower bound for all such quantile-skepticism values. Finally, we see that this lower bound is achieved under the full-revealing signal structure.

Given $(\Theta,\pi)$, we explicitly write down the induced quantile-skepticism value. For every $Q\in[0,1]$, let $k'$ be such that $\sum_{k=1}^{k'-1}\mu^k_0\leq Q\leq\widetilde{Q}:=\sum_{k=1}^{k'}\mu^k_0$. Then, the sequential self-conquering strategy gives $v^N(Q)=\E[\theta|\mu^N_Q]$ where the belief $\mu^N_Q\in\Delta(\Theta)$ is given by, for all $k$:
\begin{equation}\label{B-quantile-skepticism belief}
    \mu_Q^N(\theta^k)=
    \begin{cases}
        \frac{\mu^k_0}{(\widetilde{Q}-Q)+\sum_{j=k'+1}^{|S|}\mu^j_0}=\frac{\mu^k_0}{1-Q} & \text{if }k>k'; \\
        \frac{\widetilde{Q}-Q}{(\widetilde{Q}-Q)+\sum_{j=k'+1}^{|S|}\mu^j_0}=\frac{\widetilde{Q}-Q}{1-Q} & \text{if }k=k'; \\
        0 & \text{if }k<k'.
    \end{cases}
\end{equation}
In other words, (\ref{B-quantile-skepticism belief}) says that if the total ex-ante induced probability is $Q$, then all types greater than $\theta^{k'}$ fully disclose, all types lower than $\theta^{k'}$ never disclose, and type $\theta^{k'}$ discloses with probability $1-(\widetilde{Q}-Q)/\mu^{k'}_0$. We observe a necessary condition for the quantile-skepticism value: $(1-Q)\mu^N_Q\leq\mu_0$. We can translate these beliefs over types into beliefs over states, given by for all $i=1,2,...,N$:
\begin{equation*}
    \begin{aligned}
        \nu_0(\omega^i)&:=\sum_{k=1}^K\frac{\nu^i_0\pi(\theta^k|\omega^i)}{\mu^k_0}\mu_0^k=\nu^i_0; \\
        \nu^N_Q(\omega^i)&:=\sum_{k=1}^K\frac{\nu^i_0\pi(\theta^k|\omega^i)}{\mu^k_0}\mu^N_Q(\theta^k).
    \end{aligned}
\end{equation*}
As a result, we obtain the desired necessary condition: $v^N(Q)=\E[\omega|\nu_Q^N(Q)]$ and $\nu^N_Q\leq\frac{\nu_0}{1-Q}$, which is not related to the signal structure since the state space $\Omega$ and the associated prior $\nu_0$ are fixed. Rewriting this, we have that for every signal and quantile $Q$, there exists a belief $\nu\in\Delta(\Omega)$ such that (i) $v^N(Q)=\E[\omega|\nu]$, and (ii) $\nu\leq\frac{\nu_0}{1-Q}$. Therefore, we have the following pointwise lower-bound function:
\begin{equation}\label{B-lower bound on w^N}
    \begin{aligned}
        \underline{v}^N(Q)=&\min_{\nu\in\Delta(\Omega)}\E[\omega|\nu] \\
        \text{s.t. }&\nu\leq\frac{\nu_0}{1-Q}.
    \end{aligned}
\end{equation}
Clearly, this linear problem allows us to write the optimal ``belief path'' directly, which is similar to (\ref{B-quantile-skepticism belief}). What the solution does is to prioritize filling in probabilities of low states till the constraint becomes binding for each state, and this filling-in process stops when the condition $\nu\in\Delta(\Omega)$ is satisfied. In particular, let $\widehat{i}$ be such that $\sum_{k=1}^{\widehat{i}-1}\nu^i_0\leq Q\leq\overline{Q}:=\sum_{k=1}^{\widehat{i}}\nu^i_0$, and a solution to (\ref{B-lower bound on w^N}) is, for all $i$:
\begin{equation}\label{B-optimal belief}
    \nu_Q^*(\omega^i)=
    \begin{cases}
        \frac{\nu^i}{1-Q} & \text{if }i>\widehat{i}; \\
        \frac{\overline{Q}-Q}{1-Q} & \text{if }i=\widehat{i}; \\
        0 & \text{if }i<\widehat{i},
    \end{cases}
\end{equation}
Which coincides with (\ref{B-quantile-skepticism belief}) once there we set $\Theta=\Omega$ and $\pi$ to be the full-revealing signal. Thus, the full-reveal signal structure and the sequential self-conquering strategy together reach this pointwise lower bound. Repeating our argument before, we have shown the full-revealing signal maximizes $R^*(\Theta,\pi)$.

\subsection{Proof of Proposition \ref{proposition-producer surplus}}

\emph{Proof.} As in the proof of Proposition \ref{proposition-market power}, $\mu_k^N(q)$ is the consumer skepticism belief associated with $w^N_k(q)$. We further define $v_k^N(q):=\E[\theta|\mu_k^N(q)]$ as the associated posterior mean. Here, we only consider types strictly larger than the tipping-point type, so $w_k^N(q)=v_k^N(q)-c$. Since $v_k^N(0)=\E[\theta|\theta\leq\theta^k]<\theta^k$, the Bayes' rule implies that increasing the disclosure probability of $\theta^k$ strictly worsens skepticism, namely $v_k^N(.)$ is strictly decreasing. Also, note that $v_{k}^N(1)=v_{k+1}^N(0)$, so the skepticism posterior mean is strictly decreasing along the sequential self conquering path. Hence, we obtain $v_k^N(q)<v_1^N(0)=v_0$ (and thus $w_k^N(q)<v_0-c$) for all $(k,q)\neq(1,0)$. For $\theta^k>\theta^{k^*}$, $W(\theta^k)+c=\int_0^1v_k^N(q)dq<\int_0^1v_k^N(0)=v_k^N(0)=\E[\theta|\theta\leq\theta^k]$ and $W(\theta^k)+c=\int_0^1v_k^N(q)dq>\int_0^1v_{k+1}^N(0)=v_{k+1}^N(0)=\E[\theta|\theta\leq\theta^{k+1}]$. Moreover, for $k=k^*$, $W(\theta^{k^*})=\int_0^{q^*}[v_{k^*}^N(q)-c]dq\leq\int_0^{q^*}[v_{k^*}^N(0)-c]dq\leq v_{k^*}^N(0)-c=\E[\theta|\theta\leq\theta^{k^*}]-c$. Eventually, $W(\theta^k)=0$ for all $\theta^k<\theta^{k^*}$. At this point, one can see the result from all the above.

\newpage

\section{Online Appendix}

\subsection{Random Menus}

\cite{halr2021} point out that introducing contractual uncertainty may be able to improve the principal's payoff guarantee when contracting in the presence of externalities. This section illustrates that, in our framework, however, making menus random does not improve the platform's revenue guarantee. This contends that our platform wields pricing power that is sufficient for extracting surplus maximally. On the other hand, the format of random menus provides an alternative implementation of the robustly optimal menu profile: the platform offers a type-dependent random price for full disclosure alone. This implementation may be more applicable to settings in which the platform finds it difficult to commit to partial disclosure. For example, certification agencies only decide whether to issue certification for the buyer; whereas, the disclosure decision is within the buyer's discretion.
%; Also, an advertising platform may find it difficult to precisely implement an interior probability in the presence of, for instance, decentralized information dissemination among consumers whereas full disclosure may be easier to achieve.

By introducing uncertainty in menu offers, we refer to the following setup. The platform still chooses a profile of type-dependent menus, but now each menu is potentially stochastic. In particular, let $\mathcal{M}^k$ be the set of all deterministic menus of the form (\ref{menu profile}), and let $\Delta(\mathcal{M}^k)$ collect all the probability distributions over deterministic menus. The platform now chooses a \emph{random menu profile} $\widetilde{M}^{\Theta}=(\widetilde{M}^k)_{k=1}^K$. That is, for all $\theta^k\in\Theta$, $\widetilde{M}^k\in\Delta(\mathcal{M}^k)$. We call a menu profile \emph{deterministic} if, for all $\theta^k\in\Theta$, $\widetilde{M}^k$ is supported by a single deterministic menu, and by abusing the notation, we also refer to $\widetilde{M}^k$ as its deterministic support. Finally, given a random menu $\widetilde{M}$, let $\E_{M\sim\widetilde{M}}[.]$ denote the expectation regarding the distribution $\widetilde{M}$, where $M$ denotes a generic menu realization.

To keep the presentation compact, the timing is stated according to the equivalent model where the state space is $\Theta$ and the signal is full-revealing (see Section \ref{simplifying disclosure equilibria}). First, the platform publicly chooses a random menu profile. Next, importantly, both a type $\theta^k$ and a deterministic menu $M^k$ are realized, with the latter realized from $\widetilde{M}^k$ the menu exclusive to the type. The producer observes her type and picks a plan from the realized menu. Disclosure, price payment, belief updating, and trading happen accordingly. While both the platform and the producer observe everything, the consumer cannot observe the type, the realized menu, or the producer's choice.

Given a random menu profile, a disclosure equilibrium consists of the producer's choices, contingent on both type and menu realization, and the consumer's skepticism belief. We say two disclosure equilibria (induced by potentially different menu profiles) are \emph{outcome-equivalent} if (i) they have the same skepticism belief; and (ii) conditional on each type, they induce identical expected payoffs for the platform and the producer. We say two random menu profiles are \emph{outcome-equivalent} if, for every equilibrium induced by either profile, there is an outcome-equivalent equilibrium induced by the other profile.

\begin{definition}\label{determinized menu profile}
    \emph{For every random menu profile $\widetilde{M}^{\Theta}$, its }determinized menu profile $M^{\Theta}$\emph{ is deterministic and given by the following. For all $\theta^k\in\Theta$, $M^k$ contains the plan $(q,p)$ if and only if there exist $w\geq0$ and two functions of menus $q^k$ and $p^k$ such that (i) their expectations $\E_{M\sim\widetilde{M}^k}[(q^k(M),p^k(M)]$ equal $(q,p)$; and (ii) $(q^k(M),p^k(M))$ is type $k$'s optimal choice in each menu $M$ given skepticism belief $w$, namely, $(q^k(M),p^k(M))\in\arg\max_{(q',p')\in M}q'\theta^k+(1-q')w-p'$.}
\end{definition}

In words, given each random menu, the associated determinized menu contains all the expectation of plans that the producer would choose contingent on different menu realizations, if the skepticism value had been some $w\geq0$. We then establish that every random menu profile is outcome-equivalent to its determinized menu profile

%In words, the determinized menu profile contains all plans that are the expectation of optimal plans in the supporting menus when fixing some skepticism belief. We then establish that every random menu profile is outcome-equivalent to a deterministic menu profile

\begin{theorem*}[\ref{proposition-random menus}]
    Every random menu profile is outcome-equivalent to its determinized menu profile.
\end{theorem*}

The key to such equivalence is the fact that when computing the skepticism belief, only each type's expected disclosure probability matters. An immediate implication of Proposition \ref{proposition-random menus} is that it suffices to achieve the maximal revenue guarantee by using deterministic menu profiles, as in the main model.

Nonetheless, allowing for random menus enriches the possible forms of robustly optimal menu profiles. One particularly interesting case is that the platform only provides full-disclosure but with a type-dependent random price. To imagine how this format can correspond to a useful implementation, consider a platform that publicly sets a single price for guaranteeing that the consumer is informed while privately offering dispersed discounts to all producers.
%These discounts must follow a pre-committed distribution within each producer type group.
The following delivers such a robustly optimal random menu profile

\begin{theorem*}[\ref{proposition-bang bang price dispersion}]
    A robustly optimal random menu profile $\widetilde{M}^{*\Theta}$ is supported by deterministic menus $M_p=\{(0,0),(1,p)\}$, and each random menu $\widetilde{M}^k$ induces a cumulative distribution $H^k$ of price $p$:
    
    (i) if $\theta^k>\theta^{k^*}$, $H^k(p)=({p_k^*}')^{-1}(p)$ for all $p\in[{p_k^*}'(0),{p_k^*}'(1)]$;

    (ii) if $\theta^k=\theta^{k^*}$, $H^{k^*}(p)=\begin{cases}
        ({p_{k^*}^*}')^{-1}(p) &\text{ if }p\in[{p_{k^*}^*}'(0),\theta^{k^*}-c) ;\\
        \text{a jump to 1} &\text{ at }p=\theta^{k^*}-c;
    \end{cases}$

    (iii) if $\theta^{k^*}>\theta^k>c$, $H^k$ assigns all probability to $\theta^k-c$;

    (iv) if $\theta^k\leq c$, $H^k$ assigns all probability to 0.
\end{theorem*}

Proposition \ref{proposition-bang bang price dispersion} is an immediate corollary of Proposition \ref{proposition-random menus} because the robustly optimal menu profile $M^{*\Theta}$ given in Theorem \ref{main result} is exactly the determinized menu profile of $\widetilde{M}^{*\Theta}$. As an illustration, consider the binary-type case discussed in Section \ref{binary type example}. The associated optimal random menu for the high type is represented by the following price distribution:
\begin{equation}\label{binary price dispersion}
    H_{\text{binary}}(p)=\frac{1}{\overline{\mu}}\left(1-\frac{\overline{\theta}-\E[\theta]}{p}\right)\text{, for all }p\in\left[\overline{\theta}-\E[\theta],\overline{\theta}-\underline{\theta}\right].
\end{equation}

Recall that, in the benchmark where the platform does not account for strategic uncertainty (Section \ref{benchmark without strategic uncertainty}), the platform offers the high type a single plan that prices full disclosure at $\overline{\theta}-\underline{\theta}$. Comparing this with $H_{\text{binary}}(.)$, we learn that taking into consideration strategic uncertainty gives rise to downward price dispersion since $\overline{\theta}-\E[\theta]<\overline{\theta}-\underline{\theta}$. 
The intuition of why $H_{\text{binary}}$ also induces a unique equilibrium where the high type fully discloses resembles the logic behind the self-conquering process. First, suppose we are in the bad equilibrium where nobody discloses, the skepticism belief must equal the prior. However, the sufficiently low price $\overline{\theta}-\E[\theta]$ occurs with a positive probability so that the high type deviates to disclose, breaking the equilibrium. Likewise, the consumer believing that the high type discloses with a higher probability worsens his skepticism, reducing the producer's outside value of not disclosing. In this way, the platform is able to induce more disclosure while charging a slightly higher price. By iteratively adding increasing price tiers, the platform gradually lures the high type to play full disclosure. \\

\emph{Proof of Proposition \ref{proposition-random menus}}
Let $\widetilde{M}^{\Theta}$ and $M^{\Theta}$ be a random menu profile and its determinized menu profile, respectively. First, we take any equilibrium $\widetilde{g}$ induced by the random menu profile. In $\widetilde{g}$, denote the skepticism belief by $\widetilde{\mu}^N$, and let the producer's choice when type $\theta^k$ and menu $M$ are realized be $(\widetilde{q}^k_M,\widetilde{p}^k_M)$. Further set $\widetilde{q}^k=\E_{M\sim\widetilde{M}^k}[\widetilde{q}^k_M]$ and $\widetilde{p}^k=\E_{M\sim\widetilde{M}^k}[\widetilde{p}^k_M]$. Then, we show that, given $M^{\Theta}$, the producer strategy profile $(\widetilde{q}^k,\widetilde{p}^k)_{k=1}^K$ and the skepticism belief $\widetilde{\mu}^N$ form an outcome-equivalent equilibrium of $\widetilde{g}$. By construction, $(\widetilde{q}^k,\widetilde{p}^k)\in M^k$ for all $k$. The skepticism value is $\widetilde{w}^N=\max\{\E[\theta|\widetilde{\mu}^N]-c,0\}$. Next, for all $k$, take any $(q,p)\in M^k$, and by construction of $M^k$, there are $q^k(M)$ and $p^k(M)$ such that $(q^k(M),p^k(M))\in M$, $q=\E_{M\sim\widetilde{M}^k}[q^k(M)]$, and $p=\E_{M\sim\widetilde{M}^k}[p^k(M)]$. Then, $(\widetilde{q}^k,\widetilde{p}^k)$ is optimal among plans in $M^k$ since
\begin{equation}
    \begin{aligned}
        q\max\{\theta^k-c,0\}&+(1-q)\widetilde{w}^N-p \\
        &=\E_{M\sim\widetilde{M}^k}[q^k(M)\max\{\theta^k-c,0\}+(1-q^k(M))\widetilde{w}^N-p^k(M)] \\
        &\leq\E_{M\sim\widetilde{M}^k}[\widetilde{q}_M^k\max\{\theta^k-c,0\}+(1-\widetilde{q}_M^k)\widetilde{w}^N-\widetilde{p}_M^k] \\
        &=\widetilde{q}^k\max\{\theta^k-c,0\}+(1-\widetilde{q}^k)\widetilde{w}^N-\widetilde{p}^k.
    \end{aligned}
\end{equation}
Moreover, the belief $\widetilde{\mu}^N$ assigns the following probability to type $\theta^k$:
\begin{equation}\label{A-bayes rule random deterministic}
    \widetilde{\mu}^N_k=\frac{\mu_0^k\E_{M\sim\widetilde{M}^k}[1-\widetilde{q}_M^k]}{\sum_{j=1}^K\mu_0^j\E_{M\sim\widetilde{M}^j}[1-\widetilde{q}_M^j]}=\frac{\mu_0^k(1-\widetilde{q}^k)}{\sum_{j=1}^K\mu_0^j(1-\widetilde{q}^j)}.
\end{equation}
Hence, $\widetilde{\mu}^N$ is consistent with the Bayes' rule under profile $(\widetilde{q}^k,\widetilde{p}^k)_{k=1}^K$, and the producer's type-dependent choices are optimal given value $\widetilde{w}^N$, which gives us an equilibrium denoted by $g$. One can easily see $g$ and $\widetilde{g}$ are outcome-equivalent since they have the same skepticism belief and $\widetilde{p}^k=\E_{M\sim\widetilde{M}^k}[\widetilde{p}^k_M]$.

Second, we take any equilibrium $g$ induced by the deterministic menu profile. In $g$, denote the skepticism belief by $\mu^N$ and the producer choices by $(q^k,p^k)_{k=1}^K$. The skepticism value is $w^N=\max\{\E[\theta|\mu^N]-c,0\}$. By the construction of $M^k$, there are $w^k\geq0$ and random variables $q^k(M)$ and $p^k(M)$ such that $q^k=\E_{M\sim\widetilde{M}^k}[q^k(M)]$, $p^k=\E_{M\sim\widetilde{M}^k}[p^k(M)]$, and $(q^k(M),p^k(M))\in\arg\max_{(q',p')\in M}q'\theta^k+(1-q')w^k-p'$. Suppose $w^k$ cannot be $w^N$. Let $S^k_M=\arg\max_{(q',p')\in M}q'\theta^k+(1-q')w^N-p'$ for all $k$ and $M$. Then, we must have $\Pr_{M\sim\widetilde{M}^k}((q^k(M),p^k(M))\notin S_M^k)>0$ since, otherwise, we can have $w^k=w^N$, contradicting the supposition. We thus can construct a new plan $(q,p)=(\E_{M\sim\widetilde{M}^k}[q_M],\E_{M\sim\widetilde{M}^k}[p_M])$ where $(q_M,p_M)\in S_M^k$, so $(q,p)\in M^k$. Hence, given $M^k$ and $w^N$, type $\theta^k$ strictly prefers $(q,p)$ to $(q^k,p^k)$, a contradiction to $(q^k,p^k)$ being the equilibrium choice. Therefore, we can set $w^k=w^N$. Now, we show the producer contingent choices $(q^k(M),p^k(M))_{k,M}$ and the skepticism belief $\mu^N$ compose an equilibrium induced by $\widetilde{M}^{\Theta}$. The Bayes consistency is again given by (\ref{A-bayes rule random deterministic}), and the optimality of contingent producer choices is by the construction of each $(q^k(M),p^k(M))$ and the fact that $w^k=w^N$. Also, the outcome-equivalency is easy to see. We thereby complete the proof.

\subsection{Rationalizable Outcomes}\label{rationalizable outcomes formal}

In this appendix we present the formal definitions and analysis for Section \ref{iterated deletion of dominated strategies}.
In any equilibrium of the producer--consumer game defined in Section \ref{iterated deletion of dominated strategies}, the consumer's strategy takes the following form:
\begin{gather*}
   \begin{cases}
       a(\theta^k,p) &= 1_{\theta^k \geq p}, \quad \forall \theta^k\in\Theta; 
       \\
       a(N,p) & = 1_{\E[\theta|\mu^N] \geq p}, 
   \end{cases} 
\end{gather*}
Where $\mu^N$ is the consumer's skepticism belief defined by (\ref{bayes}).\footnote{Here we assume the consumer breaks ties in favor of the producer to simplify the exposition. This does not matter as the producer can break the tie with $\varepsilon$ perturbation.} Note that this class of strategy is parameterized by a single variable $v^N:=\E[\theta|\mu^N]$. We denote this strategy as $a^{v^N}$. The notation $v^N$ is abused a little bit as it is also known as the quantile-skepticism value (\ref{quantile-skepticism value}), which also refers to the consumer's skepticism posterior mean.

For any producer strategy $r^{\Theta}=(q^{\Theta},p^{\Theta})$, the consumer's best response is exactly $a^{v^N}$ where $v^N=\E[\theta|\mu^N(q^{\Theta})]$. This means any consumer action that does not take the form of $a^v$ cannot be a best response to any producer strategy. Therefore, to discuss rationalizable outcomes, the set of consumer strategies that we need to consider is $\{a^v:v\in [\theta^K,\theta^1]\}$, where $\theta^K$ and $\theta^1$, respectively, refer to the lower and upper bounds of a skepticism posterior mean. We identify the strategy $a^v$ simply with the threshold $v$. Formally, given a menu profile $M^{\Theta}$, we define
\begin{equation}
    \Sigma^P_0:=\{r^{\Theta}:r^k\in M^k\text{ for all }k\}; \qquad \Sigma^C_0:=[\theta^K,\theta^1].
\end{equation}
These two sets constitute the ``strategy space'' of the producer and the consumer in a simultaneous move game. Lastly, $u(r^{\Theta},v^N)=\sum_{k=1}^K\mu_0^k\left[q^k\max\{\theta^k-c,0\}+(1-q^k)\max\{v^N-c,0\}-p^k\right]$ is the utility function of the producer. To avoid unnecessary discussion, we consider $c>\theta^{\underline{k}+1}$, which means we do not have a type that equals exactly $c$ and there is some type strictly lower than $c$.

\begin{definition}
    \emph{The process of \emph{iterated deletion of strictly dominated strategies} proceeds as follows. For $n\geq1$, define $\Sigma_n^P$ and $\Sigma_n^C$ recursively by
    \begin{equation}
        \begin{aligned}
            \Sigma_n^P&=\{r^{\Theta}\in\Sigma^P_{n-1}:\text{there is no }\widetilde{r}^{\Theta}\in\Sigma^P_{n-1}\text{ such that }u(\widetilde{r}^{\Theta},v^N)>u(r^{\Theta},v^N)\text{ for all }v^N\in\Sigma^C_{n-1}\}; \\
            \Sigma_n^C&=\{v^N\in\Sigma^C_{n-1}:v^N=\E[\theta|\mu^N(q^{\Theta})]\text{ for some }r^{\Theta}\in\Sigma^P_{n-1}\}.
        \end{aligned}
    \end{equation}
    Define $\Sigma_{\infty}^P=\cap_{n=0}^{\infty}\Sigma_n^P$ and $\Sigma_{\infty}^C=\cap_{n=0}^{\infty}\Sigma_n^C$. These are the sets of producer and consumer strategies that survive the iterated deletion of strictly dominated strategies.}
\end{definition}

Since the robustly optimal menu profile $M^{*\Theta}$ given by Theorem \ref{main result} is defined as the limit of a sequence of profiles, to deal with strict dominance, we define (for small $\varepsilon>0$) the \emph{$\varepsilon$-optimal menu profile} $M^{*\Theta}_{\varepsilon}$ as follows: For all $k$, $M^{*k}_{\varepsilon}=\{(q,p-\varepsilon q):(q,p)\in M^{*k}\}$. In other words, we focus on the approximating sequence in which the prices are slightly lower than in the optimal solution. The following proposition establishes that given the $\varepsilon$-optimal menu profile, the survivors of iterated strict dominance are given by the unique equilibrium induced by the menu profile:

\begin{theorem*}[\ref{proposition-iterated strict dominance}]
    Given $M^{*\Theta}_{\varepsilon}$, $\Sigma^P_{\infty}$ contains a single strategy where all types with $\theta^k>c$ fully disclose while all types with $\theta^k\leq c$ have no disclosure, and $\Sigma^C_{\infty}$ contains a single value of $\E[\theta|\theta\leq c]$.
\end{theorem*}

By definition, the maximal revenue guarantee among all equilibria naturally serves as an upper bound on the revenue guarantee among all rationalizable outcomes. Now that Proposition \ref{proposition-iterated strict dominance} has demonstrated that the menu profile achieves this upper bound by inducing a unique rationalizable outcome, our menu profile remains optimal under a different robust criterion. \\

\emph{Proof of Proposition \ref{proposition-iterated strict dominance}.} To start with, note that all inefficient types find it dominant to choose $(0,0)$, namely all producer strategies $r^{\Theta}$ with $q^k>0$ for some $\theta^k<c$ will be soon deleted in the process, so below we focus on $\theta^k\geq c$ (i.e., only $k\leq\underline{k}$). We take any strategy $r^{\Theta}=(q^k)_{k=1}^K\in\Sigma_0^P$ and show that, regardless of consumer skepticism, the producer obtains strictly higher expected utility with another strategy profile of the form $\widetilde{r}^{\Theta}=(\widetilde{q}^k)_{k=1}^K=(1,...,1,q,0,...,0)$. To construct $\widetilde{r}^{\Theta}$, we find $j\in\Theta$ and $q\in[0,1]$ such that $\sum_{k=1}^K\mu_0^kq^k=\sum_{k=1}^K\mu_0^k\widetilde{q}^k=\sum_{k=1}^{j}\mu_0^k+\mu_0^{j+1}q$. Thus, the two profiles induce the same total ex-ante probability. Note that $\sum_{k=1}^{j}\mu_0^k+\mu_0^{j+1}q$ ranges from 0 ($j=0$ and $q=0$) to 1 ($j=K$), so such a $\widetilde{r}^{\Theta}$ always exists. If $r^{\Theta}=\widetilde{r}^{\Theta}$, we are done. We then consider $r^{\Theta}\neq\widetilde{r}^{\Theta}$, which implies that there is some $k^r$ such that $\widetilde{q}^k\geq q^k$ for all $k\leq k^r$, $\widetilde{q}^k\leq q^k$ for all $k>k^r$, and there must be some inequalities being strict. Recall that the menu for $\theta^k$ in the $\varepsilon$-optimal menu profile has the price envelope $p_k(q)=q(\theta^k-c)-\int_0^qw_k^N(s)ds-\varepsilon q$ where $w_k^N(.)$ is defined by (\ref{w^N_k(q)}). Now, we take any $v^N\in\Sigma_0^C$ and let the associated skepticism value be $w^N=\max\{v^N-c,0\}$, so the expected utility is improved by:
\begin{equation}
    \begin{aligned}
        u(&\widetilde{r}^{\Theta},w^N)-u(r^{\Theta},w^N) \\
        &=\sum_{k=1}^{\underline{k}}\mu_0^k(\widetilde{q}^k(\theta^k-c)+(1-\widetilde{q}^k)w^N-p_k(\widetilde{q}^k))-\sum_{k=1}^{\underline{k}}\mu_0^k(q^k(\theta^k-c)+(1-q^k)w^N-p_k(q^k)) \\
        &=\sum_{k=1}^{\underline{k}}\mu_0^k\left((1-\widetilde{q}^k)w^N+\int_0^{\widetilde{q}^k}w_k^N(s)ds+\varepsilon\widetilde{q}^k\right)-\sum_{k=1}^{\underline{k}}\mu_0^k\left((1-q^k)w^N+\int_0^{q^k}w_k^N(s)ds+\varepsilon q^k\right) \\
        &=\sum_{k=1}^{k^r}\mu_0^k\int_{q^k}^{\widetilde{q}^k}w_k^N(s)ds-\sum_{k=k^r+1}^{{\underline{k}}}\mu_0^k\int_{\widetilde{q}^k}^{q^k}w_k^N(s)ds+(\varepsilon-w^N)\left(\sum_{k=1}^{\underline{k}}\mu_0^k\widetilde{q}^k-\sum_{k=1}^{\underline{k}}\mu_0^kq^k\right) \\
        &=\sum_{k=1}^{k^r}\mu_0^k\int_{q^k}^{\widetilde{q}^k}w_k^N(s)ds-\sum_{k=k^r+1}^{{\underline{k}}}\mu_0^k\int_{\widetilde{q}^k}^{q^k}w_k^N(s)ds \\
        &>\sum_{k=1}^{k^r}\mu_0^k\int_{q^k}^{\widetilde{q}^k}w_{k^r}^N(1)ds-\sum_{k=k^r+1}^{{\underline{k}}}\mu_0^k\int_{\widetilde{q}^k}^{q^k}w_{k^r}^N(1)ds \\
        &=w^N_{k^r}(1)\left(\sum_{k=1}^{\underline{k}}\mu_0^k\widetilde{q}^k-\sum_{k=1}^{\underline{k}}\mu_0^kq^k\right)=0.
    \end{aligned}
\end{equation}
Here, the second equality replaces the price functions with the previous integral forms. The fourth and the last equalities use the construction assumption that the new strategy $\widetilde{r}^{\Theta}$ preserves the total ex-ante probability. The inequality exploits the fact that $w_k^N(q)$ is strictly decreasing in both $k$ and $q$. As a result, only those producer strategies where the producer fully [never] discloses if her type is higher [lower] than a threshold can survive the first round of deletion.

Next, we characterize the skepticism beliefs that survive the second round. Note that every producer strategy $r^{\Theta}$ surviving the first round is monotonic, namely $q^{k_1}\geq q^{k_2}$ whenever $\theta^{k_1}>\theta^{k_2}$. Thus, we now show that the skepticism belief $\mu^N$ given by the Bayes' rule (\ref{bayes}) with such a monotonic $r^{\Theta}$ must be weakly first-order stochastically dominated by $\mu_0$, namely $\frac{\sum_{k=j}^K\mu_0^k(1-q^k)}{\sum_{k=1}^K\mu_0^k(1-q^k)}\geq\sum_{k=j}^K\mu_0^k$ for all types $j$. We first observe that this is true for $j=K$ and $j=1$,
\begin{equation}
    \begin{aligned}
        \frac{\mu_0^K(1-q^K)}{\sum_{k=1}^K\mu_0^k(1-q^k)}\geq\frac{\mu_0^K(1-q^K)}{\sum_{k=1}^K\mu_0^k(1-q^K)}&=\mu_0^K; \\
        \frac{\sum_{k=1}^K\mu_0^k(1-q^k)}{\sum_{k=1}^K\mu_0^k(1-q^k)}=1&=\sum_{k=1}^K\mu_0^k.
    \end{aligned}
\end{equation}
Thus, suppose the dominance condition does not hold, then we can find two types $k_1>k_2$ such that:
\begin{equation}
    \frac{\mu_0^{k_1}(1-q^{k_1})}{\sum_{k=1}^K\mu_0^k(1-q^k)}<\mu_0^{k_1}\text{ and }\frac{\mu_0^{k_2}(1-q^{k_2})}{\sum_{k=1}^K\mu_0^k(1-q^k)}\geq\mu_0^{k_2}.
\end{equation}
However, this implies $1-q^{k_1}<\sum_{k=1}^K\mu_0^k(1-q^k)\leq1-q^{k_2}$, which violates monotonicity $q^{k_1}\leq q^{k_2}$. Thus, $\mu^N$ is weakly first-order stochstic dominated by $\mu_0$, and an immediate implication is that the skepticism value $v^N=\E[\theta|\mu^N]$ is no greater than the prior mean $v_0$. As a result, $\Sigma_1^C\subseteq[\theta^K,v_0]$.

After this, we show that in the following rounds, the producer strategies are deleted in a sequential self-conquering manner until the process hits the tipping point $(k^*,q^*)$. From above, we know that for every $n\geq3$, the largest skepticism posterior mean surviving round $n-1$, $\overline{v}^N:=\sup\Sigma_n^C$, is no greater than $v_0$. In fact, for every such $\overline{v}^N$, let the skepticism value be $\overline{w}^N:=\max\{\overline{v}^N-c,0\}$, so there is some $\overline{k}$ and $\overline{q}$ such that $w_{\overline{k}}^N(\overline{q})=\overline{w}^N$. Thus, the fact that $\overline{w}^N$ is the largest surviving value implies all the strategies of the form $(1,...,1,q',0,...,0)$ with $k'-1$ ones, and either $k'<\overline{k}$, or $k'=\overline{k}$ yet $q'<\overline{q}$ must already be deleted because $w_{k'}^N(q')>\overline{w}^N$. Now, we show the existence of some small $\delta>0$ such that, in this round $n$, the strategies where type $\theta^{\overline{k}}$ chooses probability $q\in[\overline{q},\overline{q}+\delta)$ are strictly dominated by choosing probability $\overline{q}+\delta$. If $\overline{q}=1$, then the same argument below also works for showing that $\delta$ dominates type $\theta^{\overline{k}+1}$'s strategies on $[0,\delta)$. One can see this from the construction of the price function. That is, for all $w^N\leq\overline{w}^N$, type $\theta^{\overline{k}}$ deviating from $\overline{q}$ to some $\overline{q}+\delta$ increases her payoff by:
\begin{equation}\label{marginal deletion}
    (\theta^{\overline{k}}-c-\overline{w}^N)\delta-(p_{\overline{k}}(\overline{q}+\delta)-p_{\overline{k}}(\overline{q}))={p^*_{\overline{k}}}'(\overline{q})\delta-(p^*_{\overline{k}}(\overline{q}+\delta)-p^*_{\overline{k}}(\overline{q}))+\varepsilon \delta\approx \varepsilon \delta+o(\delta^2).
\end{equation}
Here, we again abuse the notation a little and let the price envelop of $M^{*k}$ be $p_k^*$. We use above ${p^*_k}'(q)=\theta^k-c-w^N_k(q)$ and $p_k(q)=p_k^*(q)-\varepsilon q$. Therefore, by choosing $t>0$ to be small, this improvement is strict and the mentioned strategies are deleted. In particular, the second-order Tylor expansion of (\ref{marginal deletion}) yields the following criterion:
\begin{equation}\label{delta criterion}
    \varepsilon\delta-\frac{1}{2}{p^*_{\overline{k}}}''(\overline{q})\delta^2+o(\delta^3)\geq0.
\end{equation}
Thus, due to the strict convexity of prices before the tipping point $(k^*,q^*)$, there exists a strictly positive $\delta^*<\frac{2\varepsilon}{\max_{(k,q):\text{ before }(k^*,q^*)}{p^*_k}''(q)}$ such that (\ref{delta criterion}) is satisfied regardless of $\overline{k}$ and $\overline{q}$. Our analysis above hence suggests that, in every round $n\geq3$, there will be at least ``$\delta^*$ amount of producer's strategies'' being deleted. As a result, within $\frac{k^*-1+q^*}{\delta}+3$ rounds, the deletion process will rule out all producer strategies of the form $(1,...,1,q,0,...,0)$ with $k-1$ ones, where $(k,q)$ is before $(k^*,q^*)$.

Therefore, in the next round, by the construction of $(k^*,q^*)$, all skepticism posterior means that are higher than $c$ will be deleted, so the only skepticism value that survives is 0. And this means that every efficient type $\theta^k$ who has not fully disclosed (i.e., $\theta^{k^*}\geq\theta^k>c$) will find it dominant to fully disclose. In particular, take a producer strategy of the form $(1,...,1,q,0,...,0)$ with $k-1$ ones and $(k,q)$ is after the tipping point, so the marginal price for a higher disclosure probability is now a constant $\theta^k-c-\varepsilon$. Thus, deviating to full disclosure given the skepticism value 0 helps this type to earn $\varepsilon>0$.

In summary, the equilibrium strategy of the producer is the only one surviving the deletion process. Moreover, the unique skepticism posterior mean consistent with this strategy is $\E[\theta|\theta\leq c]$.

At this point, we complete the proof.

\subsection{Disclosure Cost}
In our previous analysis, we assumed that disclosure is costless for the platform. We consider an extension where the platform incurs a probability-dependent cost $c(q)$ when implementing a plan with disclosure probability $q\in[0,1]$. We do not impose any structure on the costs except that $c(.)$ must be continuous to guarantee solution existence. We hence consider the \emph{maximal profit guarantee}:
\begin{equation}
    P^*:=\sup_{M^{\Theta}\in\mathcal{M}}\inf_{g\in\mathcal{E}(M^{\Theta})}\sum_{k=1}^K\mu_0^k(p^k-c(q^k)).
\end{equation}

The following result shows that most features of our main result hold but the robustly induced disclosure outcome now could be partial disclosure:

\begin{proposition}\label{proposition-convex solution with costs} 
    Consider a platform with disclosure cost $c(q)$. There is some probability profile $q^{*\Theta}$ and a robustly optimal menu profile $M^{*\Theta}$ given by, for all $k<K$,
    \begin{equation}\label{multiple type with costs}
        \begin{aligned}
            M^{*k}&=\{(q,\widetilde{p}_k(q)):q\in[0,q^{*k}]\}\text{, where } \\
            \widetilde{p}_k(q)&=\int_0^q\left[\theta^k-w^N(q^{*1},...,q^{*k-1},\widetilde{q},0,...,0)\right]d\widetilde{q}.
        \end{aligned}
    \end{equation}
    Moreover, this profile induces a unique equilibrium where each type $\theta^k$ chooses its maximal plan $(q^{*k},\widetilde{p}_k(q^{*k}))$. The maximal revenue guarantee is therefore $P^*=\sum_{k=1}^{K-1}\mu_0^k\left[\widetilde{p}_k(q^{*k})-c(q^{*k})\right]$.
\end{proposition}

To see this, simply observe that the proof of Lemma \ref{lemma-upper bound} shows that, fixing a robustly induced probability profile, it is optimal to use the sequential self-conquering strategy. In particular, the strategy starts by moving the highest type's disclosure from 0 to $q^{*1}$, and then moving the second-highest type's disclosure from 0 to $q^{*2}$, and so on sequentially for the other types. Therefore, both the revenue and the cost can be written as functions of the probability profile.

Proposition \ref{proposition-convex solution with costs} not only demonstrates that sequential self-conquering remains central despite the presence of costs, but it also highlights that the prediction that producers will face convex advertising prices even when the platform's disclosure technology is nonconvex is robust.

\end{document}